\documentclass[twocolumn]{aastex61}

\newcommand{\uv}{$\{u, \mathrm{v}\}$}

\newcommand{\modif}[1]{#1}
\newcommand{\degree}{\ensuremath{^\circ}}
\usepackage{color}

\received{Ocotber 19, 2017}
\revised{January 18, 2018}
\accepted{January 31, 2018}
\submitjournal{ApJ}

\shorttitle{MWC614}
\shortauthors{Kluska et al.}

\begin{document}
%
\title{A multi-instrument and multi-wavelength high angular resolution study of MWC614: quantum heated particles inside the disk \modif{cavity} \footnotemark[1]}

\footnotetext[1]{Based on observations made 
with the Keck observatory (NASA program ID N104N2) and with ESO telescopes
at the Paranal Observatory (ESO program IDs 073.C-0720, 077.C-0226, 077.C-0521, 083.C-0984, 087.C-0498(A), 190.C-0963, 095.C-0883) and with the CHARA observatory.
}

\correspondingauthor{Jacques Kluska} %
\email{jkluska@astro.ex.ac.uk}

\author[0000-0002-9491-393X]{Jacques Kluska}
\affiliation{Astrophysics Group, School of Physics, University of Exeter, Stocker Road, Exeter, EX4 4QL, UK}

\author[0000-0001-6017-8773]{Stefan Kraus}  
\affiliation{Astrophysics Group, School of Physics, University of Exeter, Stocker Road, Exeter, EX4 4QL, UK}

\author[0000-0001-9764-2357]{Claire L. Davies}
\affiliation{Astrophysics Group, School of Physics, University of Exeter, Stocker Road, Exeter, EX4 4QL, UK}

\author{Tim Harries}
\affiliation{Astrophysics Group, School of Physics, University of Exeter, Stocker Road, Exeter, EX4 4QL, UK}

\author{Matthew Willson}
\affiliation{Astrophysics Group, School of Physics, University of Exeter, Stocker Road, Exeter, EX4 4QL, UK}

\author{John D. Monnier}
\affiliation{Department of Astronomy, University of Michigan, Ann Arbor, MI 48109, USA}

\author{Alicia Aarnio}
\affiliation{University of Colorado Boulder, 3665 Discovery Drive, Boulder, CO 80303 USA}

\author{Fabien Baron}
\affiliation{Department of Physics and Astronomy, Georgia State University, Atlanta, GA, USA}

\author{Rafael Millan-Gabet}
\affiliation{Infrared Processing and Analysis Center, California Institute of Technology, Pasadena, CA, 91125, USA}
\affiliation{NASA Exoplanet Science Institute, California Institute of Technology, 770 S. Wilson Ave., Pasadena, CA, 91125, USA}

\author{Theo ten Brummelaar}
\affiliation{The CHARA Array of Georgia State University, Mount Wilson Observatory, Mount Wilson, CA 91203, USA}

\author{Xiao Che}
\affiliation{Department of Astronomy, University of Michigan, Ann Arbor, MI 48109, USA}

\author{Sasha Hinkley}
\affiliation{Astrophysics Group, School of Physics, University of Exeter, Stocker Road, Exeter, EX4 4QL, UK}

\author{Thomas Preibisch}
\affiliation{Universit\"ats-Sternwarte M\"unchen, Ludwig-Maximilians-Universit\"at, Scheinerstr. 1, 81679, M\"unchen, Germany}

\author{Judit Sturmann}
\affiliation{The CHARA Array of Georgia State University, Mount Wilson Observatory, Mount Wilson, CA 91203, USA}

\author{Laszlo Sturmann}
\affiliation{The CHARA Array of Georgia State University, Mount Wilson Observatory, Mount Wilson, CA 91203, USA}

\author{Yamina Touhami}
\affiliation{Center for High Angular Resolution Astronomy, Department of Physics and Astronomy, Georgia State University, P.O.\ Box 4106, Atlanta, GA 30302-4106, USA}




\begin{abstract}
High angular resolution observations of young stellar objects are required to study the inner astronomical units of protoplanetary disks in which the majority of planets form.
As they evolve, gaps open up in the inner disk regions and the disks are fully dispersed within $\sim$10 Myrs.
\object{MWC~614} is a pre-transitional object with a $\sim$10\,au radius gap.
We present a set of high angular resolution observations of this object including SPHERE/ZIMPOL polarimetric and coronagraphic images in the visible, KECK/NIRC2 near-infrared aperture masking observations and VLTI (AMBER, MIDI, and PIONIER) and CHARA (CLASSIC and CLIMB) long-baseline interferometry at infrared wavelengths.
We find that all the observations are compatible with an inclined disk ($i\sim$55$^\circ$ at a position angle of $\sim$20-30$^\circ$).
The mid-infrared dataset confirms \modif{the disk inner rim to be at} 12.3$\pm$0.4\,au \modif{from the central star}.
\modif{We determined an upper mass limit of 0.34\,M$_\odot$ for a companion inside the cavity.}
Within the \modif{cavity}, the near-infrared emission, usually associated with the dust sublimation region, is unusually extended ($\sim$10\,au, 30 times larger than the theoretical sublimation radius) and \modif{indicates} a high \modif{dust} temperature (T$\sim$1800\,K).
As a possible result of companion-induced dust segregation, quantum heated dust grains could explain the extended near-infrared emission with this high temperature.
Our observations confirm the peculiar state of this object where the inner disk has already been accreted onto the star exposing small particles inside the \modif{cavity} to direct stellar radiation.
\end{abstract}



\keywords{techniques: high angular resolution, interferometers, polarimetric; stars: pre-main sequence,variables: T Tauri, Herbig Ae/Be; object: MWC614}

\section{Introduction}

The key for understanding the observed diversity of planetary systems is hidden in the initial conditions of their formation, which are inherently linked to the physical conditions at play in protoplanetary disks.
These disks have been extensively studied around the \modif{low-mass} T\,Tauri stars using photometry, leading to the classification of objects from full, gas-rich protoplanetary disks to cold debris disks that exhibit only large dust grains and, in some cases, show evidence for an already formed planetary system \citep{Lada1987,Andre1994}.
Between these two evolutionary stages, the primordial disk has $\sim$10 million years to form \modif{giant} planets before its gas is dispersed \citep[e.g.][]{Sicilia2005}\modif{, even though this timescale is debated} \citep{Pfnalzer2014}.
Full disks are characterized by their strong infrared (IR) and millimeter emission, far in excess of what would be expected from pure photospheric emission.
However, some objects display a lack of emission in the near and mid-infrared (NIR and MIR) compared to full disk systems' spectral energy distributions \citep[SED; e.g.][]{Calvet2005}.
These objects have a disk with a large dust-depleted inner cavity that has been spatially resolved with sub-mm interferometry for individual objects \citep[e.g.][]{Andrews2011,vanderMarel2013,Canovas2015,Pinilla2015,vanderMarel2016,Dong2017,Pinilla2017,Sheehan2017} and are called transition disks because they are believed to undergo the process of disc clearing. 
Between the full disk and the transitional disk stage, pre-transitional objects show a depletion in the MIR but still having an excess of NIR emission, likely caused by the presence of an inner disk interior to the cavity, forming a gapped disk \citep[e.g.][]{Espaillat2014,Kraus2017}.

For intermediate-mass, Herbig Ae/Be stars, another classification has been suggested, where ``Group I'' objects have a positive MIR slope with wavelengths (or can be fitted with a $\sim$200\,K black-body function) and ``Group II'' have a negative one \citep{Meeus2001}.
\citet{Meeus2001} interpreted the Group~I/II as an evolutionary sequence, where Group~I objects represent flared disks that evolve into geometrically flat disks with a Group~II-like SED.
Alternatively, \citet{Maaskant2013} proposed another scheme, where a primordial flared disk evolves either into A (Group~I) or B (Group~II).
Accordingly, in this scheme, Group~I disks feature extended gaps, similar to the (pre-)transitional discs.
However it is not clear whether the lack of infrared emission indicates density-depleted gaps caused by planets \citep{Crida2006,Papaloizou2007}, photo-evaporation \citep[e.g.][]{Hollenbach1994}, self-shadowed region \citep[][]{Dullemond2004,Dong2015} \modif{or the dead zone inside the disk \citep{Pinilla2016}}. 
\modif{In some cases, the disk gap is also only partially cleared.  For instance, the gap around the Herbig Ae star \object{V1247\,Orionis} contains considerable amount of optically thin dust grains that still dominate the mid-infrared emission \citep{Kraus2013,Kraus2017}.}

The photo-evaporation mechanism can explain transition disks with small cavities and low accretion rates \citep{Owen2011}.
On the other hand, transition disks with large holes and high accretion rates can not be explained by photo-evaporation alone and planets inside the cavity are potential candidates to sustain the accretion up to the star \citep{Zhu2012,Pinilla2012,Owen2014}.
Interestingly, these planets can also create a pressure bump at the outer disk inner edge where the largest grains ($\sim1$mm) are confined, while the small grains ($\lesssim1\mu$m) can pass through and continue to accrete towards the star.
\modif{Looking for} companion\modif{s} in the disk while the disk is in a dispersal process is therefore important to constrain these theories.

Another feature displayed by gapped disks is a high ionization degree of polycyclic aromatic hydrocarbon (PAH) \citet{Maaskant2014}.
PAHs, and more generally all quantum heated particles (QHPs), can be quantum heated when directly exposed to the ultra-violet (UV) flux from the central star \citep{Purcell1976,Draine2001}.
As such, they can reach a temperature which is higher than the equilibrium temperature at a certain radius from the star.
Recently, \citet{Klarmann2016} found that extended flux seen by long-baseline interferometry in the NIR can be explained by QHPs localized in the disk gap and was demonstrated by reproducing the NIR interferometric observations for \object{HD100453}.

To extend our knowledge on the evolutionary phases of transition disks, we need to focus on some peculiar objects that have just started to clear the inner parts of their disk.
MWC~614 (alias HD~179218) is suggested to be one such object.
It is a Herbig Ae star with 100\,L$_\odot$ \citep{Menu2015}, whose SED was classified as Group~I, consistent with either a flared or a gapped disk.
\modif{Throughout this paper we will use the Gaia distance of 293$^{+34}_{-27}$\,pc \citep{Lindegren2016}.}
The disk was spatially resolved by NIR interferometry showing very extended emission \citep[\textgreater 3.1\,au;][]{Monnier2006} and is the most resolved object in the survey of 51 Herbig AeBe stars observed by VLTI/PIONIER \citep{Lazareff2016}.
The emitting material is much more extended than the theoretical dust sublimation radius predicted for a star with MWC~614's luminosity, a dust sublimation temperature of 1500\,K and grain absorption efficiency (Q$_\mathrm{abs}$) of unity \citep[$\sim$0.35\,au;][]{Monnier2002}.
Emission from the dust sublimation radius has been found to dominate the emission at these wavelengths for other Herbig objects \citep[e.g.][]{Monnier2005}. 
\citet{Fedele2008} studied the \modif{MWC614} disk structure in the MIR continuum using MIR \modif{interferometry with the MIDI instrument on} four baselines. 
They deduced that the dust is mainly located in an outer disk (starting at 14.5\,au) and also in a marginally resolved region (\textless\,3.2\,au).
However, the baselines of these observations do not cover the lower spatial frequencies that set the global size and orientation of the object.
\citet{Fedele2008} interpreted the lack of emission between the two regions could be either due to a gap or a shadow cast by a puffed-up inner disk.

CO emission ($^\mathrm{12}$CO and $^\mathrm{13}$CO isotopologues) near 4.7$\mu$m has been observed in the spectrum of MWC~614 \citep{Banzatti2015,vanderPlas2015}.
Similar features have been observed in the spectra of the known gapped disk systems HD97048, HD100546, \modif{and V1247\,Orionis} \citep{Kraus2013,vanderPlas2015} and are thought to arise due to the direct illumination of gas by UV stellar photons \citep{Thi2013}.
This suggests that the usual sheltering provided by dust grains is absent in certain regions of the disk. 
Thus, the existence of these lines in the infrared spectrum of MWC~614 supports the idea that the disk around MWC~614 features a gap or gaps.
No companion has been detected so far to explain this structure.

In this paper we describe our multi-wavelength and multi-technique observational campaign to constrain the inner regions of MWC~614 (Sect.\ref{sec:obs}).
We present \modif{the} geometric \modif{modelling} and image reconstruction techniques that we used to constrain the brightness distribution in thermal light and scattered light as well as our companion search (Sect.\,\ref{sec:geometry}).
We then discuss our findings and their implications on the evolutionary state of this special object in Sect.\,\ref{sec:discussion} and summarize our results in Sect.\,\ref{sec:conclusion}.

\section{Observations}
\label{sec:obs}

\subsection{VLT/SPHERE polarimetric imaging in visible light}

\begin{table*} 
\begin{center}
\caption{Observing log. \label{tab:obslog} }
\begin{tabular}{cccccccc}
\tableline
\tableline
  Instrument & Date [UT]  & Telescope(s)/ & Mode & Filter & NDIT x DIT & \# of pointings \\
  &&Configuration\\
\tableline
    SPHERE/ZIMPOL & 2015-06-10 & UT3 & P2 & N\_R & 48 x 45\,s & 1\\
\tableline
    NIRC2 & 2013-11-16 & Keck-II & SAM & H & 2 x 25 x 0.845\,s & 1 \\
\tableline
    PIONIER & 2013-06-06 & A1-G1-J3-K0 & GRISM & H &- & 1  \\
     & 2013-07-04 & A1-B2-C1-D0 & GRISM & H &-& 2\\
\tableline
    AMBER & 2013-06-13 & UT2-UT3-UT4 & Low Res. & K & 5000 x 26\,ms &1\\
\tableline
    CLASSIC & 2010-07-20 & E1-W1 & - & K & - &2\\
   & 2010-07-21 & E1-S1 & - & K & - &2\\
   & 2011-06-12 & W1-W2 & - & K & - &1\\
\tableline
    CLIMB & 2011-06-12 & E1-W1-W2 & - & K & - &2\\
    & 2011-06-14 & E1-W1-W2 & - & K & - &3\\
    & 2011-06-16 & E1-W1-W2 & - & K & - &3\\
    & 2011-06-21 & E1-W1-W2 & - & K & - &2\\
    & 2011-06-25 & E2-S1-W2 & - & K & - &1\\
    & 2011-06-26 & S1-W1-W2 & - & K & - &3\\
    & 2011-08-04 & E2-S2-W2 & - & K & - &3\\
    & 2012-06-28 & E1-E2-S1 & - & K & - &3\\
    & 2012-06-30 & E1-S1-S2 & - & K & - &1\\
\tableline
    MIDI & 2003-06-16 & UT1-UT3 & PRISM & N &-& 3 \\
    &2004-04-10 &UT2-UT3 &PRISM & N &-& 2\\
    &2006-05-15 &UT2-UT3 &GRISM &N & -&1 \\
    &2006-05-16 &UT1-UT3 &GRISM &N &-& 2\\
    &2006-05-17 &UT3-UT4 &PRISM &N &-& 1\\
    &2006-06-11 &UT3-UT4 &PRISM &N &-& 1\\
    &2006-06-14 &UT1-UT2 &PRISM &N &-& 2\\
    &2006-07-09 & UT3-UT4 &PRISM &N & -&1\\
    &2006-07-13 &UT1-UT2 &GRISM &N &-& 2\\
    &2009-08-14 &E0-G0 &PRISM &N & -&7\\
    &2009-08-15 &H0-G0 &PRISM &N & -&3 
\end{tabular}
\end{center}
\end{table*}

MWC\,614 was observed on the night of 2015 June 10 with SPHERE \citep{Beuzit2008} as part of ESO observing program 095.C-0883 (PI: S. Kraus).
SPHERE is a high performance adaptive optics system \citep{Fusco2014} installed on UT3 at the Very Large Telescope (VLT) on the top of Cerro Paranal in Chile.
We used ZIMPOL \citep{Roelfsema2014}, the polarimetric imaging instrument of SPHERE operating in the visible and with two cameras.
The $R$-band filters ($\lambda_\mathrm{c}$=645.9\,nm, $\Delta\lambda$=56.7\,nm) were selected for both cameras and the slow-polarimetric mode (P2) was used (see Table\,\ref{tab:obslog}).
The observations were performed using a Lyot coronagraph (\texttt{V\_CLC\_M\_WF}) with an on-sky-projected diameter of 155\,mas.

A polarimetry observing sequence includes observations with the half-wave plate (HWP) rotated along position angles of 0, 90, 45 and 135$^{\circ}$ to measure the Stokes $Q$ parameter and of 22.5, 112.5, 67.5 and $157.5^{\circ}$ for the Stokes $U$ parameter.
One sequence of observation includes two polarimetric cycles ($Q$ and $U$).

Because we are using a coronagraphic mask, we have repeated this observation sequence three times with three different field rotations of 0$^\circ$, 30$^\circ$ and 60$^\circ$.

 \begin{figure}
   \centering
   \includegraphics[height=4cm]{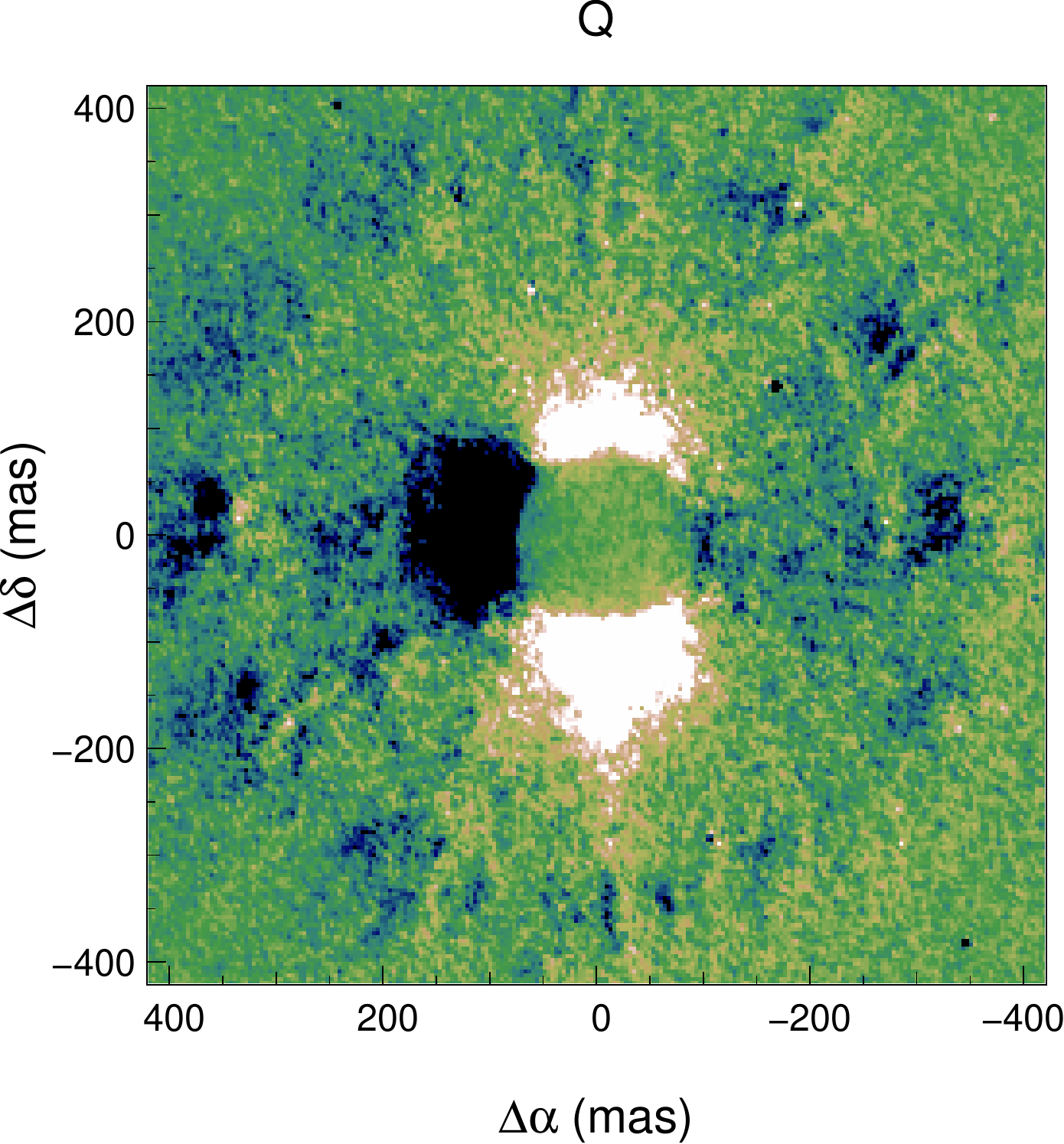}
    \includegraphics[height=4cm]{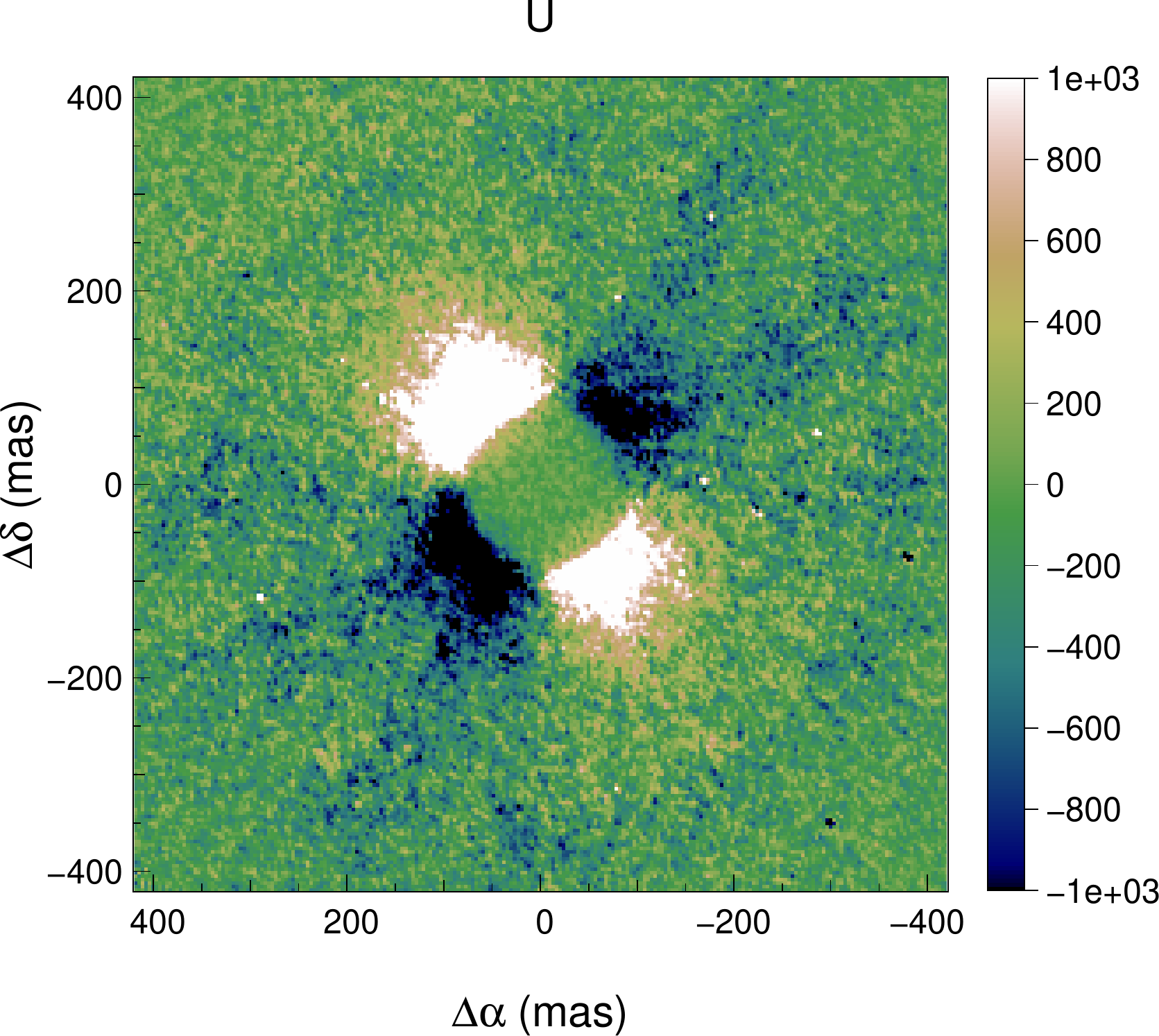}
      \caption{Raw polarized intensities from the SPHERE/ZIMPOL instrument for the STOKES parameters Q (Left) and U (Right).
              }
         \label{fig:QandU}
   \end{figure}

The raw data cubes were processed using the SPHERE/ZIMPOL data reduction pipeline (version 3.12.3).
The output of the pipeline are cubes of $Q^+$, $Q^-$, $U^+$ and $U^-$ Stokes components and their associated total flux intensities.
We then centered the images with respect to the coronagraphic mask and derotated them using custom scripts.
We constructed the Stokes $Q$ and $U$ parameters as follows:
\begin{eqnarray}
Q &=& \frac{Q^+ - Q^-}{2}\\
U &=& \frac{U^+ - U^-}{2}
\end{eqnarray}
As suggested by \citet{Avenhaus2014}, we have equalized the fluxes coming from the ordinary and extraordinary beams (linear polarization parallel and perpendicular to the optical bench, respectively) for each frame and corrected for any difference in the acquisition between the Stokes $Q$ and $U$ parameters \citep[we computed the efficiency of measurement of Stokes U of $e_U=0.986$ as in][]{Avenhaus2014}.
We then stacked all the centered and derotated frames to obtain total intensity, $Q$ and $U$ images (see Fig.\,\ref{fig:QandU}).

 \begin{figure*}
   \centering
     \includegraphics[width=6.0cm]{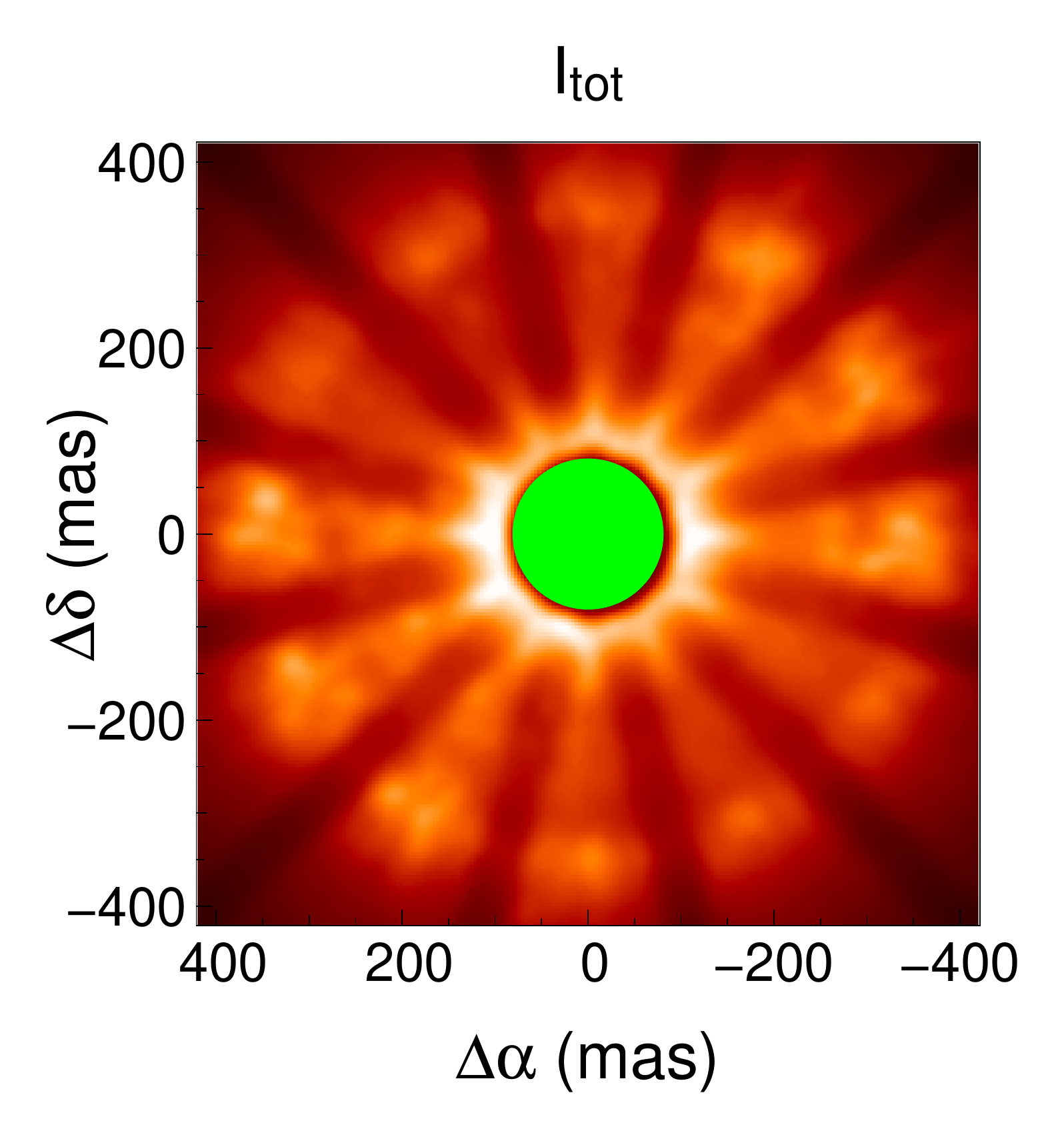}
    \includegraphics[width=6.2cm]{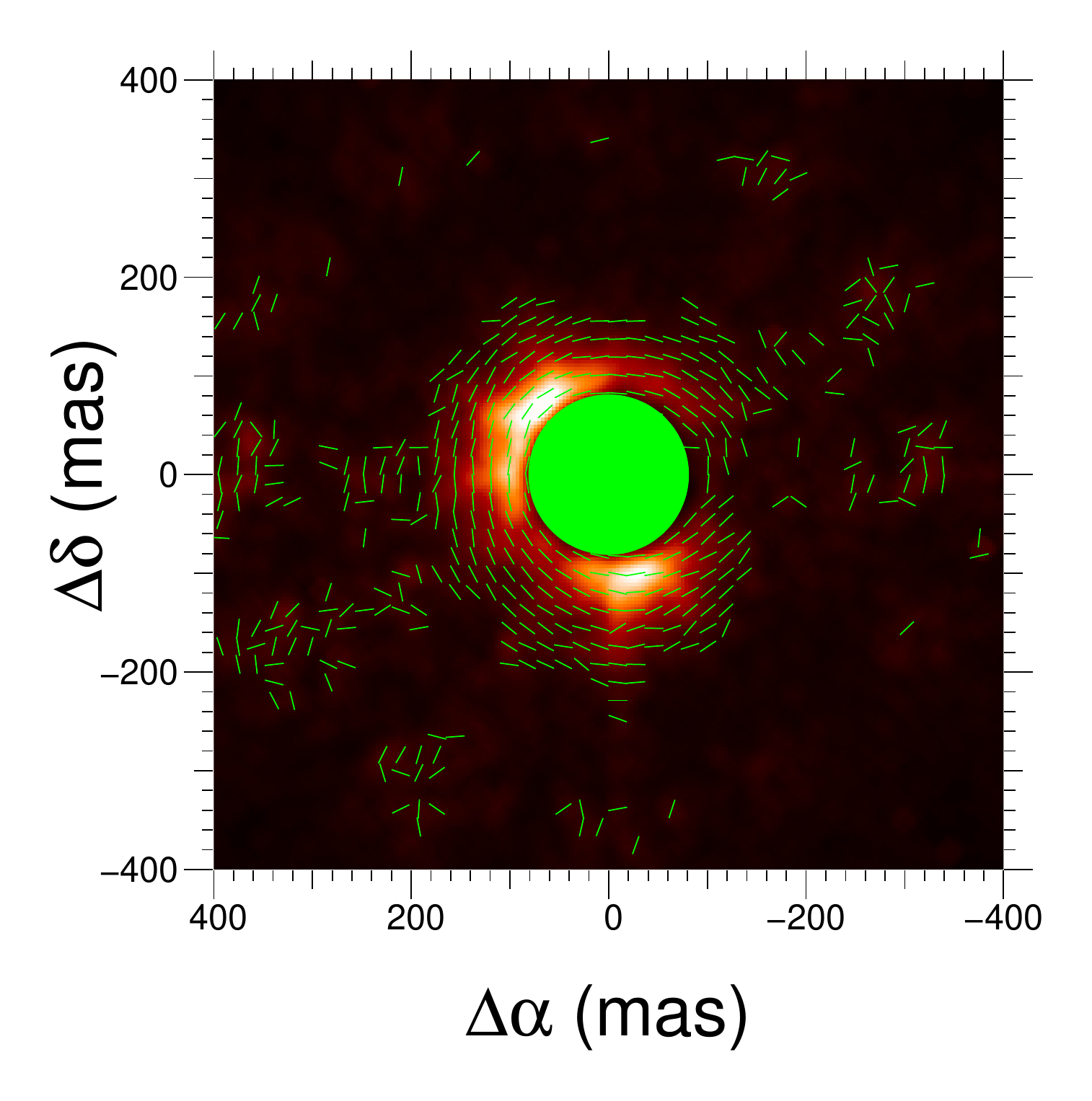}
    \includegraphics[height=6.4cm]{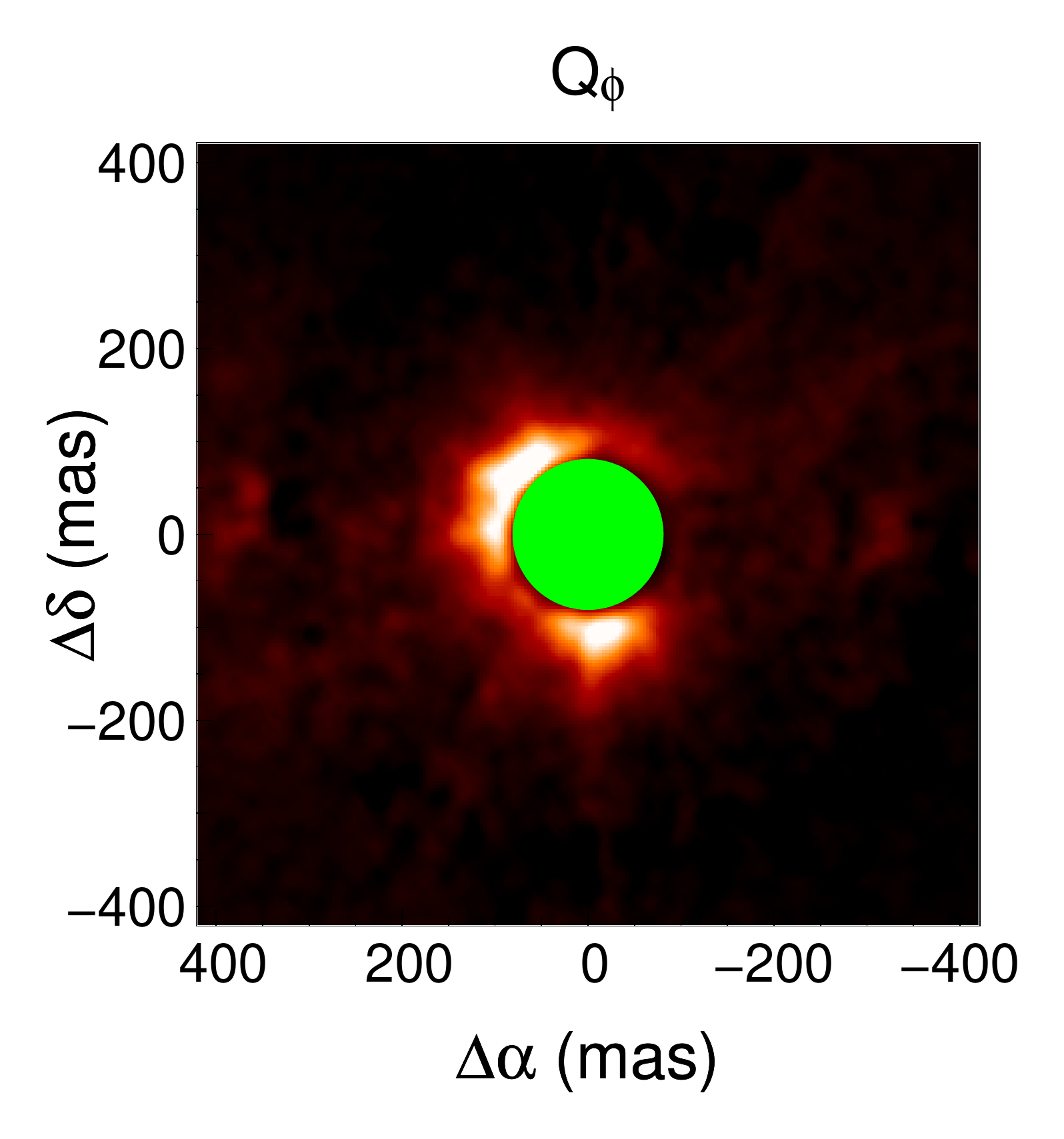}
    \includegraphics[height=6.4cm]{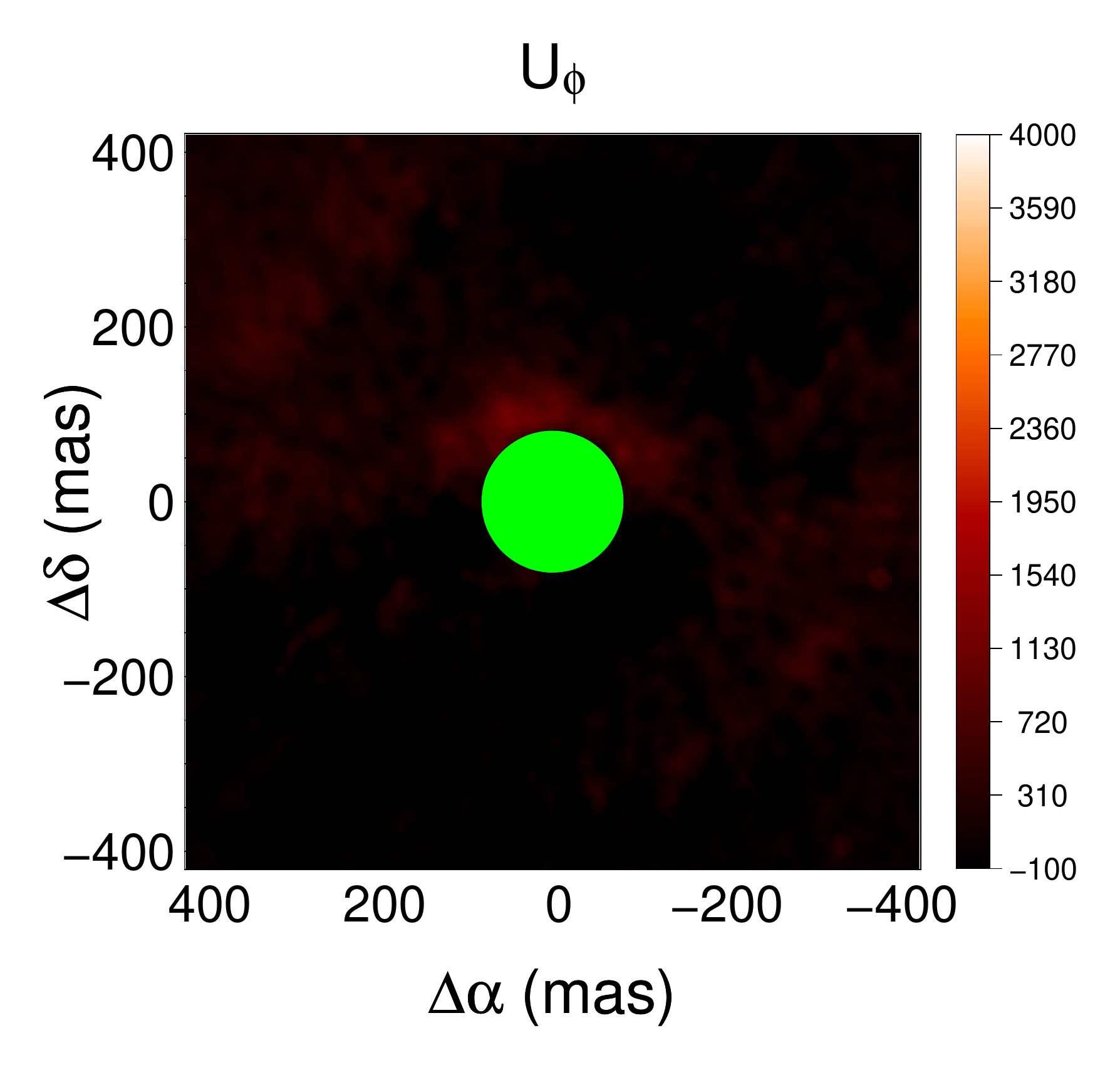}
      \caption{SPHERE/ZIMPOL observations of MWC~614. Top left: Total intensity image, which show the coronagraphic mask \modif{in green} and the \modif{coronagraph} s\modif{pider} arms. Top right: Polarized intensity image, where the green bars represent the orientation of the polarized light. Bottom left: Q$_\phi$ intensity image. Bottom right: U$_\phi$ intensity image.}
         \label{fig:SphereImg}
   \end{figure*}

The linear polarization fraction map ($p_\mathrm{L}$) and the position angle of the electric vector ($\theta$) was built using:
\begin{eqnarray}
p_\mathrm{L} &=& \sqrt{Q^2 + U^2}\\
\theta &=& \frac{1}{2} \arctan \frac{U}{Q} 
\end{eqnarray}

\noindent We have also computed the polar Stokes components ($Q_\phi$ and $U_\phi$) as described in \citet{Avenhaus2014}:
\begin{eqnarray}
Q_\phi &=& Q \cos 2\Phi + U \sin 2\Phi\\
U_\phi &=& -Q \sin 2\Phi + U \cos 2 \Phi
\end{eqnarray}
$\phi$ refers to the azimuth in polar coordinates and $\Phi$ being the angle of a pixel ($x$, $y$) with respect to the star ($x_0$, $y_0$):
\begin{eqnarray}
\Phi &=& \arctan \frac{x-x_0}{y-y_0} + \theta,
\end{eqnarray}
with $\theta$ being an angle correcting for the instrumental polarization (we found $\theta$=1.76$^\circ$).
This decomposition of the Stokes parameters was used to have one image with the polarised flux and one image with noise estimation.
We show all the images in Fig.\,\ref{fig:SphereImg}.
This representation assumes that the polarized intensity is tangential and \citet{Canovas2015} indicated that, for special conditions, this assumption is not true and that $U_\phi$ can still contain astrophysical information.
In our case, the $Q_\phi$ image is almost identical to the $p_L$ image showing that the polarised intensity vector is mostly tangential.

\subsection{Keck/NIRC2 Sparse Aperture Masking Interferometry}

\begin{figure}
    \centering
    \includegraphics[width=5.5cm]{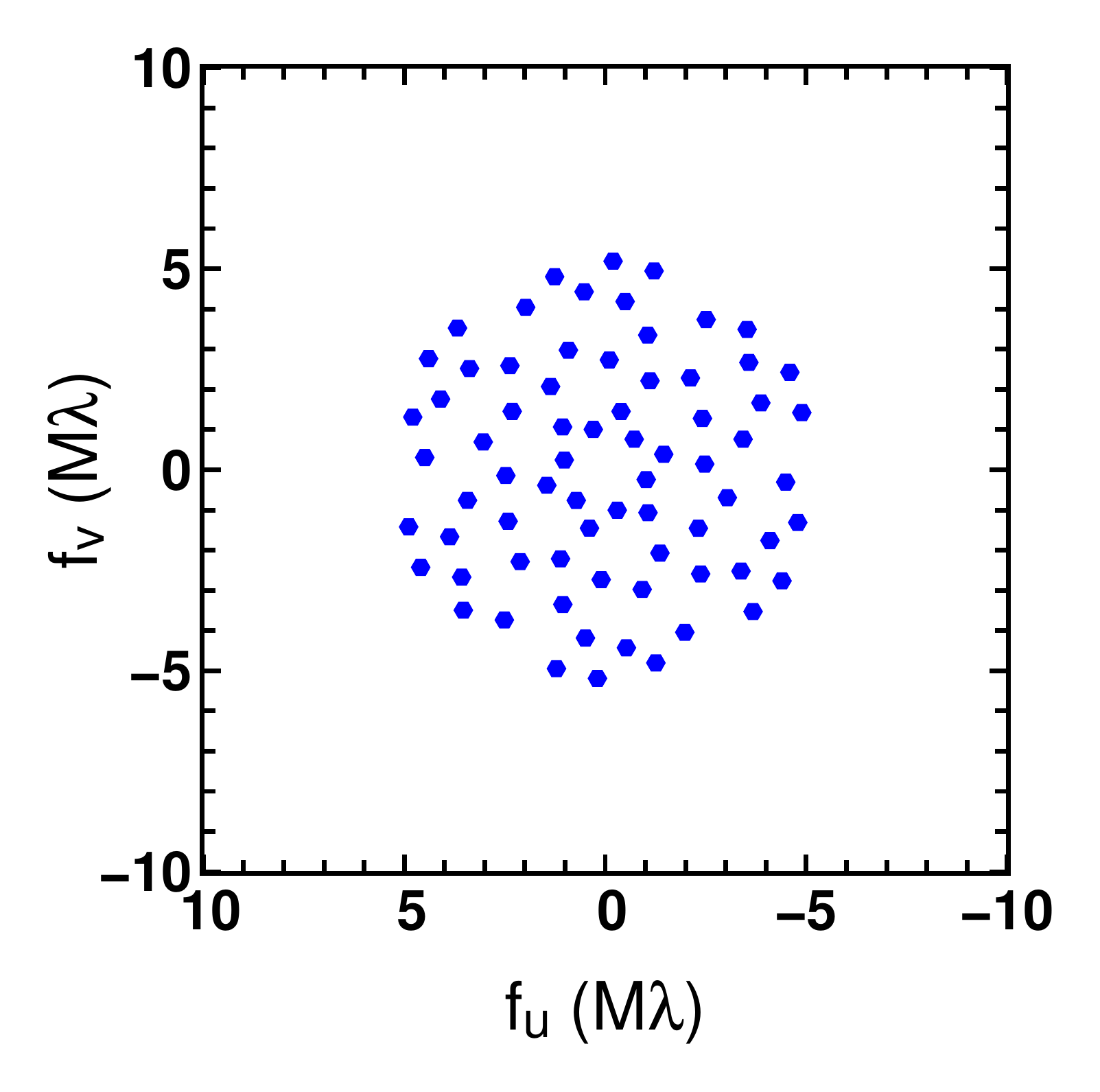}
    \includegraphics[width=4.2cm]{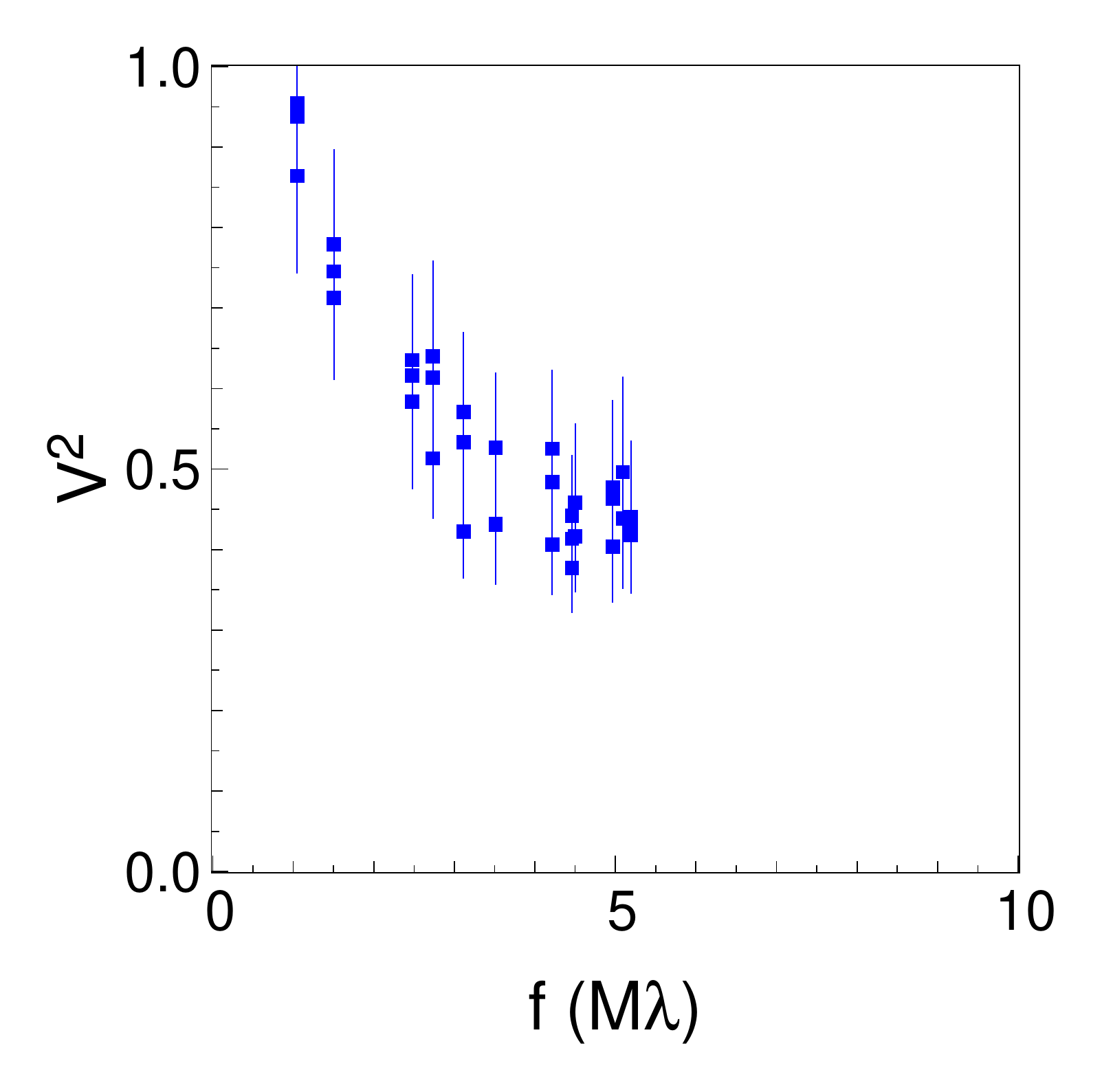}
    \includegraphics[width=4.2cm]{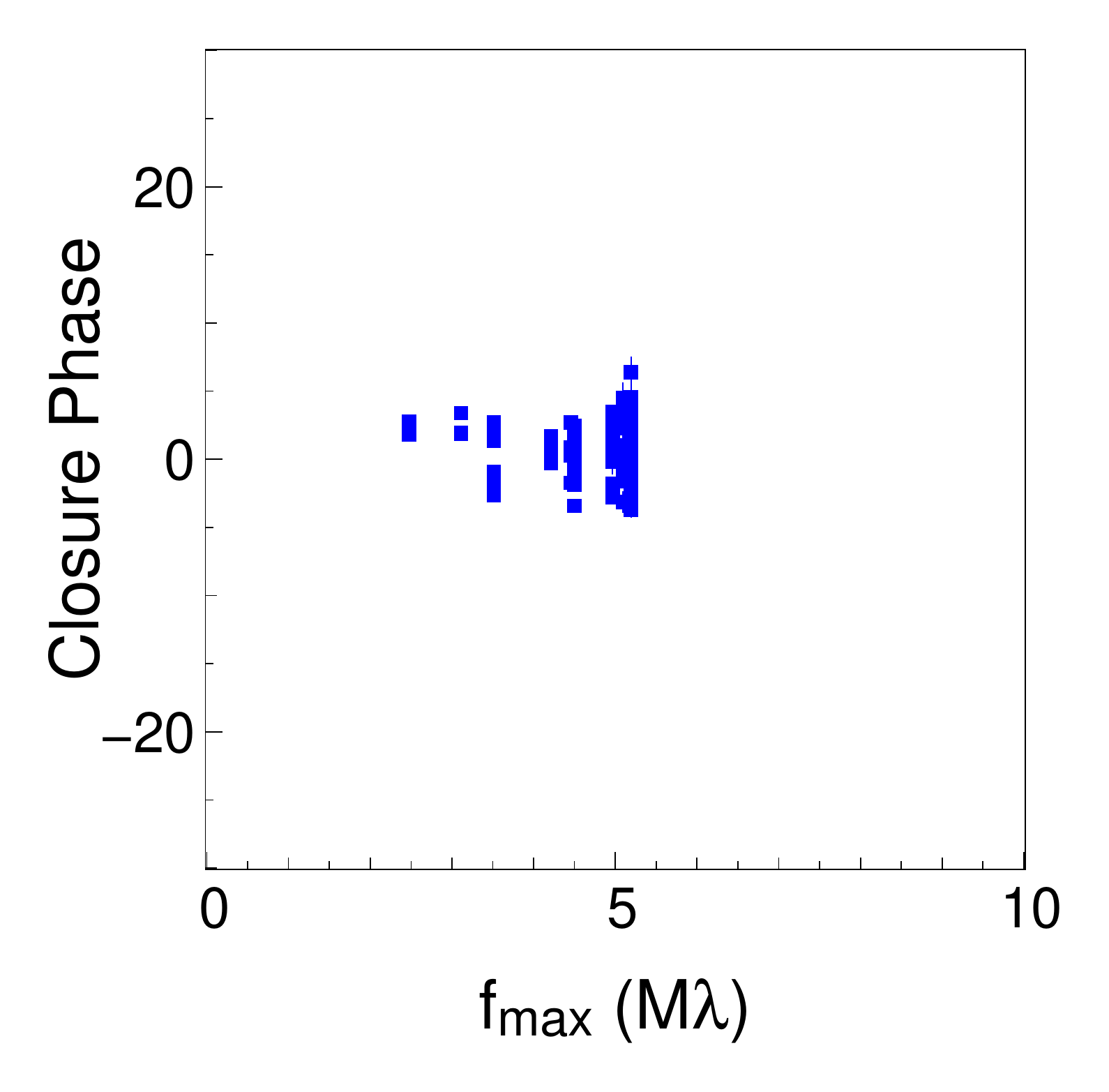}
    \caption{Aperture masking dataset from Keck-II/NIRC2 in $H$-band (1.63$\mu$m). Top: the \uv-plane. Bottom left: squared visibilities as function of the spatial frequency. Bottom right: CPs as function of the maximum spatial frequency sampled by the closed triangle of baselines.}
    \label{fig:NIRC2}
\end{figure}

The sparse aperture masking (SAM) observations were taken on 2013 November 16 using the NIRC2 instrument mounted to the 10m Keck-II telescope located on the summit of Mauna Kea, Hawaii. 
We observed MWC~614 as part of a CAL1-SCI-CAL1-SCI-CAL2 sequence where \object{HD178568} and \object{HD178332} were used as calibrators CAL1 and CAL2, respectively. 
We used the $H$-band filter ($\lambda_\mathrm{c}$=1.63,$\Delta\lambda$=0.33$\mu$m). 
The integration time on target totals 2 $\times$ 25 coadds $\times$ 0.845\,s (see Table\,\ref{tab:obslog}). 
The use of the 9-holes mask allowed us to measure 84 closure phases (CPs) and 36 squared visibilities (V$^2$; see Fig.\,\ref{fig:NIRC2}).

The absolute level of V$^2$ is unknown due to calibrations issues, while the relative values are preserved.
We describe in Section\,\ref{sec:V2cal} the method we used to calibrate these V$^2$.
In Fig.\,\ref{fig:NIRC2}, we can see a drop of V$^2$ with the spatial frequency\modif{, which indicates that the object is resolved}. 
The CP is an indication of the degree of departure from centro-symmetry of the object, \modif{where non-zero} CPs indicate a centro-symmetric object.
\modif{We measure CPs up to 5$^{\circ}$ (Fig.\,\ref{fig:NIRC2}), which clearly indicates that the object is asymmetric.}

\subsection{Infrared long baseline interferometry}

\subsubsection{VLTI/PIONIER}

One part of the interferometric dataset was taken at the VLT Interferometer (VLTI) with the PIONIER instrument \citep{JBLB2011}.
PIONIER is an optical interferometric instrument combining four telescopes in the NIR ($H$-band centered at 1.65$\mu$m).
The dataset was taken on 2013 June 06 and 2013 July 03 as part of the PIONIER Herbig~Ae/Be Large Program \citep[190.C-0963, see][]{Lazareff2016}.
It was reduced using \texttt{pndrs} \citep{JBLB2011}.

 \begin{figure}
   \centering
  \includegraphics[width=5.5cm]{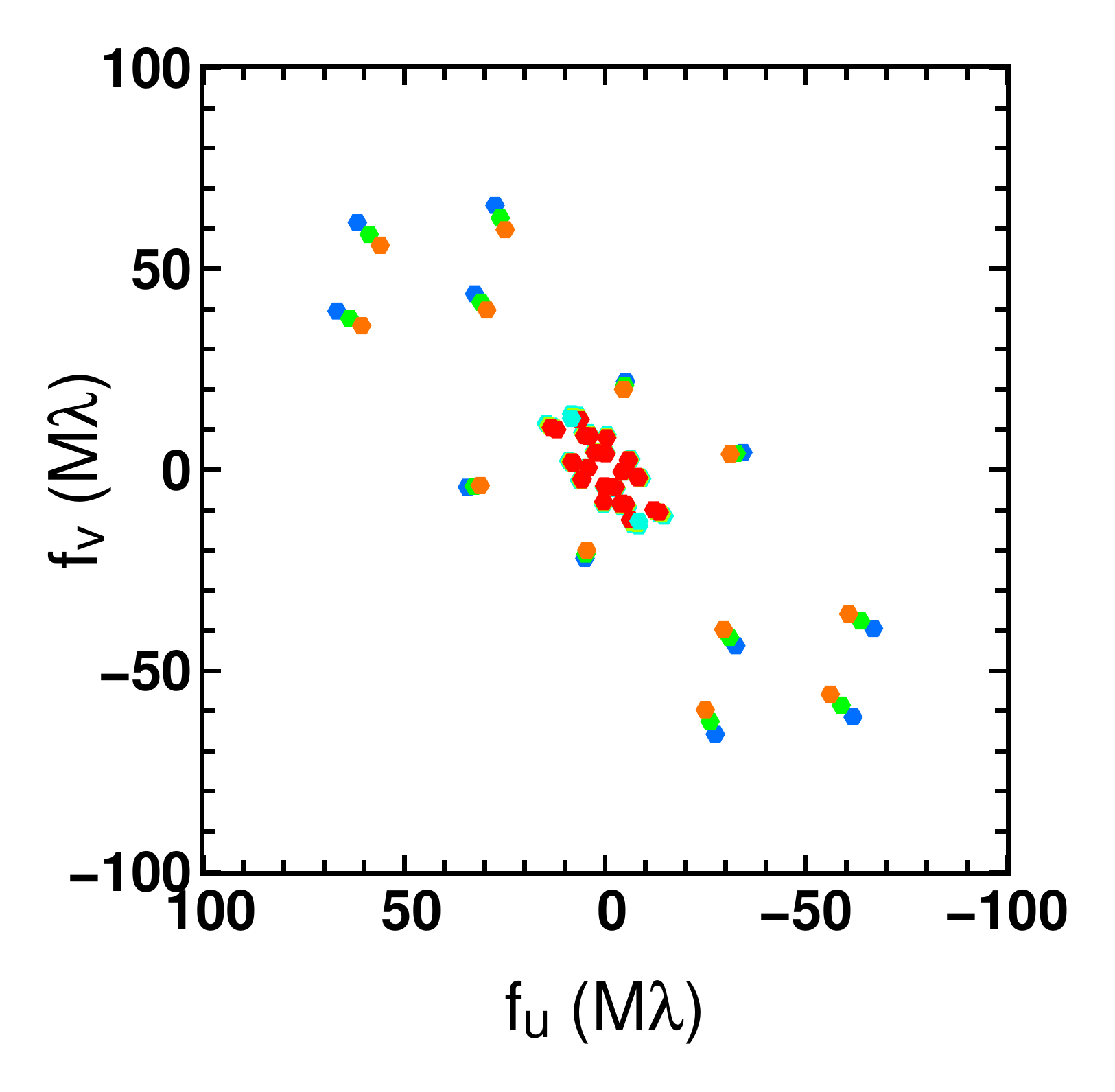}
    \includegraphics[width=4.2cm]{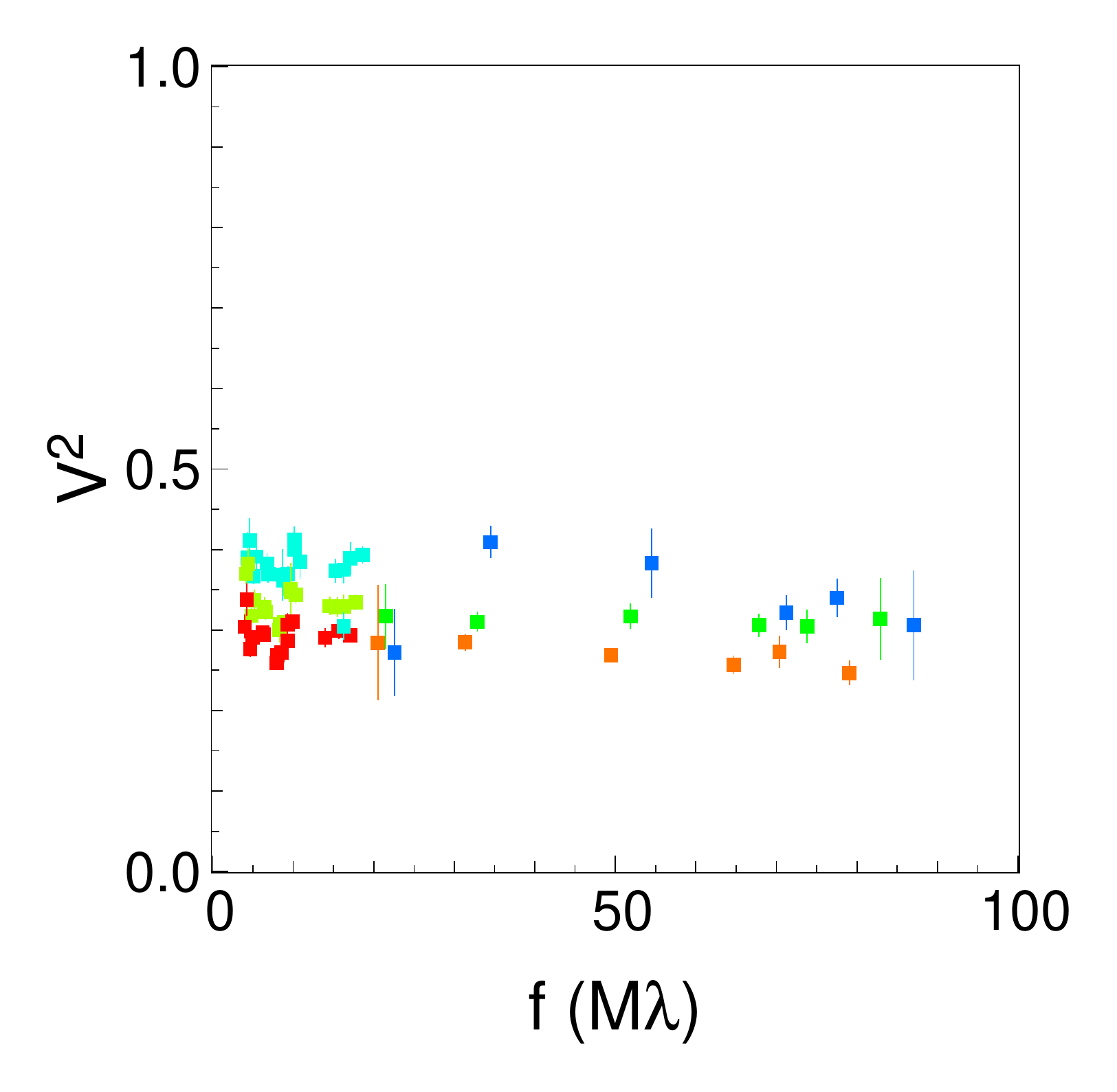}
     \includegraphics[width=4.2cm]{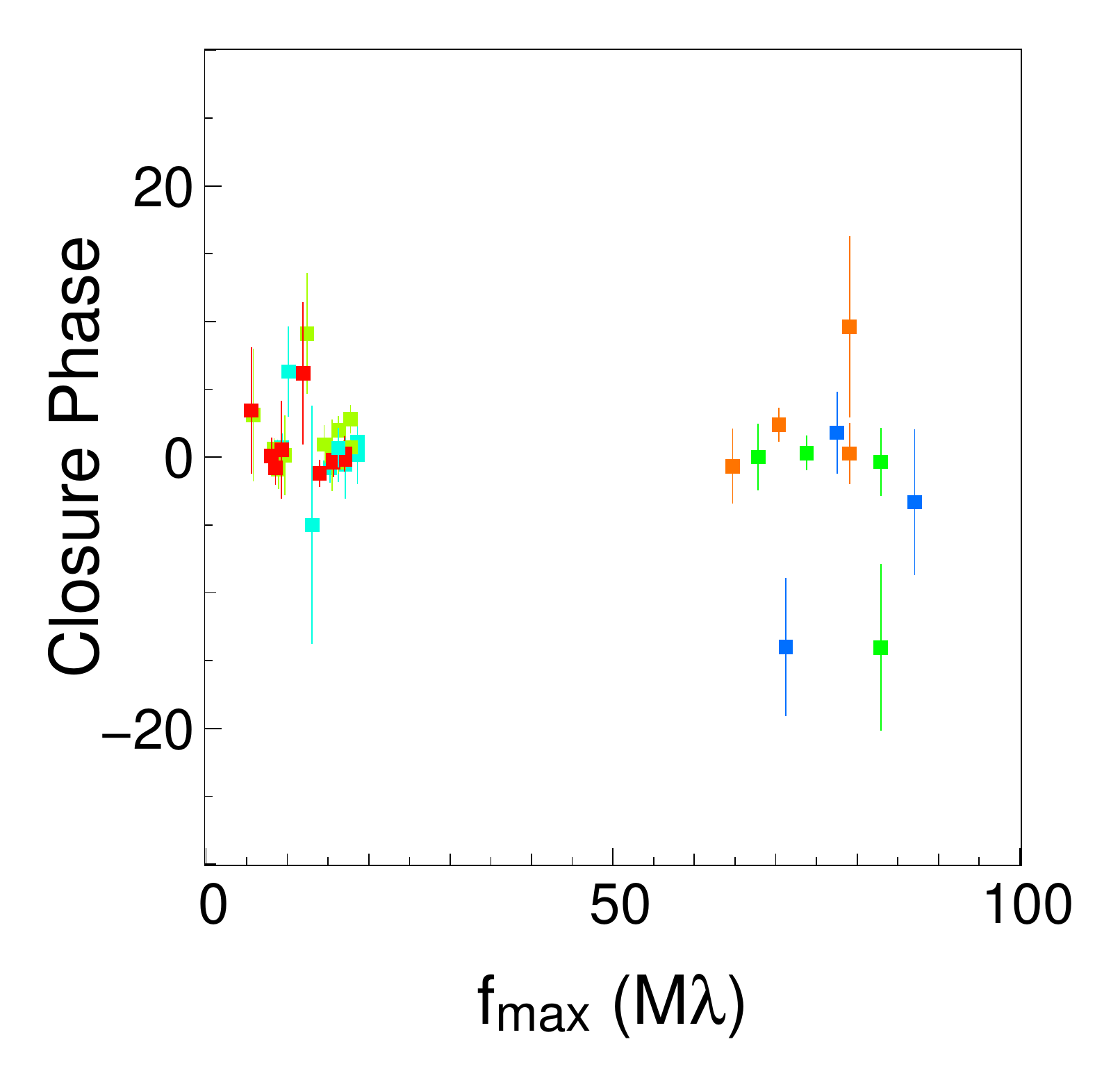}
      \caption{As Fig.\,\ref{fig:NIRC2} but for the VLTI/PIONIER dataset in $H$-band. The colors indicate the different wavelengths from 1.60$\mu$m (blue) to 1.79$\mu$m (red).
              }
         \label{fig:PIONIER}
   \end{figure}

The {\uv} coverage (top panel of Fig.~\ref{fig:PIONIER}) of the observations correspond\modif{s} to a maximum reached spatial resolution of 2.3\,milli-arcseconds (mas) and includes 66 individual measurements. 
The dataset covers baseline lengths ranging from 7 to 140\,m and is fully described in \citealt{Lazareff2016} (see also Table\,\ref{tab:obslog}).

The PIONIER V$^2$ and CP are presented in the bottom left and right panels of Fig.\,\ref{fig:PIONIER} respectively, where the different colors represent different channels (from blue, for 1.6$\mu$m, to red, for 1.83$\mu$m).
We see that at a given baseline there is a decrease of the V$^2$ with increasing wavelengths.
This is due to the chromatic effect that arises from the temperature difference between the star and its circumstellar environment \citep[][]{Kluska2014}.
Besides this, the V$^2$ measurements show a plateau indicating that the circumstellar structure is already over-resolved (larger than the smallest probed spatial frequency which corresponds to an angular resolution of 40\,mas).
The CPs signal do not seem to indicate any departure from point symmetry.

\subsubsection{VLTI/AMBER}

MWC~614 was also observed with AMBER, which is a VLTI 3-telescope beam combiner working in the $K$-band \citep[centered on 2.2$\mu$m,][]{Petrov2007}.
The observations were conducted on 2011 June 13 as part of ESO observing programme 087.C-0498(A) \modif{(PI S. Kraus)}.	
We used the 8.2\,m VLTI unit telescopes (UTs) on the UT2-UT3-UT4 configuration, which provided baseline lengths between 30 to 80\,m (see Table\,\ref{tab:obslog}).
Employing AMBER's low resolution mode, our observations cover the $K$-band with a spectral resolution of $R=30$.
We recorded a total of 5000 interferograms with a detector integration time of 26\,ms and extracted visibilities and CPs using the {\em amdlib} software \citep[Release 3;][]{tat07b,che09}.
We follow the standard AMBER data reduction procedure and select the interferograms with the 10\% best signal-to-noise ratio with the goal to minimize the effect of residual telescope jitter.

\begin{figure}
   \centering
     \includegraphics[width=5.5cm]{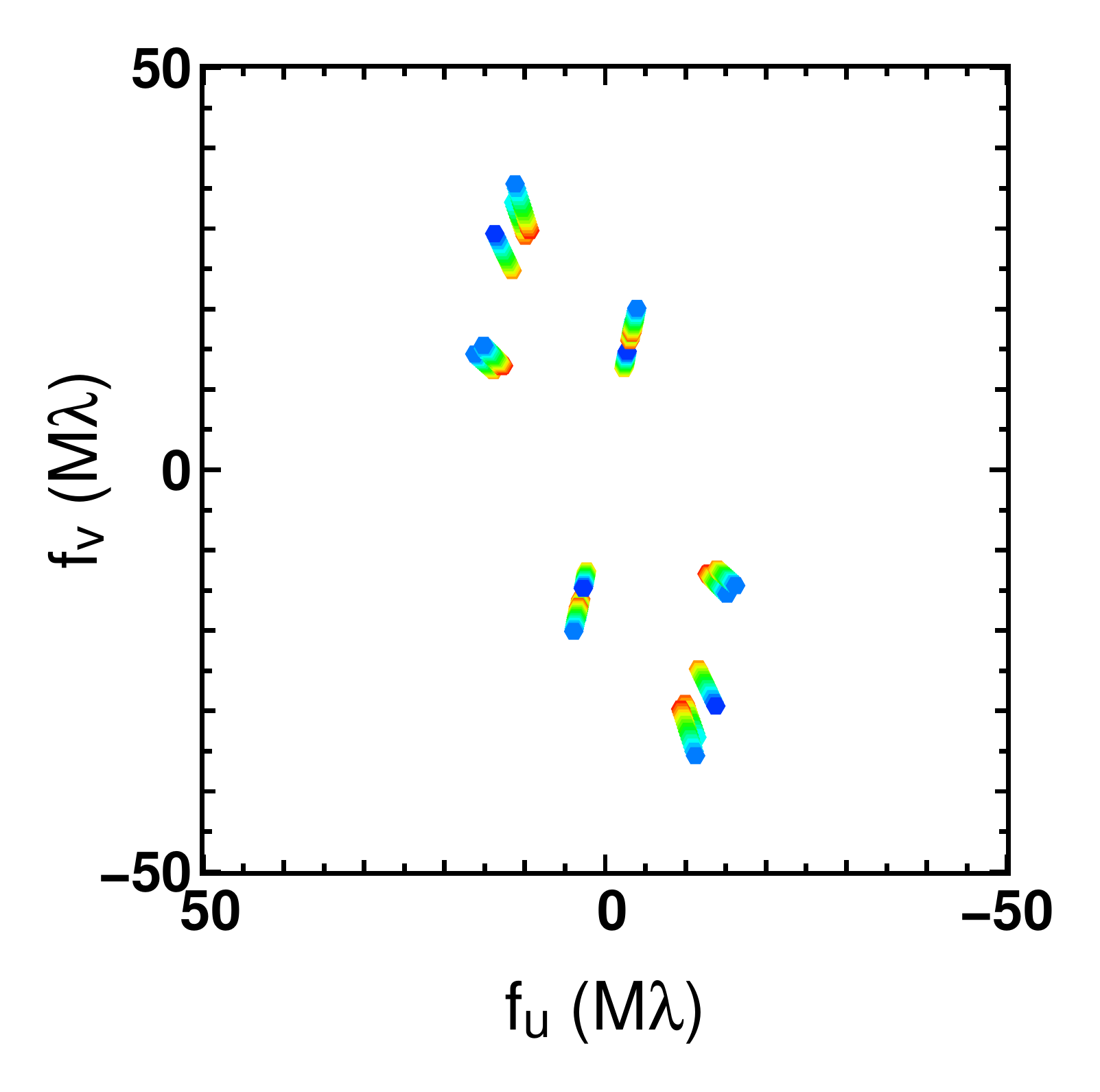}
    \includegraphics[width=4.2cm]{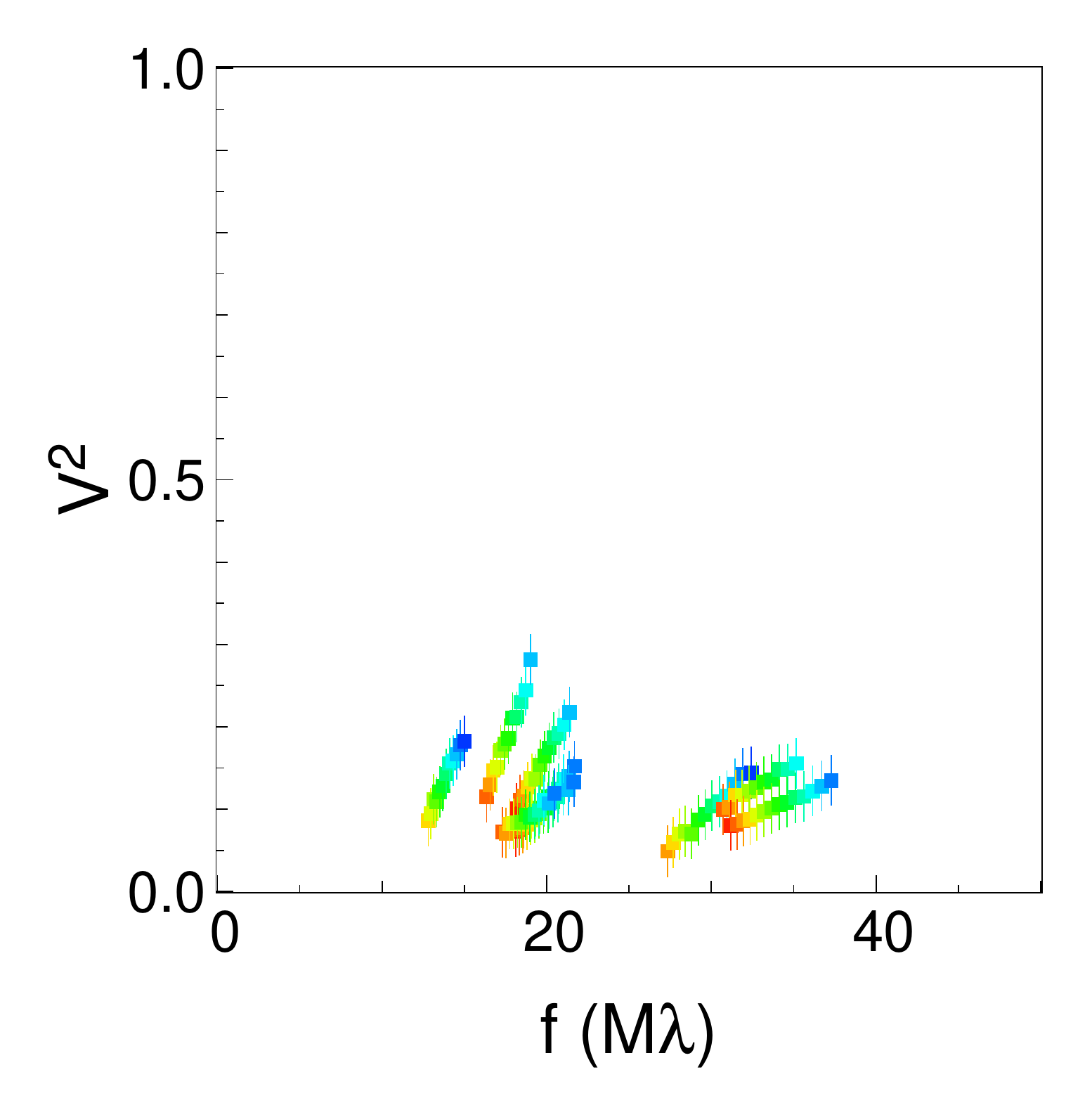}
    \includegraphics[width=4.2cm]{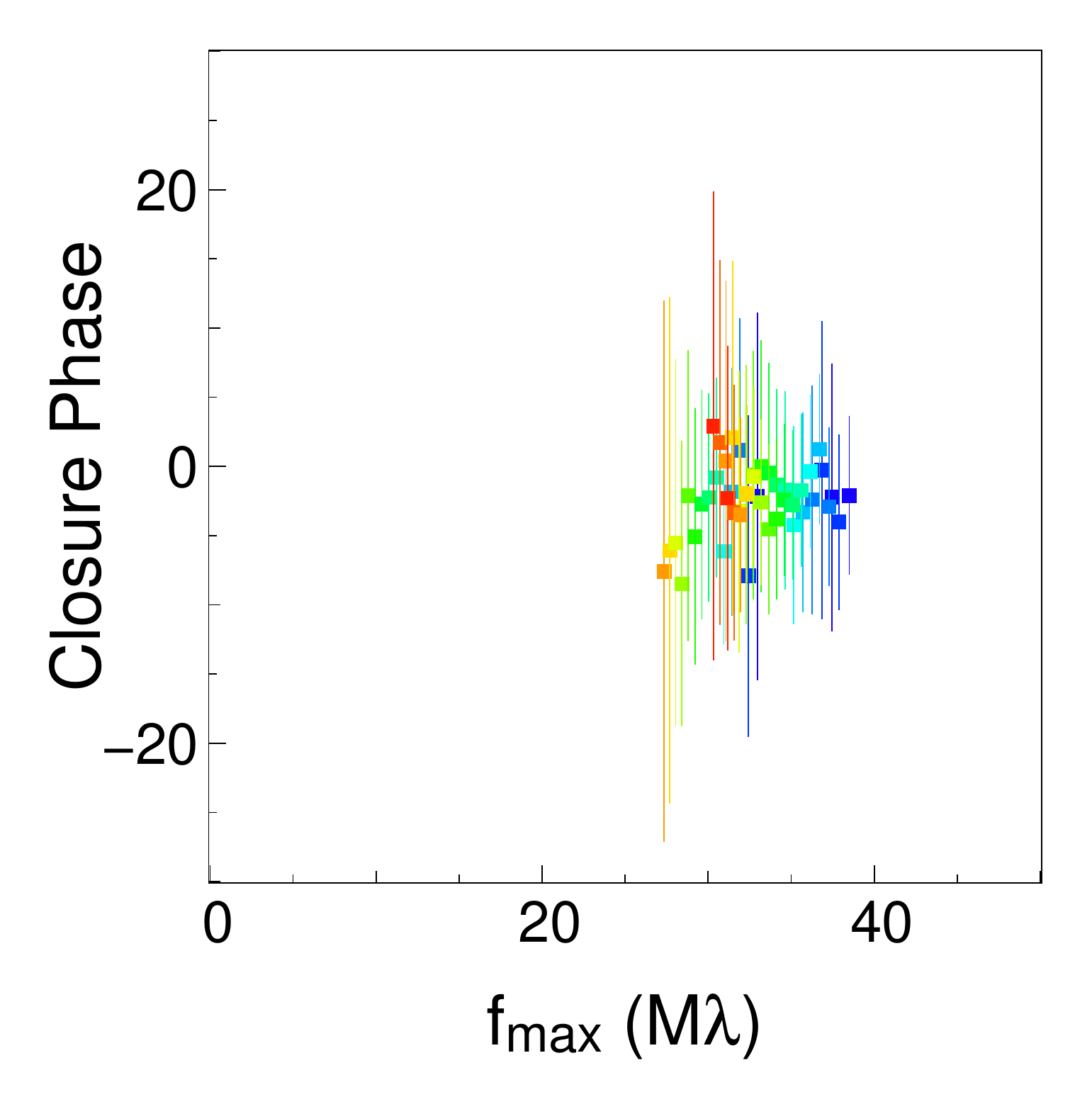}
      \caption{As Fig.\,\ref{fig:NIRC2} but for the VLTI/AMBER dataset in $K$-band. The colors indicate the different wavelengths from 2.1$\mu$m (blue) to 2.5$\mu$m (red).
              }
         \label{fig:AMBER}
   \end{figure}

The V$^2$ profile measured with AMBER shows a plateau (bottom left panel of Fig\,\ref{fig:AMBER}), indicating that the extended component that is also seen with PIONIER (Fig\,\ref{fig:PIONIER}) and NIRC2 (Fig\,\ref{fig:NIRC2}) is also overresolved in the $K$-band.

\subsubsection{VLTI/MIDI}
\label{sec:MIDI}
For our interpretation we also include archival MIDI observations on MWC~614 that were presented in \citet{Menu2015}. 
MIDI is a interferometric instrument combining light from two telescopes in the MIR \citep[8 to 13$\mu$m;][]{Leinert2003}.
This dataset consists of 27 individual observations with baseline lengths ranging from 10 to 90\,m (see Table\,\ref{tab:obslog}).

\subsubsection{CHARA/CLIMB and CHARA/CLASSIC}

\begin{figure}
   \centering
  \includegraphics[width=5.5cm]{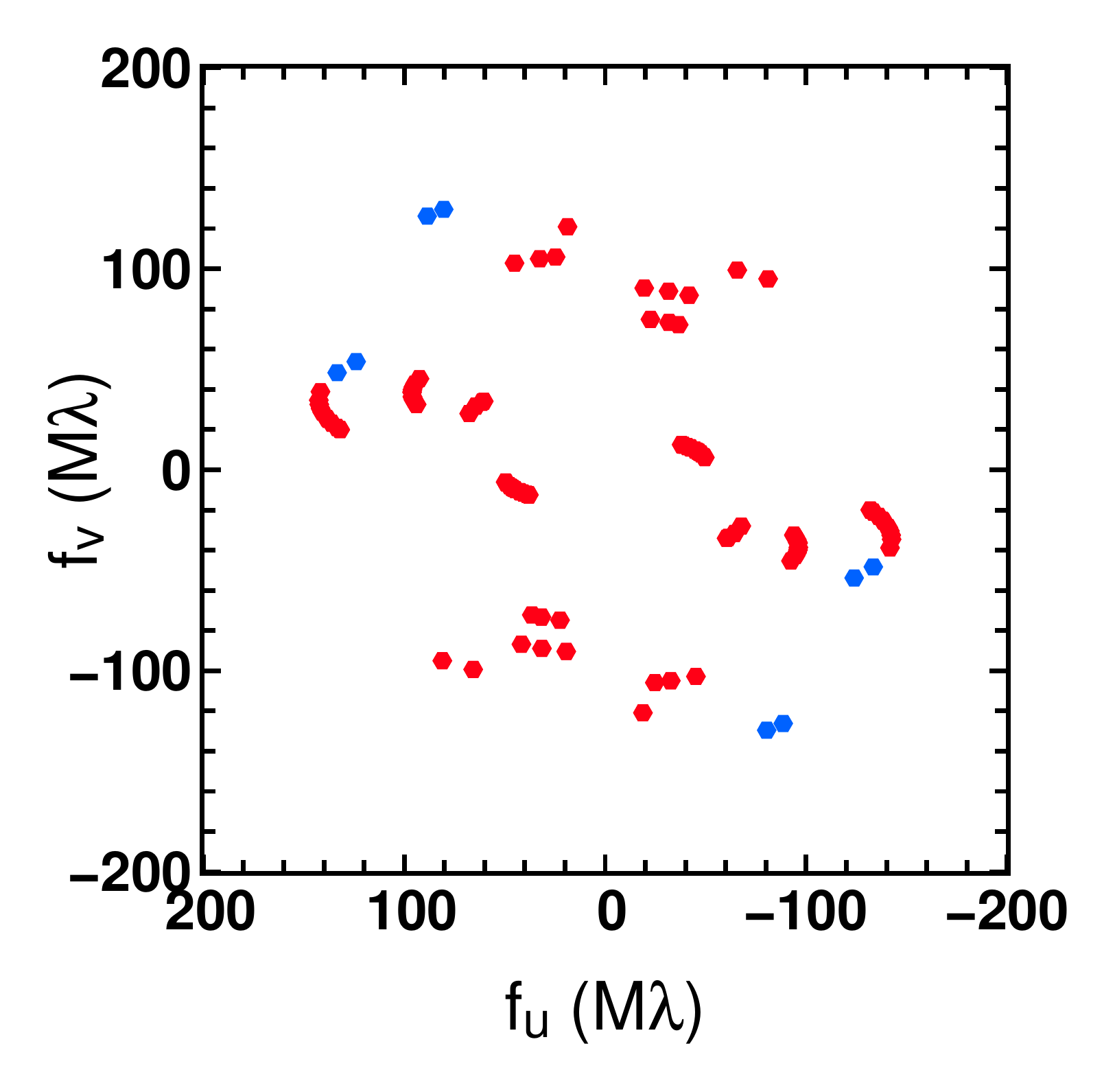}
    \includegraphics[width=4.2cm]{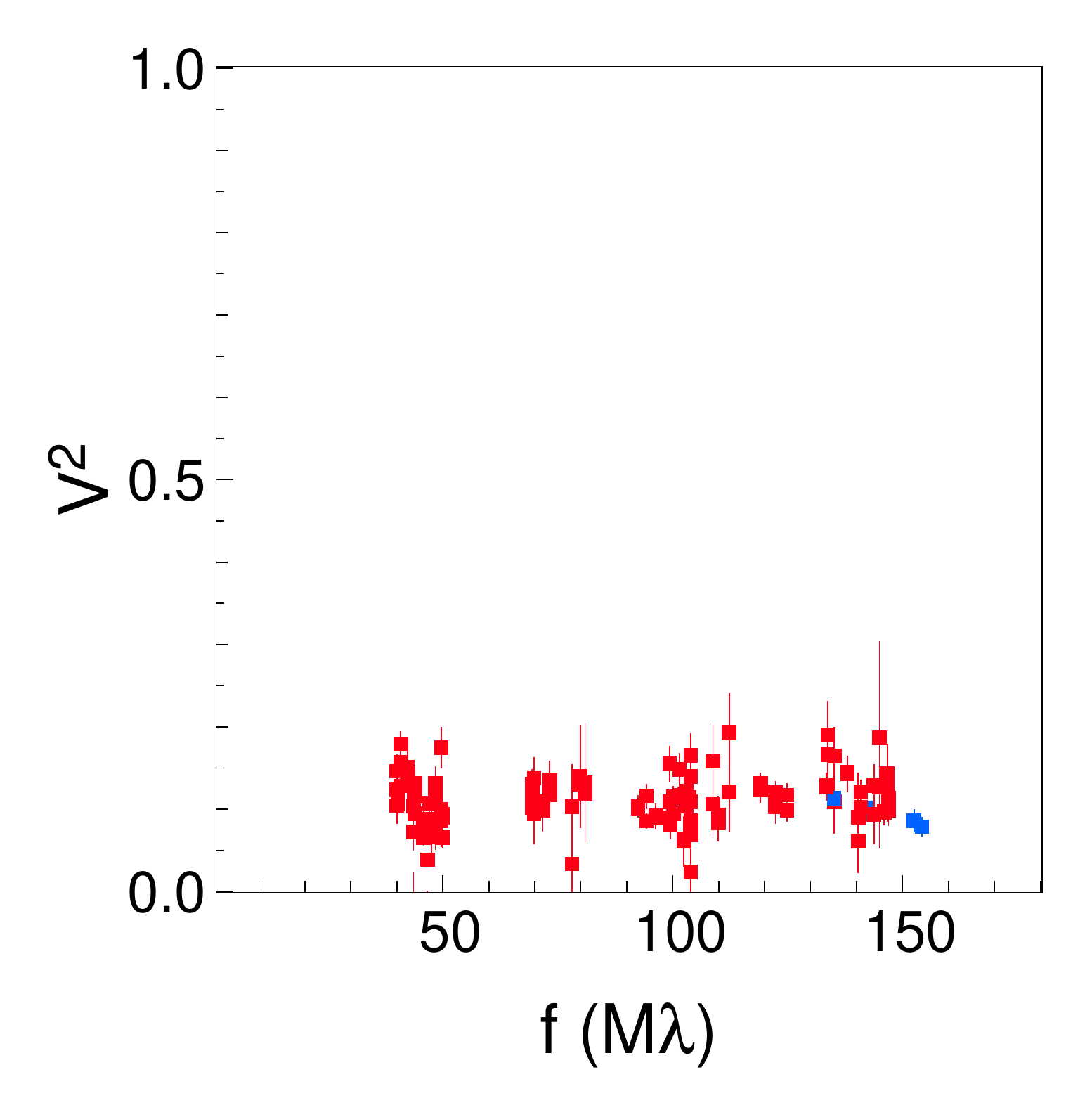}
     \includegraphics[width=4.2cm]{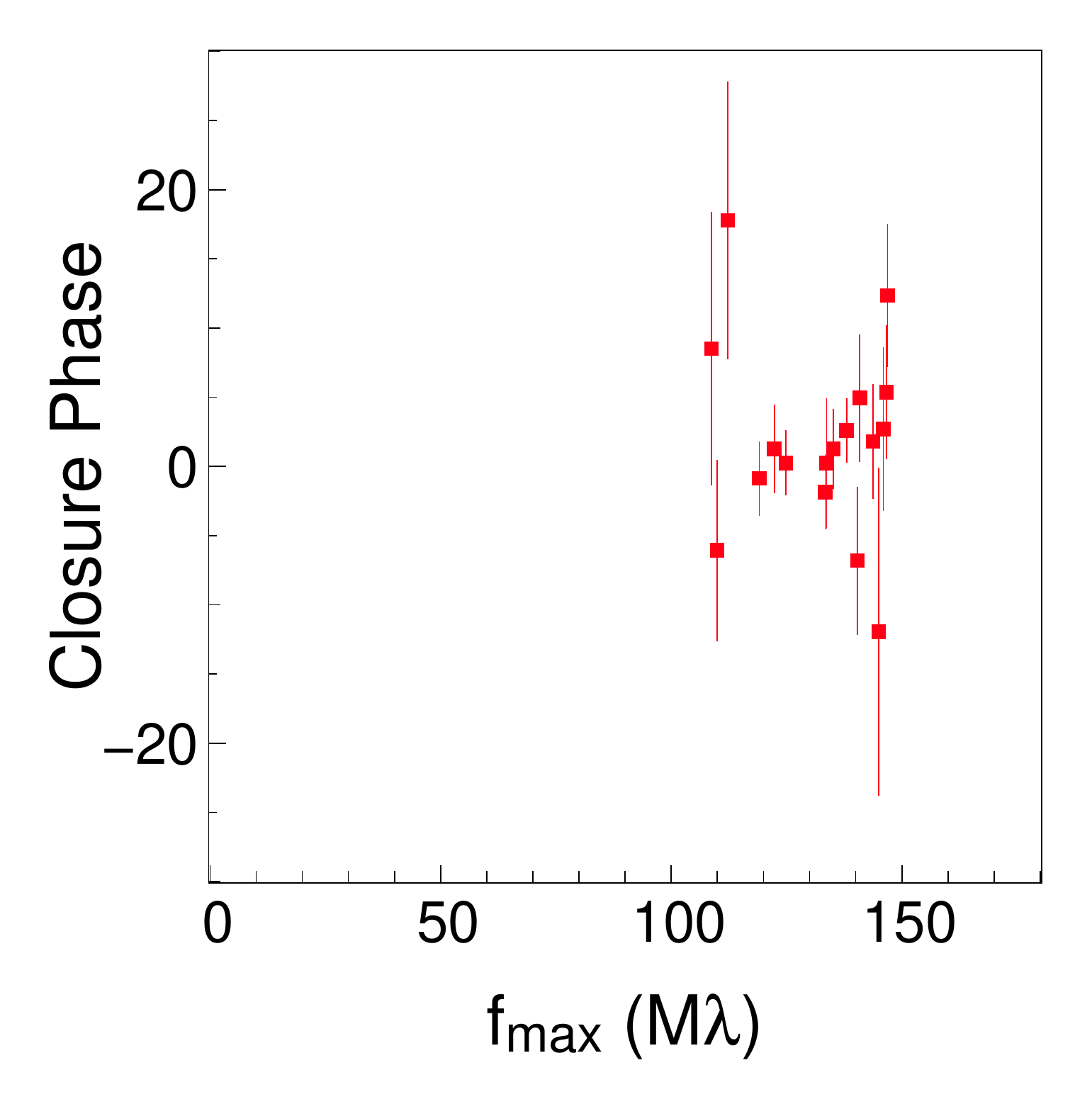}
      \caption{As Fig.\,\ref{fig:NIRC2} but for the CHARA/CLASSIC (blue) and CHARA/CLIMB (red) dataset in $K$-band. These observations are done in one channel at 2.13$\mu$m.
              }
         \label{fig:CHARA}
   \end{figure}

Longer baseline interferometric observations of MWC~614 were obtained using the Center for High Angular Resolution Astronomy (CHARA) Array over a two year period between 2010 July 20 and 2012 June 30. The CHARA Array is a Y-shaped array consisting of six $1\,$m-class telescopes located at Mount Wilson Observatory. The CLASSIC two telescope and CLIMB three telescope beam combiners \citep{tenBrummelaar2013} were used to obtain $K$-band ($\lambda_c=2.13\,\mu$m) interferometric fringes on a variety of telescope configurations covering baseline lengths from $33$ to $329\,$m. The total dataset consists of six~V$^2$ measurements from CLASSIC and 1\modif{19} V$^2$ and 1\modif{7} CP measurements from CLIMB (see Table\,\ref{tab:obslog} and Fig.\ref{fig:CHARA}). 

The CLASSIC and CLIMB data were reduced using a pipeline developed at the University of Michigan which is better suited to recover faint fringes for low visibility data than the standard reduction pipeline of \citet{Theo2012}. 
\modif{All data which showed no clear signs of being affected by instrumentation or observational effects (drifting scans or flux drop out on one or more telescopes, for example) were retained in the data reduction process.
Particular attention was given to instances where drift or low signal to noise dominated the majority of scans during data acquisition on a particular baseline pair but observation notes were clear that fringes were present during the data acquisition. 
In these cases, the affected scans were carefully flagged while the power spectrum, averaged over the retained data, was inspected for a signal. 
This procedure results in an improved noise estimation for the dataset.}
The observed visibilities and CPs were calibrated using the standard stars selected with JMMC SearchCal \modif{(HD 178568: uniform disk [UD] diameter = $0.127\pm0.01$mas; HD 177305: UD diameter = $0.25\pm0.05$mas; HD 178332: UD diameter = $0.178\pm0.013$mas; HD 178379: UD diameter = $0.216\pm0.015$mas; HD 179586: UD diameter = $0.193\pm0.014$mas; HD 181253: UD diameter = $0.238\pm0.017$mas; HD 189509: UD diameter = $0.26\pm0.02$mas).}

\section{Disk geometry and companion search}
\label{sec:geometry}

We constrain the morphology of the emission by fitting geometrical models to our extensive data set, both in polarized light, NIR thermal emission, and MIR thermal emission.
We also perform a companion search on a part of the CHARA/CLIMB dataset.

\subsection{Scattered light emission}
\label{sec:SPHEREfit}

\begin{figure*}
   \centering
  \includegraphics[width=4.2cm]{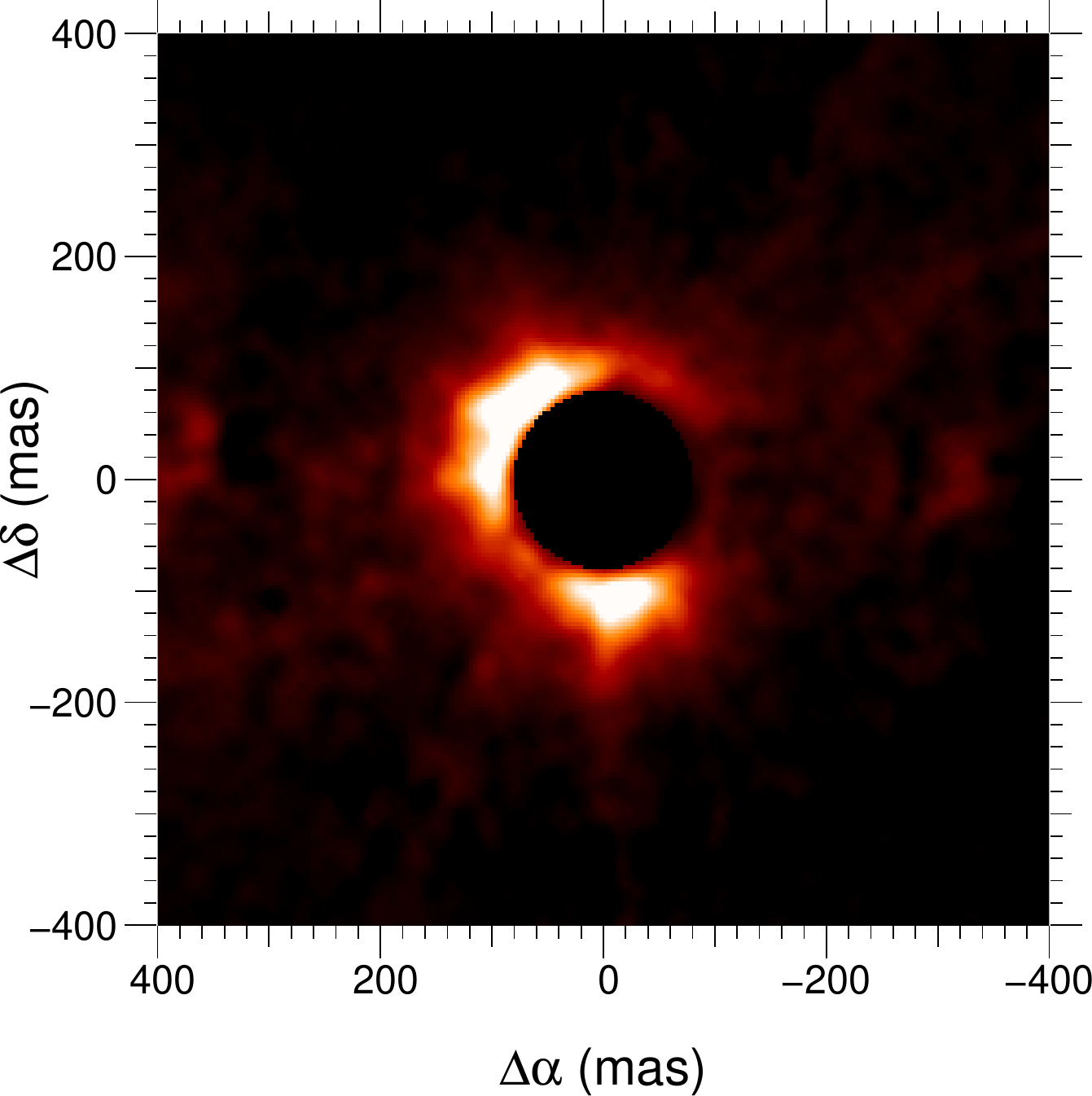}
  \includegraphics[width=4.2cm]{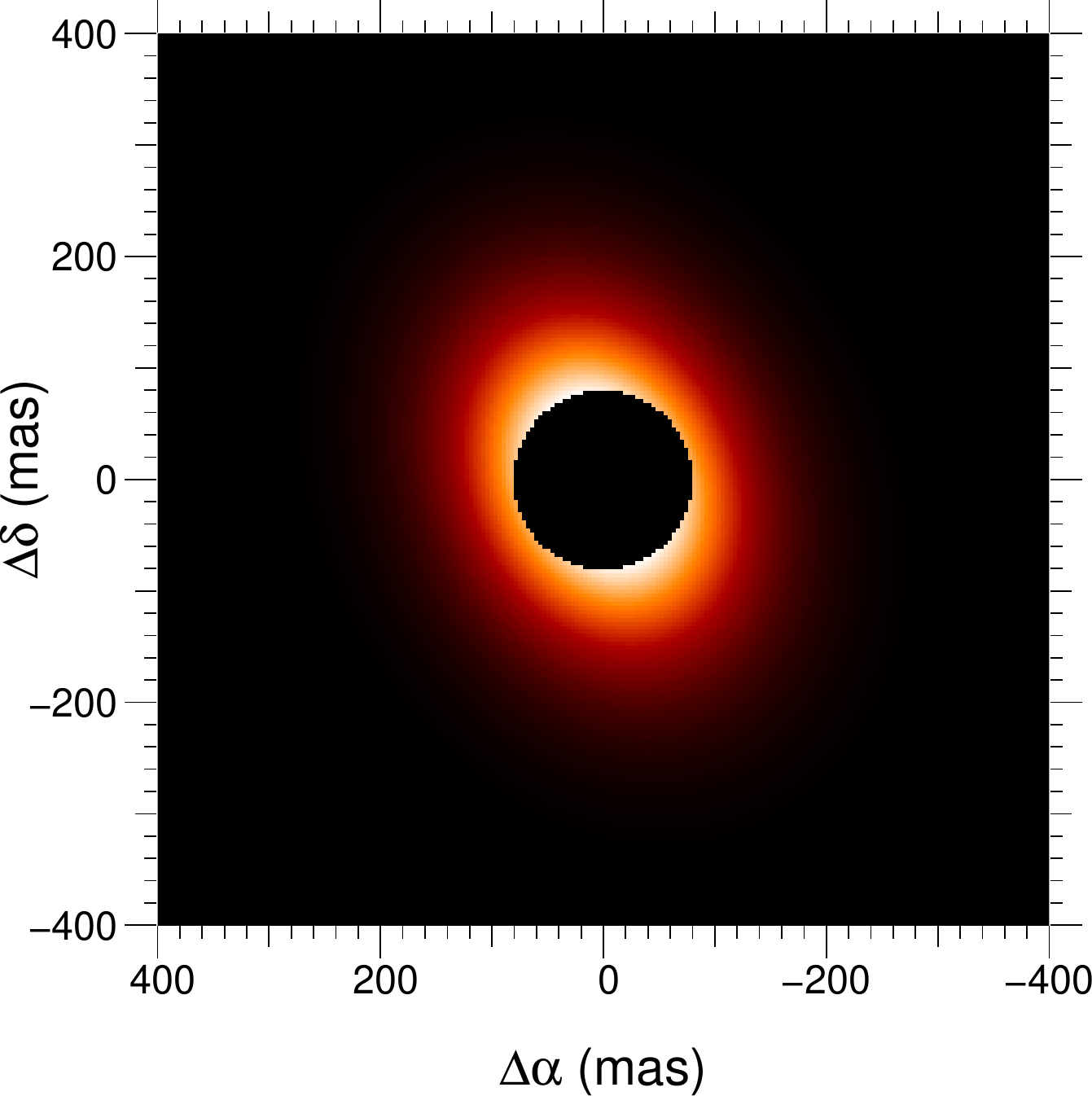}
 \includegraphics[width=4.2cm]{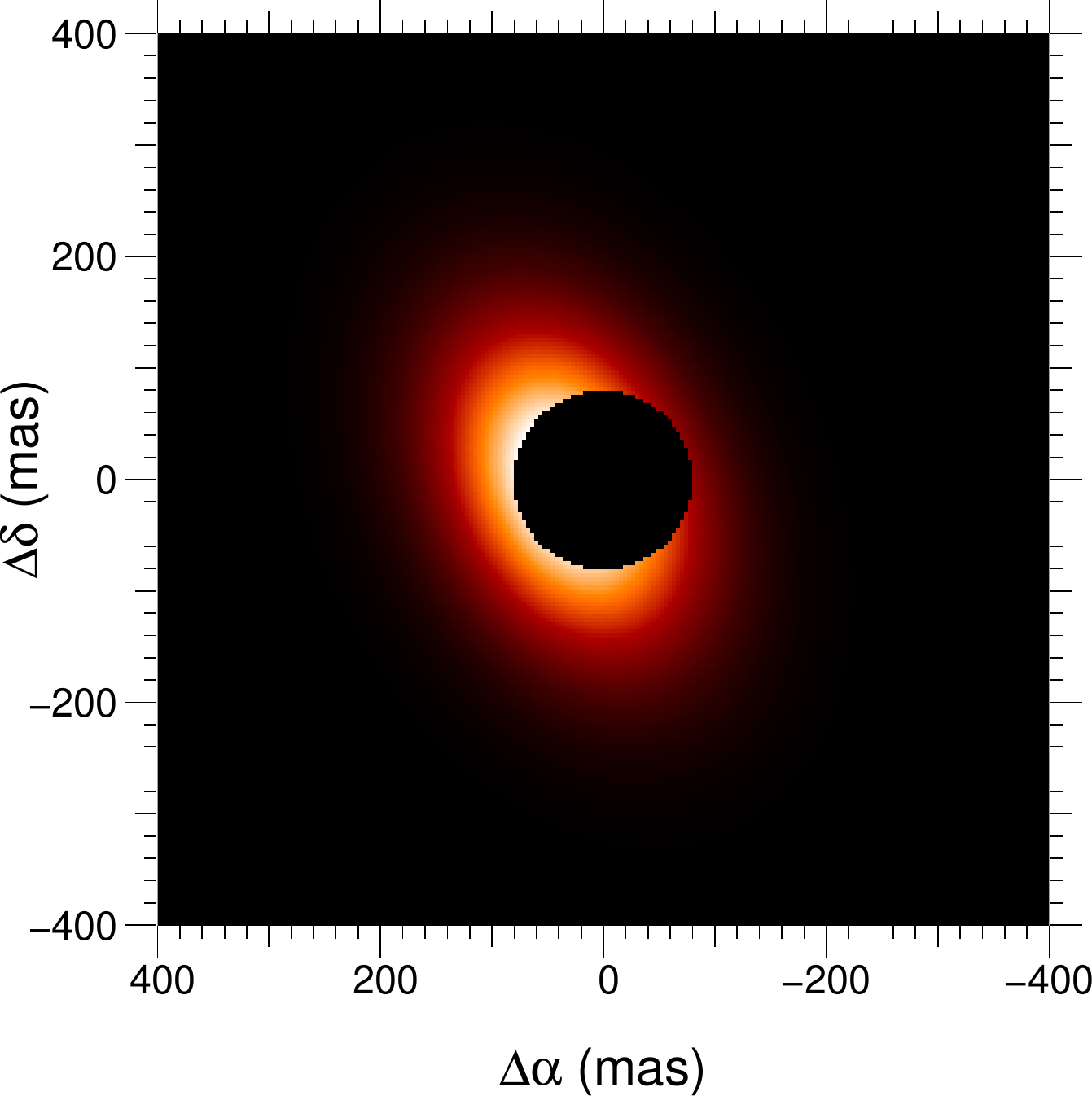}
 \includegraphics[width=4.2cm]{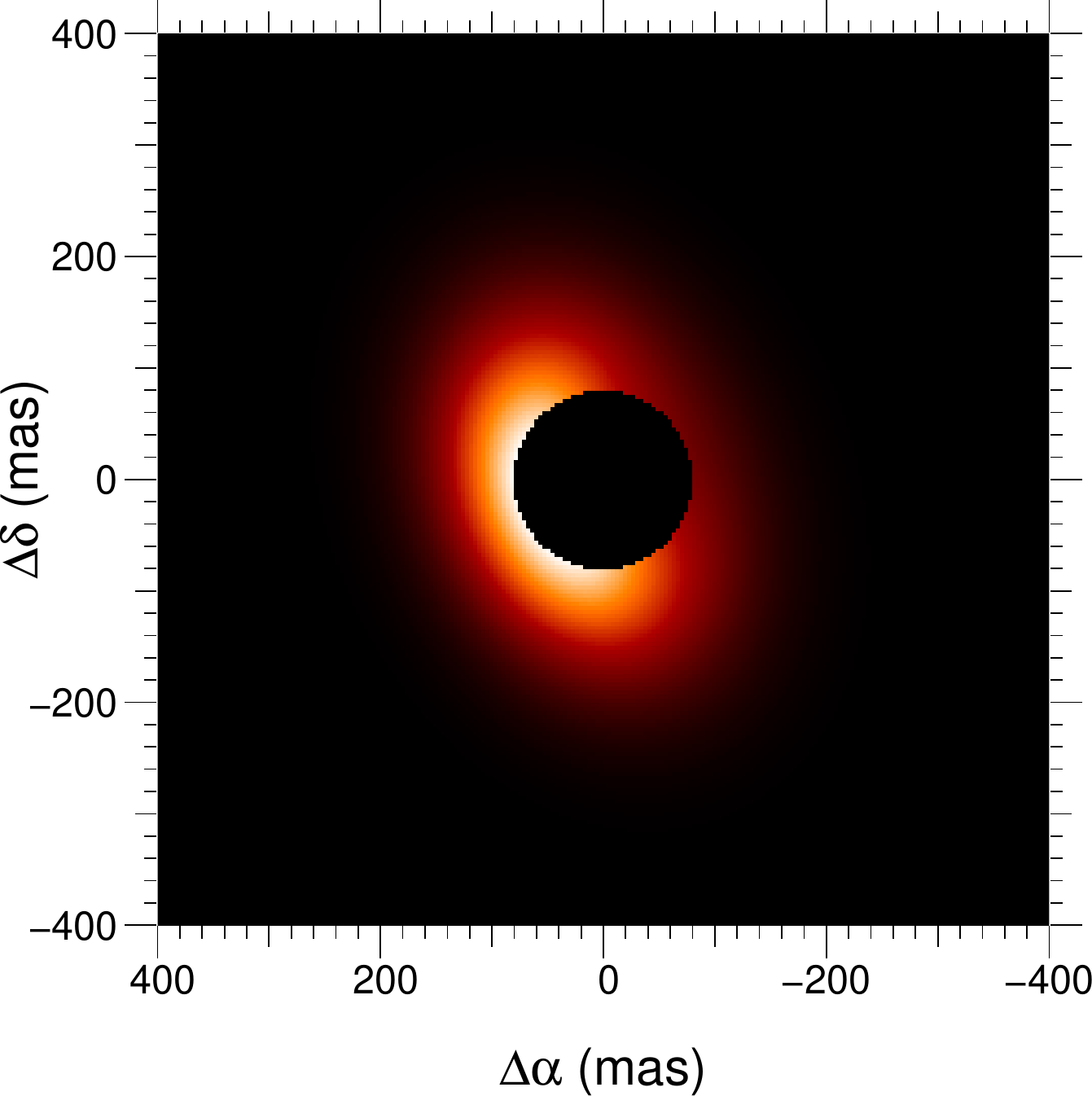}\\
 \hspace{0.7cm}
 \includegraphics[width=3.5cm]{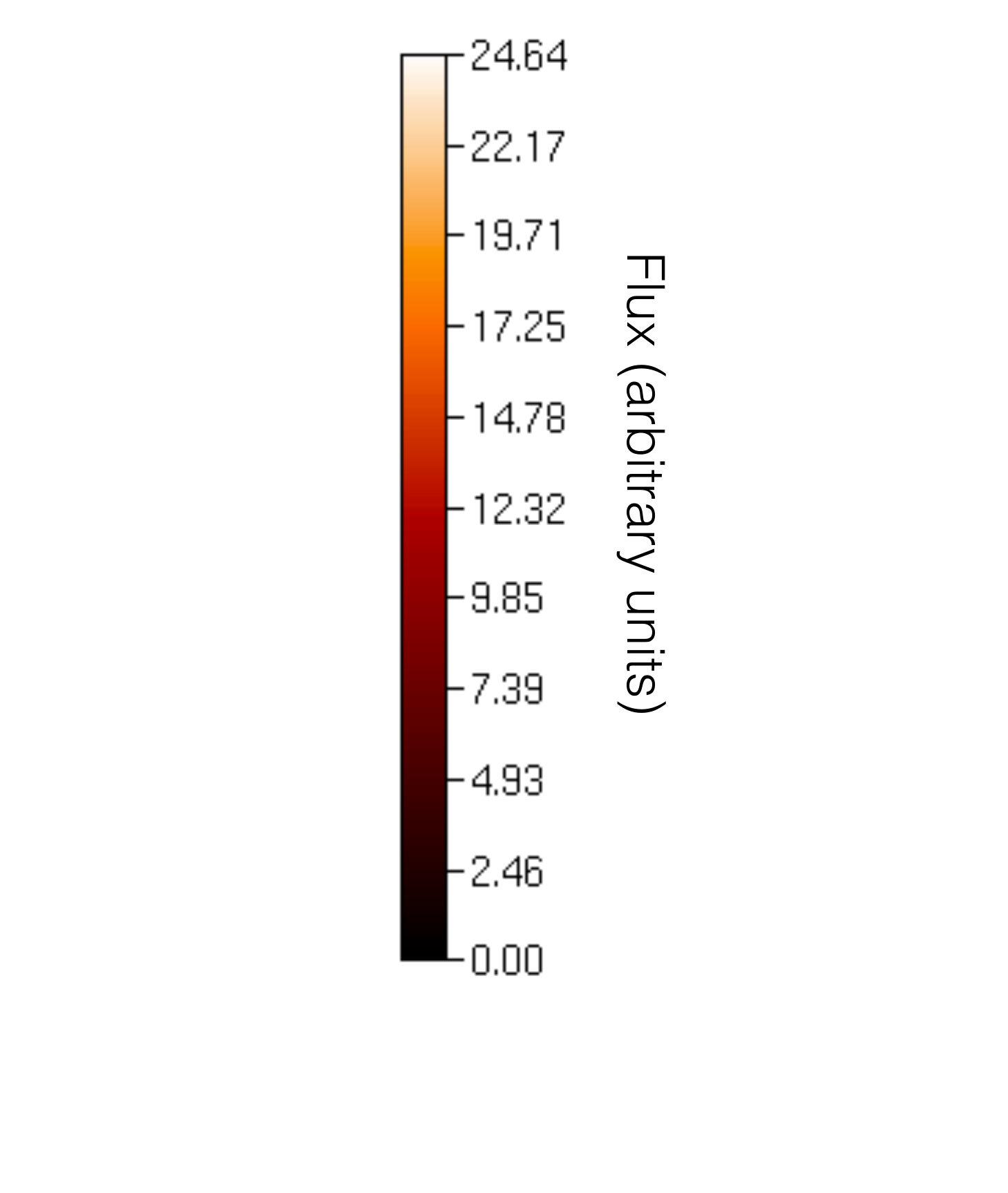}
  \includegraphics[width=4.2cm]{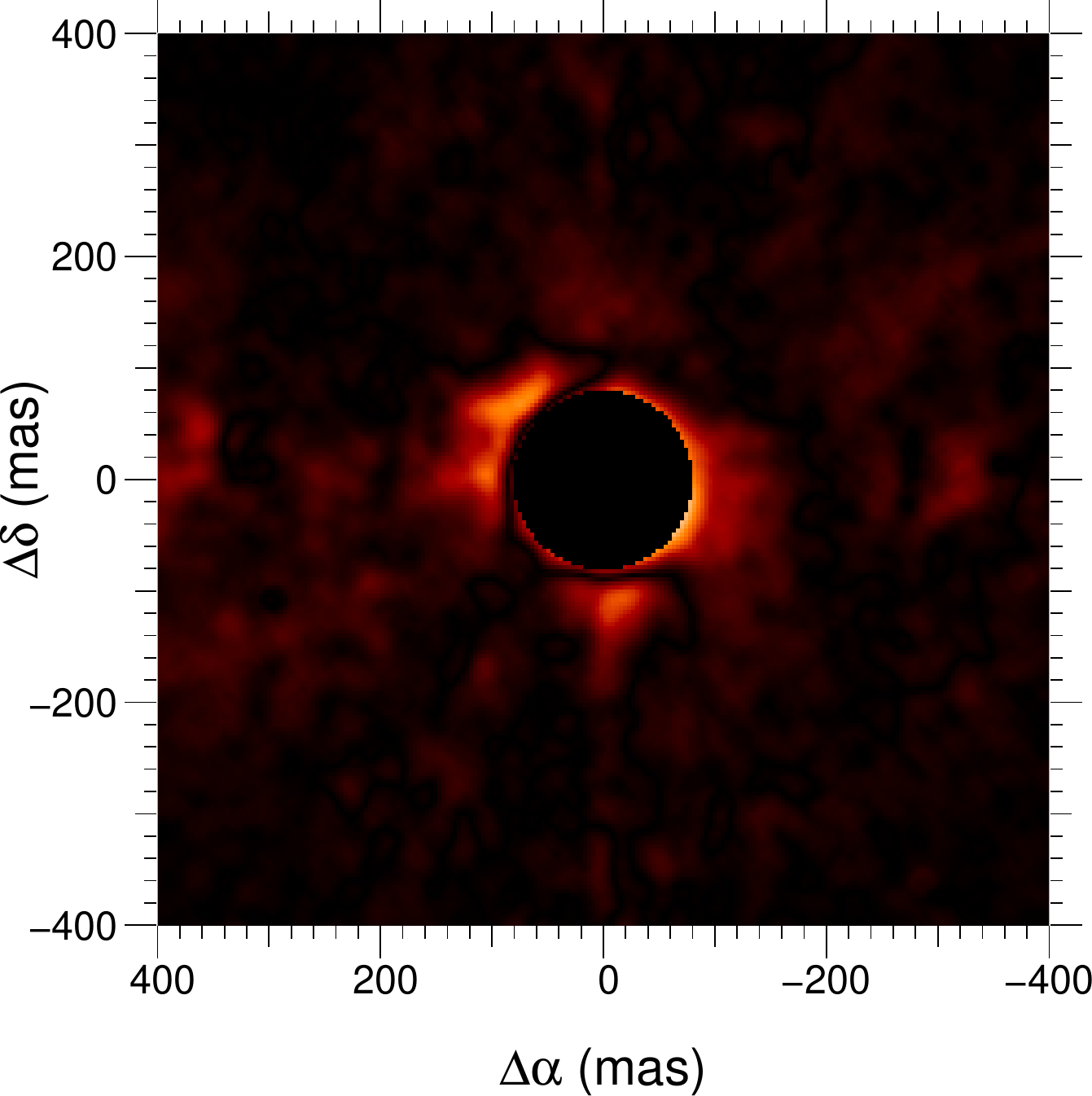}
 \includegraphics[width=4.2cm]{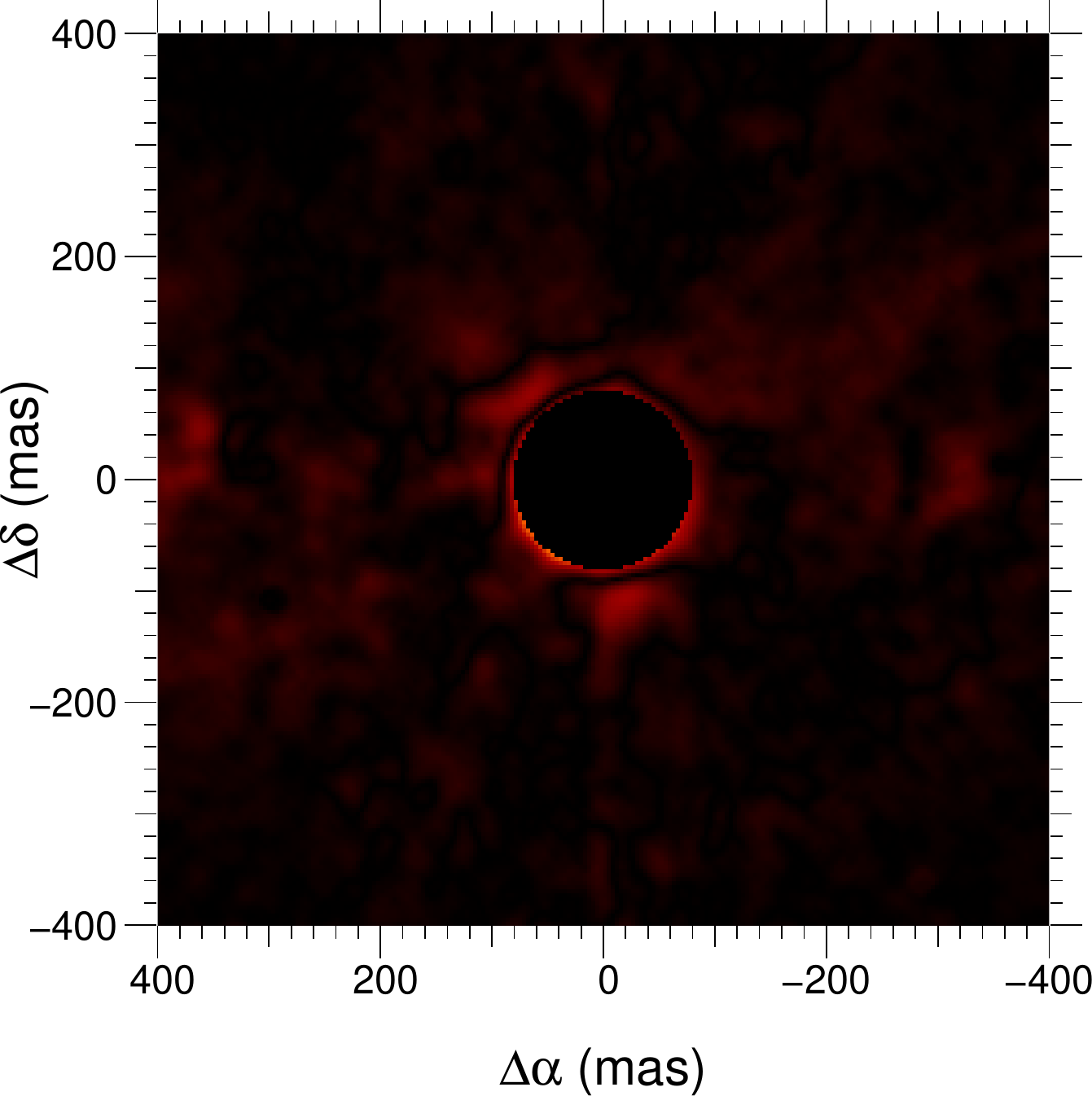}
  \includegraphics[width=4.2cm]{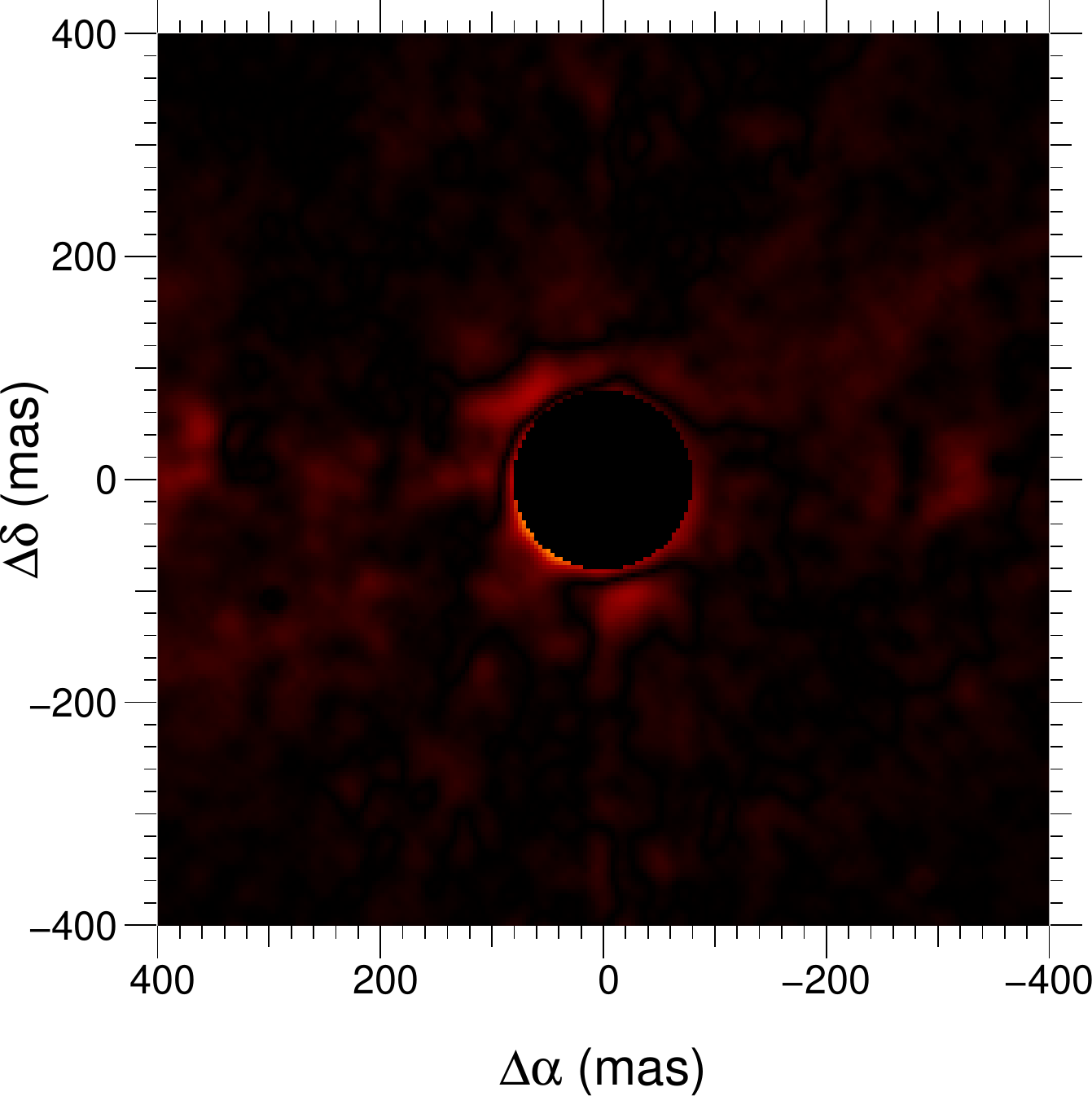}
      \caption{Images of the fit to the SPHERE data. Top row, left to right: Sphere image, best fit model images with the coronagraph for the \modif{centered} Gaussian, the \modif{off-center} Gaussian and the skewed Gaussian ring respectively.
      Bottom row, from left to right: residuals to the \modif{centered} Gaussian, the \modif{off-center} Gaussian and the skewed Gaussian ring respectively.
              }
         \label{img:SPHEREfit}
   \end{figure*}
   
    \begin{table*} 
\begin{center}
\caption{Best fit model parameters to the SPHERE image. \label{tab:SPHEREfit} }
\begin{tabular}{c|rcl|rcl|rcl}
\tableline
Model &  \multicolumn{3}{c|}{Centered Gaussian}  & \multicolumn{3}{c|}{Off-centered Gaussian} & \multicolumn{3}{c}{Skewed ring} \\
\tableline
$\chi^2_\mathrm{norm}$ & \multicolumn{3}{c|}{2.01}  &\multicolumn{3}{c|}{1.02}  & \multicolumn{3}{c}{1.00} \\
\tableline
Param. & Value& $\pm$ &Err & Value& $\pm$ &Err & Value& $\pm$ &Err \\
\hline
$R$ $[$mas$]$& & - & & & - &  &55.7&$\pm$&6.3 \\
$w$ $[$mas$]$ & 259.0 &$\pm$ & 1.7 & 252.1&$\pm$& 1.1& 205.0&$\pm$& 8.1\\
$i$ $[^\circ]$ & 40.5 & $\pm$& 0.5 &47.4 &$\pm$ &0.3 & 41.4 &$\pm$&0.7 \\
$\theta$  $[^\circ]$ & 23.7 & $\pm$& 0.9 &24.8&$\pm$& 0.4& 21.5 &$\pm$&0.4\\
$s$ & &- & & & -& & 0.53&$\pm$& 0.01\\
$x_\mathrm{*}$ $[$mas$]$ &  &- & & 29.7 &$\pm$ & 0.3 & &- \\
$y_\mathrm{*}$ $[$mas$]$& &- & & -5.4 &$\pm$ & 0.4 & &-
\end{tabular}
\end{center}
\end{table*}

The SPHERE/ZIMPOL image (Fig.\,\ref{fig:SphereImg}) shows two patches of flux coming from the East and the South just outside the \modif{coronagraphic mask}.
The East patch is more luminous and more extended.
In the context of the stellar light scattered on the disk surface these patches are related and can be interpreted as coming from an inclined disk where we see only one side (the North-West side not being illuminated).

In order to reproduce this emission geometry we used geometric models of a centered Gaussian, off-centered Gaussian and a skewed Gaussian ring to reproduce the disk surface.
The models are fully described in the Appendix\,\ref{app:modelSPHERE}.
Our Gaussian models consist of a Gaussian defined by its full width half maximum (FWHM; $w$).
This two-dimensional Gaussian model has a minor-to-major axis ratio of $\cos i$ and a position angle ($\theta$) defined from North to East.
In the off-centered case, this Gaussian can be shifted East and North with respect to the star by $x_*$ and $y_*$.
The skewed ring is defined by an infinitesimal ring with a radius ($R$), an inclination ($i$) and a position angle ($\theta$).
It is modulated azimuthally by a sine function starting at a major-axis and with an amplitude of $s$ (-1 $< s <$ 1). 
Finally, the ring is convolved by a Gaussian with a FWHM of $w$.

We show the best-fit model image in Fig.~\ref{img:SPHEREfit} and list the corresponding parameters in Table~\ref{tab:SPHEREfit}.
The coronagraphic mask sets our inner working angle to $\sim 150$\,mas, which prevents us from constraining the inner disk radius reliably.

The $\chi^2$ are normalized to the best fit model. 
The centered Gaussian model is the simplest model but also has the worst $\chi^2_\mathrm{norm}$ twice larger than the two other ones.
It means that, as expected, the flux is larger on one side of the coronagraph \modif{than} the other.
The two models reproducing this asymmetry have very similar $\chi^2_\mathrm{norm}$.
The best fitting model is the off-centered Gaussian model, with a 30\,mas separation along the direction of 105$^\circ$.
The second model which well reproduces this asymmetry is the skewed ring model.
The skewness ($s$=0.53$\pm$0.01) mimics radiative transfer effects of an inclined disk \citep[e.g.][]{Lazareff2016}.

Keeping in mind the disk interpretation, we can compare the sizes and orientations of the two best models.
The size of the Gaussian is 252.1$\pm$1.1\,mas.
The ring model has a radius of 55.7$\pm$6.3\,mas and a Gaussian width of 205$\pm$8.1\,mas.
The large error bars are due to the \modif{coronagraphic mask} that does not allow us to probe the inner parts of the disk, creating a degeneracy between the radius and the Gaussian width of the ring.
The inclinations of both the Gaussian ($i=47.4^\circ\pm0.3^\circ$) and the skewed ring ($i=41.4^\circ\pm0.7^\circ$) are different at 8.5$\sigma$.
The disk position angle ($\theta$) indicates roughly (at 8$\sigma$) the same orientation for the major axis ($\theta$=24.8$^\circ\pm$0.4$^\circ$ and $\theta$=21.5$^\circ\pm$0.4$^\circ$) for the Gaussian and the skewed ring respectively.

\subsection{MIR thermal emission}
\label{sec:MIDIfit}
\begin{figure}
   \centering
   \includegraphics[width=6.2cm]{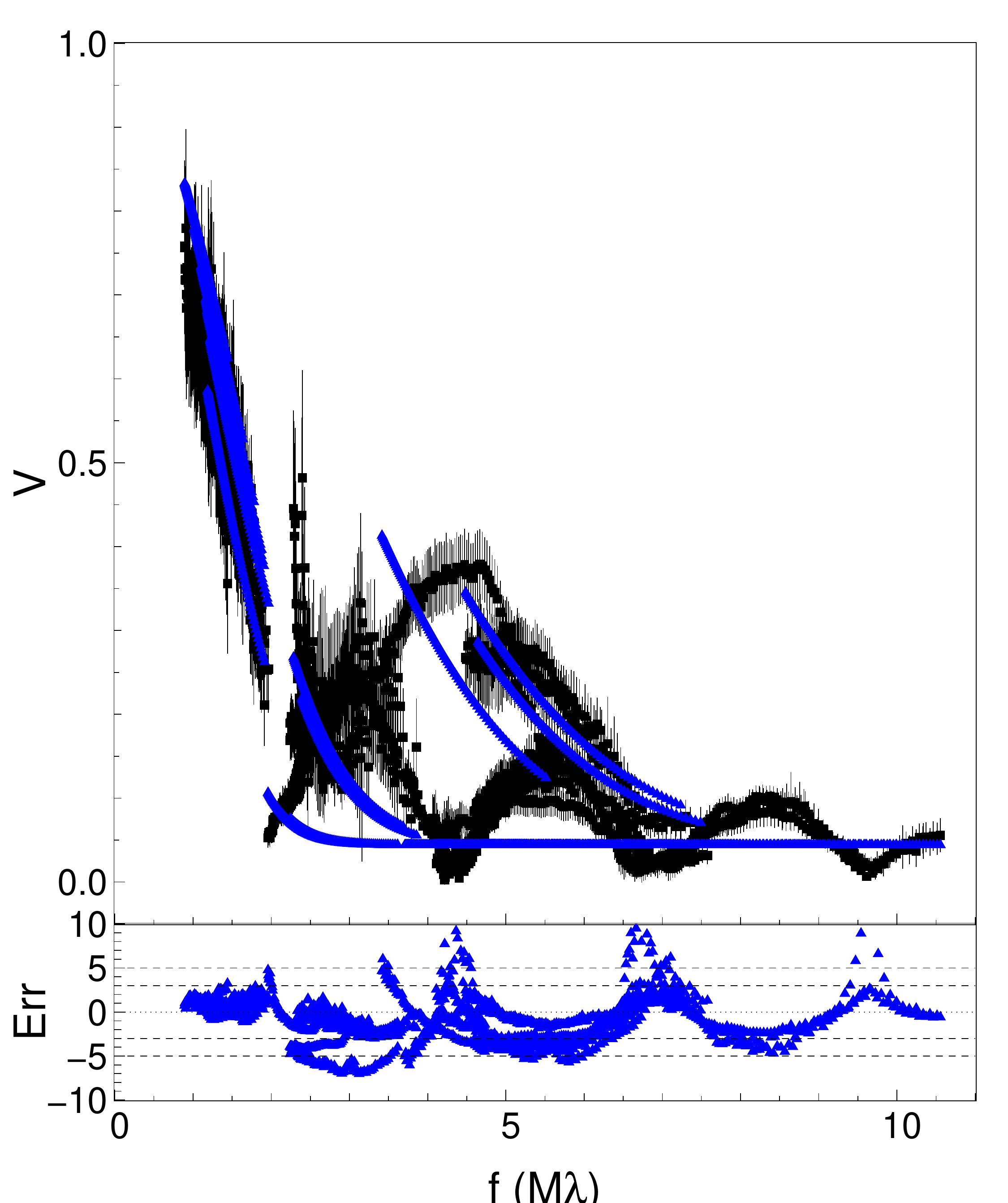}
  \includegraphics[width=6.2cm]{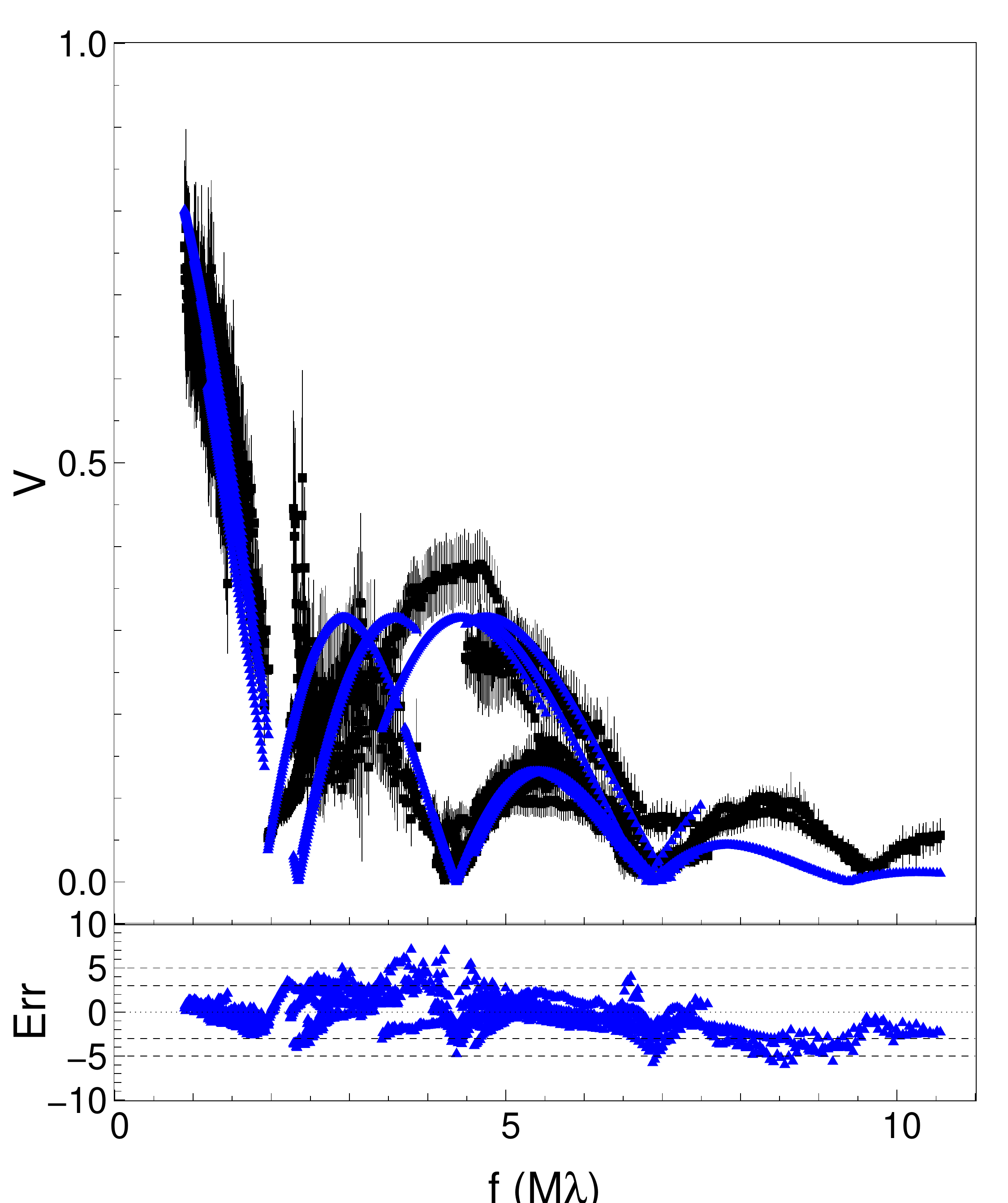}
  \caption{Fit to the MIDI visibilities between 8 and 12~$\mu$m. Left: The Gaussian model. Right: The Gaussian ring model. The dataset is in black squares and the model is in blue triangles. Bottom: the residuals. }
  \label{fig:VfitMIDI}
\end{figure}

\begin{table} 
\begin{center}
\label{tab:MIDIfit}
\caption{Best fit models to the MIDI dataset}
\begin{tabular}{c|rl|rl}
\tableline
 Model & \multicolumn{2}{c|}{Gaussian}& \multicolumn{2}{c}{Gaussian ring}  \\
\tableline
$\chi^2$ &  3.43 & & 1.61 \\
\tableline
$F_*^{10\mu\mathrm{m}}$ & 4.4\% &$\pm$1.0\% & 2.7\% & $\pm$0.9\% \\
$R$ [mas] & 0 & - & 41.8 & $\pm$0.9 \\
$w$ [mas] & 92.4 & $\pm$27.7 & 18.6 & $\pm$8.8 \\
$i$ [$^\circ$] & 74.2 & $\pm$12.1 & 52.5 & $\pm$6.1 \\
$\theta$ [$^\circ$] & 23.2 & $\pm$27.2 &26.4 & $\pm$6.2 \\
\tableline
\end{tabular}
\end{center}
\end{table}

The inner disk region is not resolved by SPHERE observations as it is inside the region covered by the coronagraph.
In order to characterize the disk gap we analyze archival MIDI data, which traces dust at temperatures of $\gtrsim 300$\,K.
This dataset is more complete than the one in \citet{Fedele2008} and we therefore expect to \modif{probe} the disk geometry in more detail.

The visibility data presents several lobes that seem inconsistent with a Gaussian structure. 
These lobes could be produced by a Bessel function that is the Fourier transform of a sharp ring.
We therefore decided to fit both a Gaussian and a Gaussian ring model.
The best fit parameters are presented in Table~\ref{tab:MIDIfit} with the error bars computed using a bootstrap method.

The Gaussian model differs from the Gaussian ring by setting the radius of the ring to 0.
The Gaussian model does not reproduce the dataset very well with a reduced $\chi^2$ of 3.43.
We can see that the orientation parameters are not well constrained (an error bar of 12.1$^\circ$ on the inclination and 27.2$^\circ$ on the position angle).

The Gaussian ring fit reproduces the visibility curve in a better way (see Fig.\,\ref{fig:VfitMIDI}) with a reduced $\chi^2$ of 1.61.
We can see that the ring radius is about 41.8$\pm$0.9\,mas.
The ring orientations ($i$=52.5$^\circ\pm$6.1$^\circ$, PA=26.4$^\circ\pm$6.2$^\circ$) are consistent with the ones derived from the SPHERE/ZIMPOL image ($i$=41.4$^\circ\pm$0.7$^\circ$, PA=32.3$^\circ\pm$1.0$^\circ$) and previous work \citep[$i$=57$^\circ\pm$2$^\circ$, PA=23$^\circ\pm$3$^\circ$;][]{Fedele2008}.
From the best fit model the central point source contribution (that can be interpreted as the stellar contribution) is very low ($2.7\%\pm0.9\%$), at 3-$\sigma$ from 0\% contribution.
This is different from the 20\% contribution of unresolved emission as fitted by \citet{Fedele2008}.

   \begin{figure}
   \centering
   \includegraphics[width=6cm]{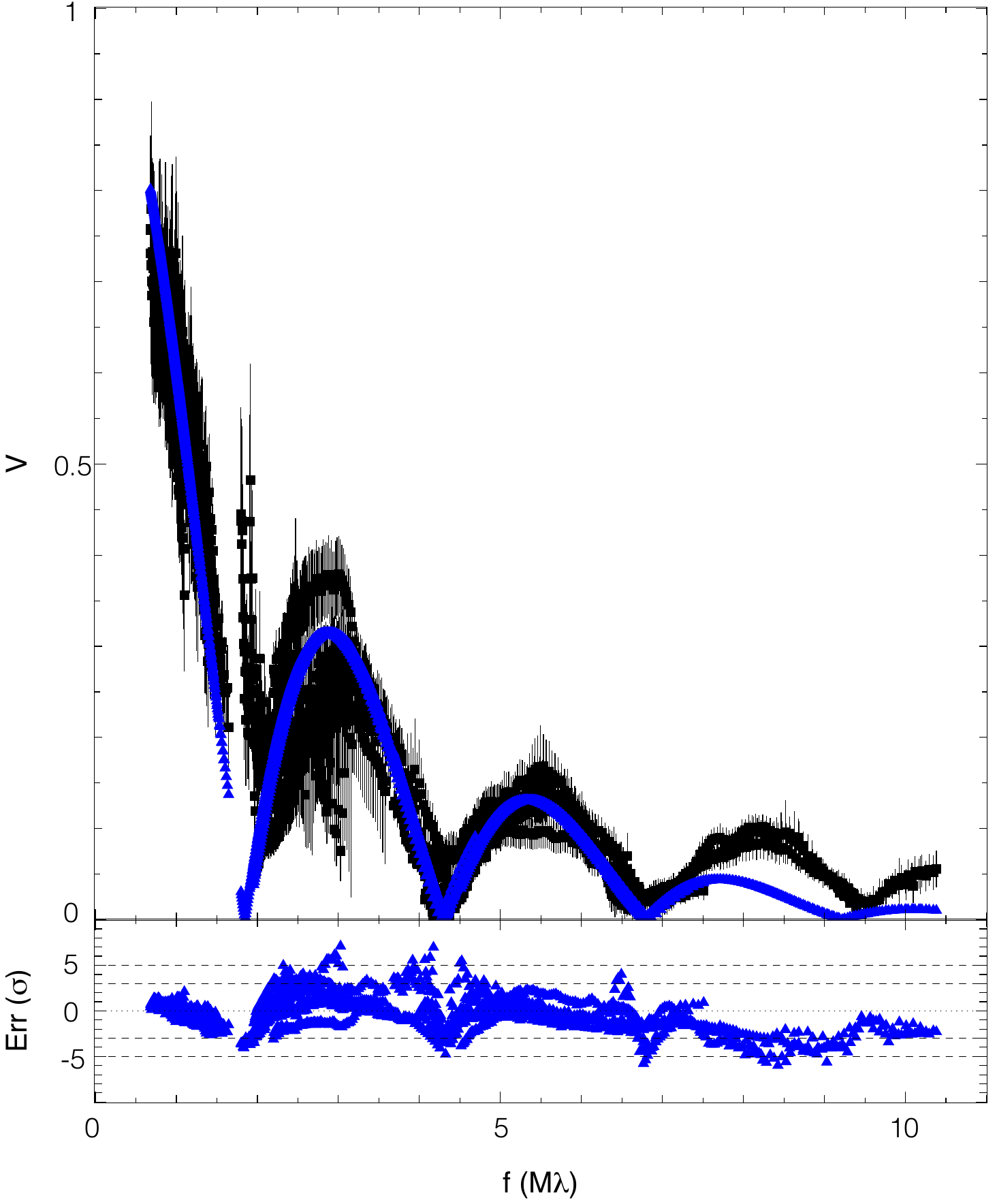}
      \caption{The MIDI visibilities (black squares) and the Gaussian ring best fit (blue triangles) vs. effective baseline, i.e. baseline that is oriented to match the inclination and position angle of the best fit Gaussian ring model ($i$=52.5$^\circ$ and PA=26.4$^\circ$). The visibility profile shows a clear Bessel-like function.
              }
         \label{fig:Beff}
   \end{figure}
We can verify the orientation of the best Gaussian ring fit by transforming the baseline length to the effective baseline length corrected for the inclination and position angle of the best fit (i.e. $i$=52.5$^\circ$ and PA=26.4$^\circ$).
In Fig.\,\ref{fig:Beff} we can see that the data points align well and form a Bessel-like function.
We also note that the best fit follows the general trend of this profile even though the visibility level from our model is too low for the longest baselines.

\subsection {NIR thermal emission}

In order to probe the geometry of the innermost regions of the disk (\textless10\,au) we use a similar model that was used in Sect.\,\ref{sec:MIDIfit} adding some parameters that are probed by the NIR dataset.
We also use the image reconstruction technique to recover the intensity distribution for the dataset with the best sampled \uv-plane which is the SAM dataset.

\subsubsection{Model Fitting}
\label{sec:modelfitting}

After normalizing the short-baseline sparse aperture masking visibility data (see Appendix\,\ref{sec:V2cal}), we fit our analytic disk model to the NIR interferometric datasets.
The model consists of a point source representing the star and a Gaussian that represents the dusty environment of the star.
The star can be shifted with respect to the Gaussian to reproduce non-zero CP signal seen in the aperture masking dataset (see Fig.\,\ref{fig:NIRC2}, bottom right).
The details of the model are presented in the Appendix \ref{app:modelVis}.

\begin{figure}
 \centering
 \includegraphics[width=4.2cm]{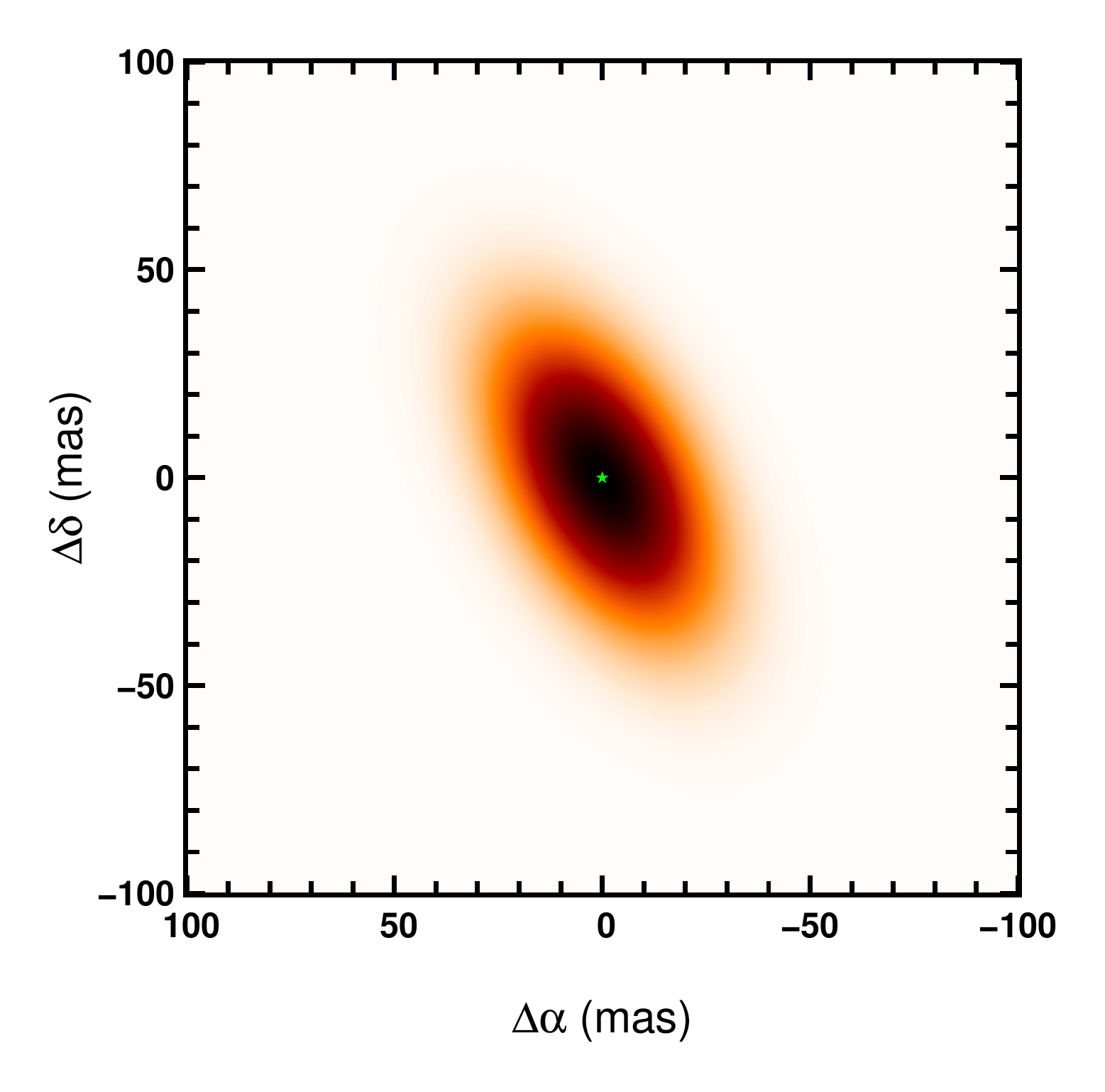}
 \includegraphics[width=4.2cm]{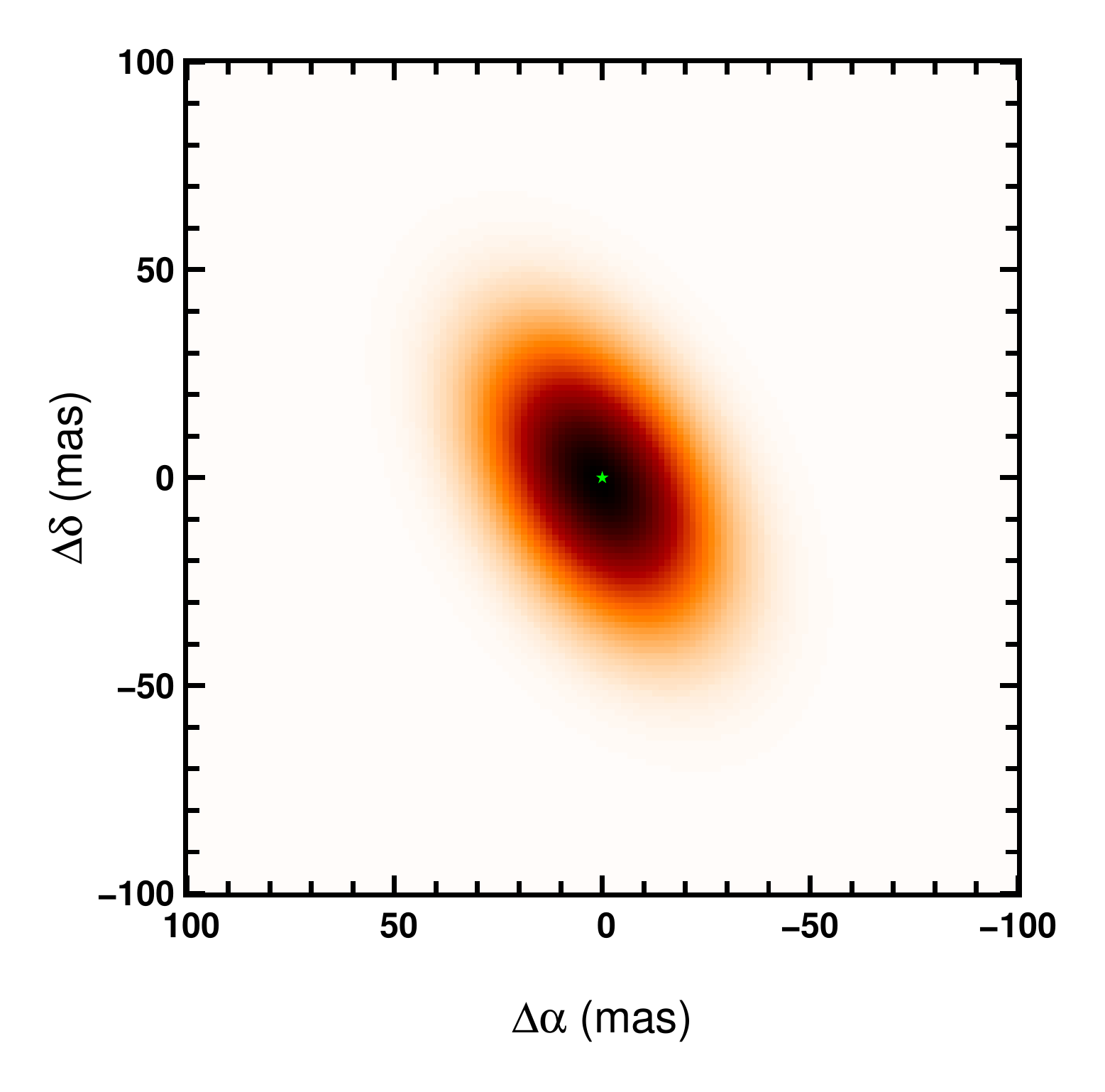}
 \includegraphics[width=4.2cm]{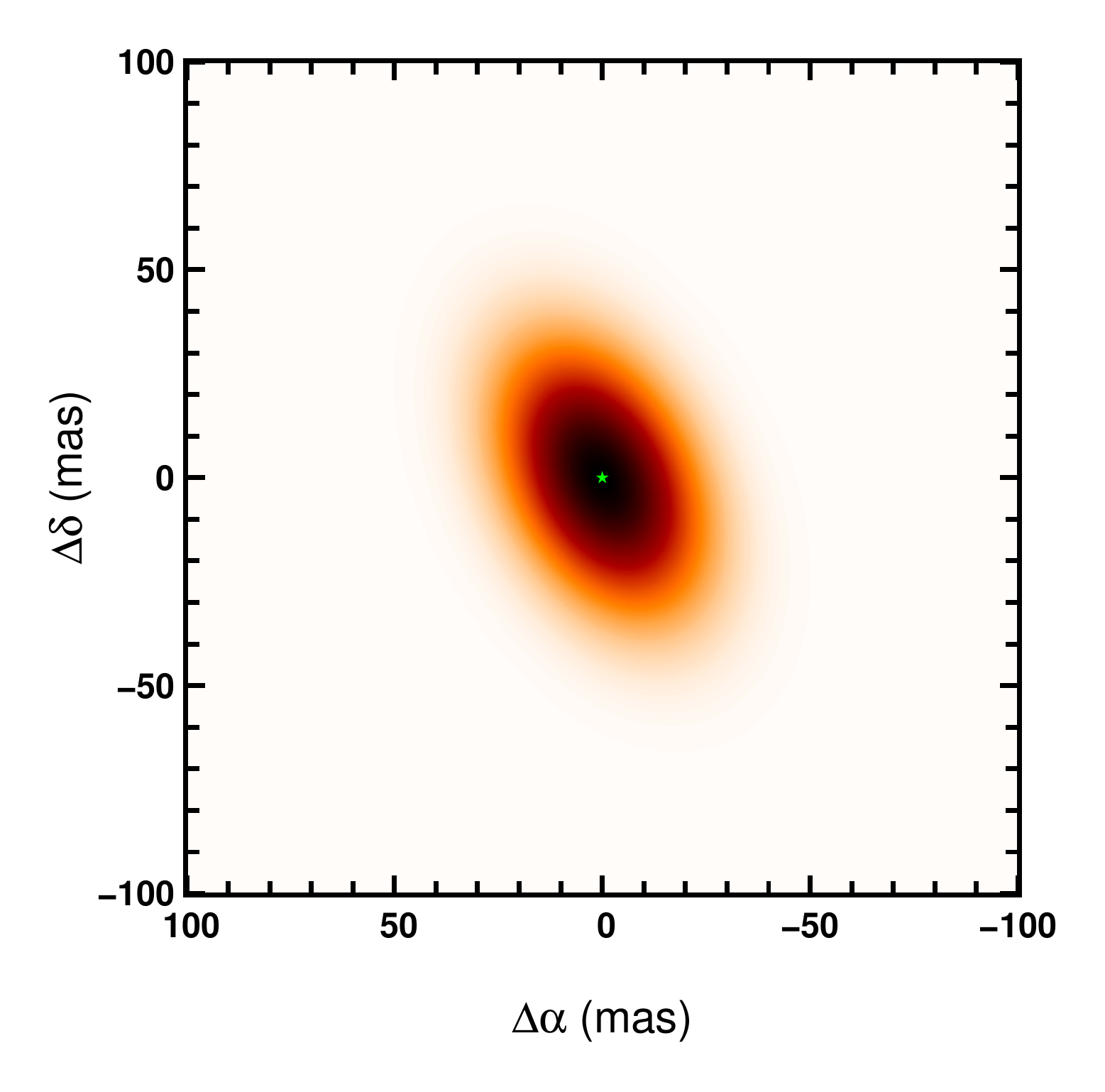}
 \includegraphics[width=4.2cm]{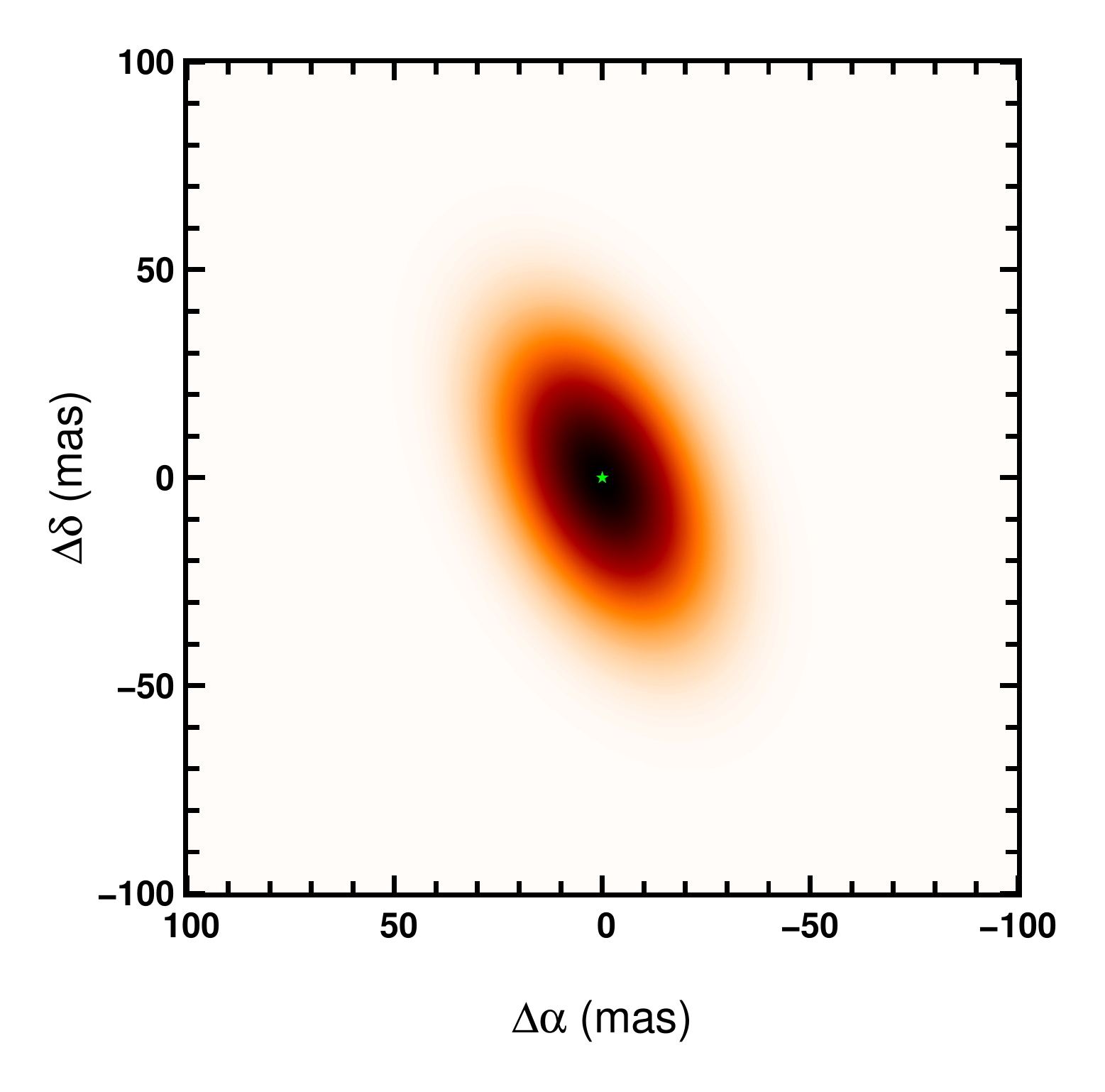}
 \includegraphics[width=4.4cm]{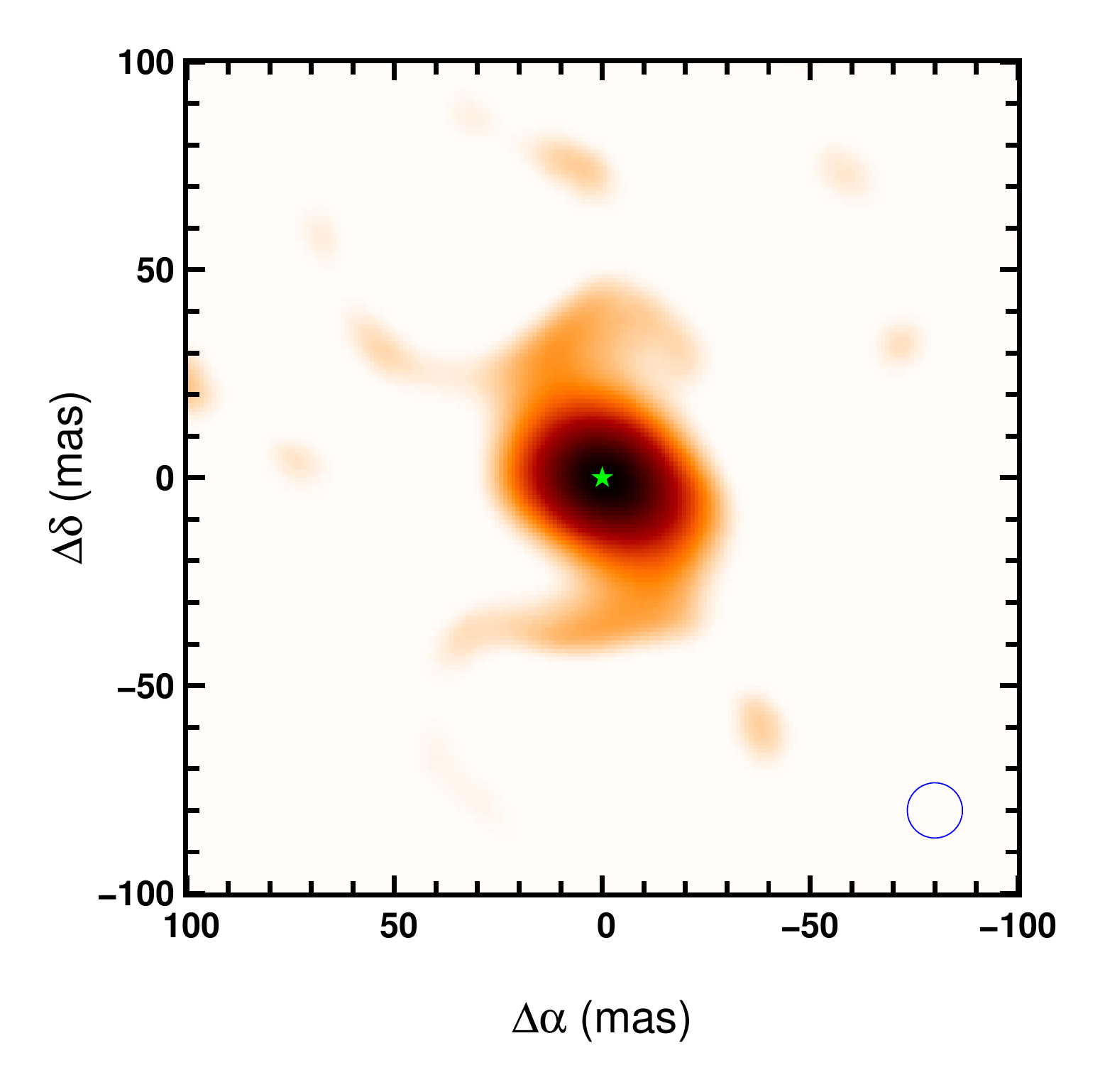}
      \caption{Model images that correspond to our best fit models fitted to the following data sets: top-left:  SAM, top-right: SAM+PIONIER, middle-left: SAM+PIONIER+AMBER. middle-right: SAM+PIONIER+AMBER+CHARA. Bottom: the image reconstruction on the SAM data. The green star represents the position of the star and the blue solid contour represents the beam size.
      }
         \label{fig:fitres}
   \end{figure}

We first applied our model to the aperture masking data only, as this data set alone constrains the size of the extended emitting region. 
We then successively added the PIONIER dataset (which covers the same band as the aperture masking data), the longer-wavelength AMBER and CHARA data as the final step.
Because the SAM V$^2$ data sets the size and orientation of the emission and because it has large error bars, we increased its weight in the fitting by a factor of 25 (corresponding to artificially reducing its error bars by a factor of 5).
The best-fit results for these three fits are presented in Table \ref{tab:fitres} and the corresponding model images in Fig.~\ref{fig:fitres}.

\begin{table*} 
\begin{center}
\caption{Parameters of the best-fit Gaussian model to our various data sets. \label{tab:fitres}}
\begin{tabular}{c|rl|rl|rl|rl}
\tableline
\tableline
Parameters & \multicolumn{2}{c|}{SAM}  & \multicolumn{2}{c|}{SAM+PIONIER} & \multicolumn{2}{c|}{SAM+PIONIER+AMBER} & \multicolumn{2}{c}{All}\\
\tableline
$\chi^2$ & 20.4 &  & 9.3 &  & 8.4 &  & 6.3\\
$\chi^2_\mathrm{V2}$ & 0.3 &  & 0.1 &  & 0.1 & & 0.1 \\
$\chi^2_\mathrm{CP}$ & 53.5 &  & 15.9 &  & 17.0 & & 14.6\\
\tableline
$F_*^{1.65\mu\mathrm{m}}$ [\%] & 48.9 & $\pm$3.1 & 59.8 & $\pm$0.6 & 59.0 & $\pm$0.4 & 59.5 &$\pm$ 0.4\\
$w$ [mas] & 55.6 & $\pm$ 22.1 & 52.0 &$\pm$6.4 & 47.7 & $\pm$2.4 & 51.0 &$\pm$2.4\\
$i$ [$^\circ$] & 58.0 & $\pm$11.3 & 52.9 & $\pm$7.7 & 51.3 & $\pm$3.6 & 54.8 & $\pm$3.0\\
$\theta$ [$^\circ$] & 28.0 & $\pm$5.9 & 31.1 & $\pm$22.1 & 26.3 & $\pm$3.6 & 25.5 & $\pm$2.6\\
$T$ [K] & - & - & 1719 & $\pm$154 & 1812 & $\pm$71 & 1435 &$\pm$28 \\
$x_\mathrm{*}$[mas] & -1.05 & $\pm$0.71 & -1.58 & $\pm$0.49 & -0.80 & $\pm$0.11 & 0.81 & $\pm$ 0.09 \\
$y_*$ [mas] & -0.61 & $\pm$1.02 & -1.38 & $\pm$0.48 & -0.12 & $\pm$0.05 & 0.11 & $\pm$0.04\\
\tableline
\end{tabular}
\end{center}
\end{table*}

We can see that a similar disk orientation is found for the different model fits: inclinations between 51$^\circ$ and 58$^\circ$ and position angles between 25$^\circ$ and 31$^\circ$.
These numbers are in very good agreement (within 1 or 2-$\sigma$ for the inclination and less than 1-$\sigma$ for the position angle) with the orientations derived from the SPHERE image in Sect.\,\ref{sec:SPHEREfit} and the MIDI dataset in Sect.\,\ref{sec:MIDIfit}.

From the squared visibility fit, a Gaussian-like geometry seems to suit the global shape of the NIR circumstellar emission.
The derived FWHM of $\approx$\,50\,mas is consistent between the different fits.
The size is well-constrained, in particular at the PIONIER, AMBER and CHARA baselines where the environment is fully resolved. 
The emission is therefore smoothly distributed and does not show a clear rim-like profile like in the MIDI observations tracing cooler disk regions at longer wavelengths (where we see the different lobes coming from a Bessel function indicating a sharp morphology; see Sect.\,\ref{sec:MIDI}).

The stellar-to-total flux ratio at 1.65$\mu$m ($F^{1.65\mu\mathrm{m}}_*$) shows a rise of about 10\% between the SAM fit and the three other fits.
This is due to the fact that this ratio is well constrained once the visibilities reach the over-resolved regime, which is the case for \modif{the} PIONIER, AMBER and CHARA datasets.
At spatial frequencies over 5M$\lambda$ (corresponding to baselines $\gtrsim 10$\,m at 1.65\,$\mu$m),  the extended component is over-resolved and the flux ratio can be constrained very tightly from the plateau level in the visibility, if the long-baseline data is included.
The best fits to the SAM+PIONIER, SAM+PIONIER+AMBER and SAM+PIONIER+CHARA data agree on a value of $F^\mathrm{1.65\mu m}_*\sim59\pm$0.6\% of unresolved flux at 1.65$\mu$m.

The spectral dispersion of the $H$-band PIONIER and $K$-band AMBER and CHARA datasets and the span across two bands for the whole long-baseline dataset allow us to probe the temperature of the environment assuming that the star is in the Rayleigh-Jeans regime.
Our fit indicates a temperature of $\sim$1800\,K, with $F^*_\lambda \propto \lambda^{-4}$ for the SAM+PIONIER and SAM+PIONIER+AMBER datasets ($T$=1719$\pm$154\,K and $T$=1812$\pm$71\,K respectively).
However the best fit to the SAM+PIONIER~+AMBER+CHARA dataset indicates a lower temperature ($T$=1407$\pm$26\,K).

Significant deviations from zero CPs are observed in the SAM data, suggesting the brightness distribution is non-axisymmetric. 
The only model parameters that produce a non-zero CP signal are the coordinates of the central star ($x_*$ and $y_*$) relative to the center of the inclined Gaussian. 
We notice the high $\chi^2$ for the fit of the SAM dataset and that the CPs are not well fitted by the four models (Fig.\,\ref{fig:fitresV2CP}). 
This suggests a more complex geometry for the NIR emission.

\subsubsection{Image reconstruction}

Another approach to determine the emission morphology is to reconstruct an image using aperture synthesis methods.
This approach remains model-independent and is more likely to be able to fit the CPs seen in the aperture masking data and that were not reproduced by the parametric model.
We reconstruct an image from the SAM dataset alone, as this dataset provides a rather uniform \uv-coverage. 
Adding the PIONIER and AMBER datasets would introduce artifacts in the image reconstruction, as the Fourier plane coverage is relatively poor in these cases.
We carry out the image reconstruction using the \texttt{MiRA} software package \citep{Thiebaut2008} together with the SPARCO technique \citep{Kluska2014} that separates the star from the image of its environment using the linearity of the Fourier transform.
We use a quadratic smoothness regularization and determine the regularization weight ($\mu$) using the L-curve method \citep{Kluska2016,Willson2016a}, which yields the best-fit value $\mu=9\times10^9$.
When determining this value, we set the star/disk flux ratio to the level determined by the parametric models (i.e.\ $F^\mathrm{1.65\mu m}_*$ = 59.0\%).

\begin{figure}
\centering
\includegraphics[width=4.2cm]{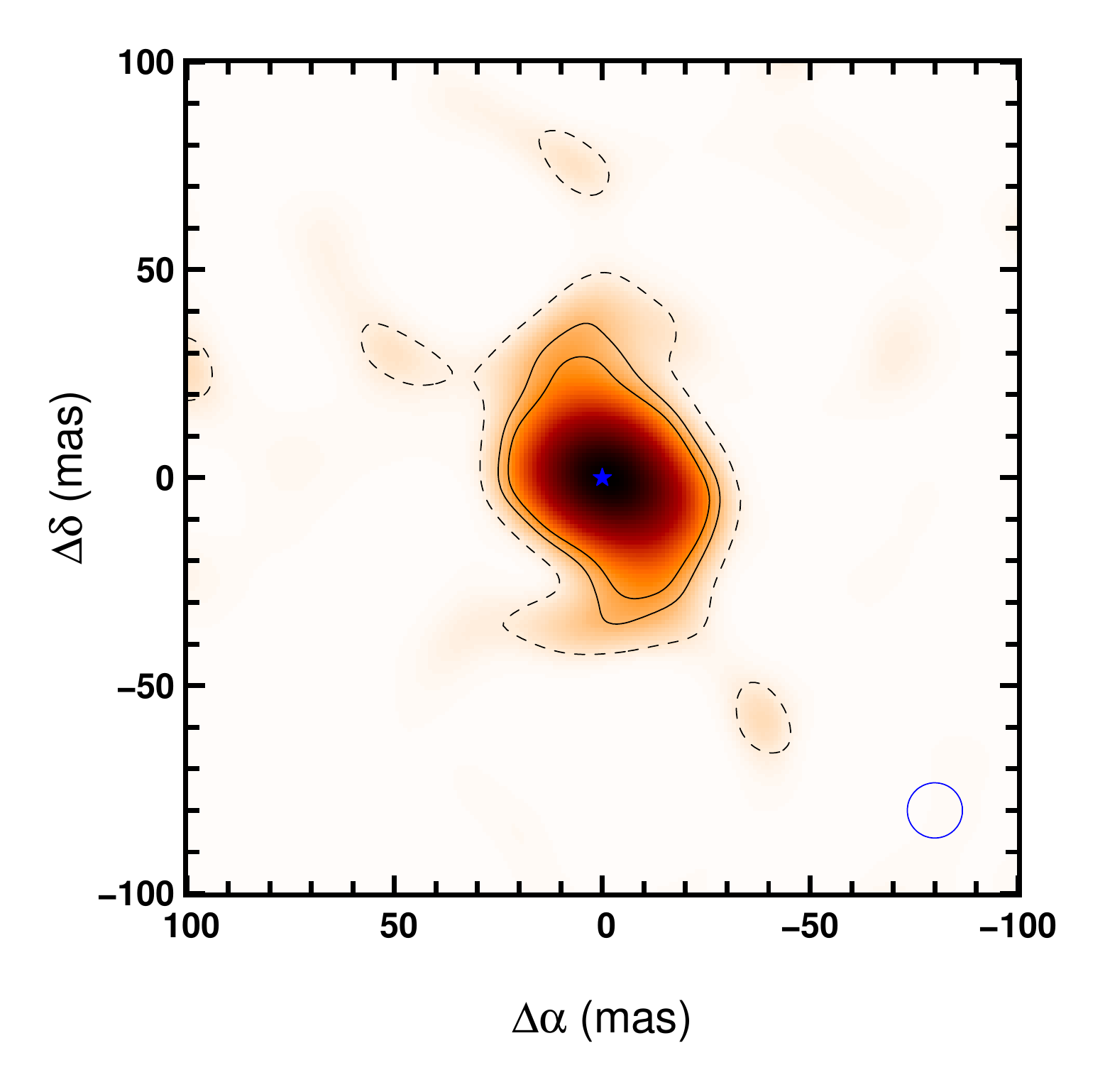}
\includegraphics[width=4.2cm]{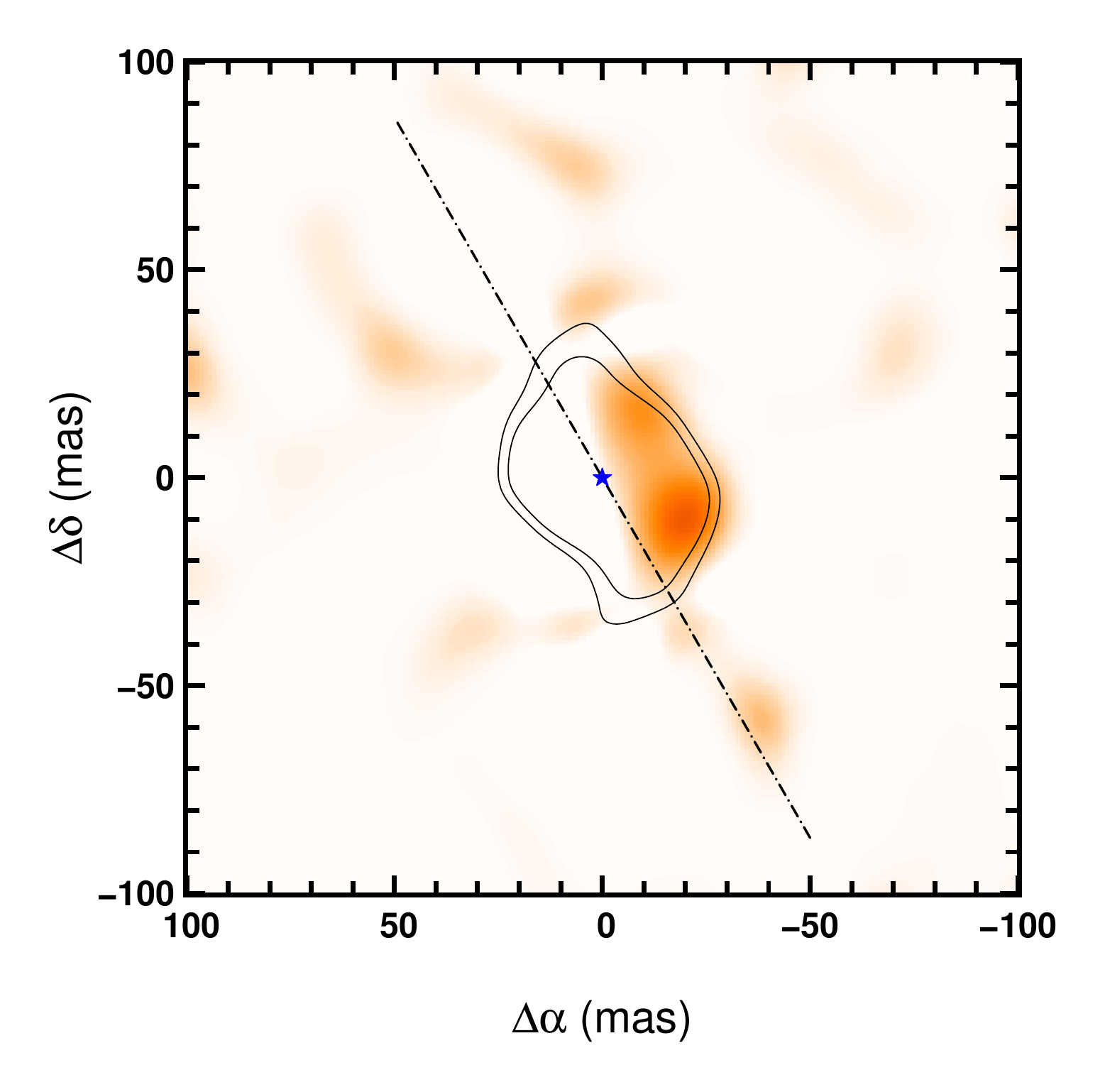}
\caption{Analysis of the reconstructed image. Left: average image reconstruction built from the bootstrap method. The green star represents the position of the star, the solid and dashed lines are the 5, 3 and 1-$\sigma$ significance contours and the blue solid contour represents the beam size.  Right: the asymmetry map (see text) with contours of the original image at 3 and 5-$\sigma$ significance overlaid. The green star represents the star and the dashed line represent the major-axis. }
\label{fig:boot}
\end{figure}

Our reconstructed image is shown in Fig.\,\ref{fig:fitres}.
The image has a reduced $\chi^2=1.2$.
The general size and orientation of the structure seen in the image is similar to the one found in the parametric models.
However, the image also reveals spiral features that seem to extend from the major axis of the disk.
To assess the significance of the features seen in the image we used the bootstrap method, where we reconstruct images from 500 new datasets built by drawing baselines and triangles from the full dataset \citep[see][]{Kluska2016}.
Fig.\,\ref{fig:boot} shows the average image computed for all bootstrap reconstructions together with contours corresponding to the pixel significance levels.
The spiral features are significant up to a level of 3-$\sigma$ making them marginally significant.

In contrast to the parametric models, the image reconstruction is able to reproduce the measured CP signals reasonably well.
To see which parts of the image influence the CP signal we produce an asymmetry map, which is computed by subtracting the 180\degree~rotated image from the original image \citep{Kluska2016}.
The asymmetry map is shown in Fig.~\ref{fig:boot} and reveals that the CP signal is mainly caused by the north-western disk region.
This part is located on one side from the major-axis indicating inclination effects, even though this axis is misaligned slightly (by $\sim$15$^\circ$) with respect to the disk major axis determined in Sect.~\ref{sec:modelfitting}.

\subsection{\modif{Companion detection limits inside the cavity}}
\label{sec:detlim}

\begin{figure}
    \centering
    \includegraphics[width=4.2cm]{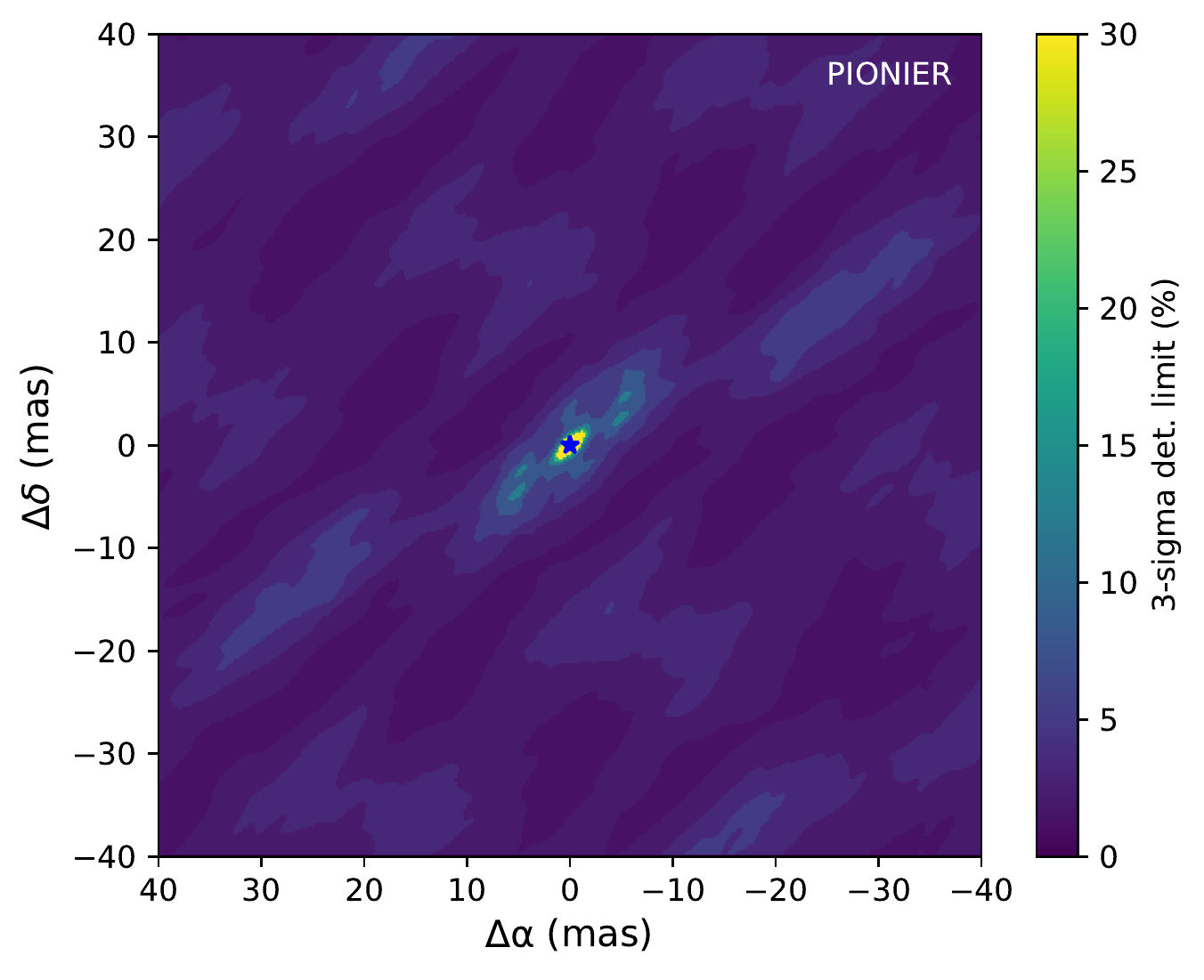}
    \includegraphics[width=4.2cm]{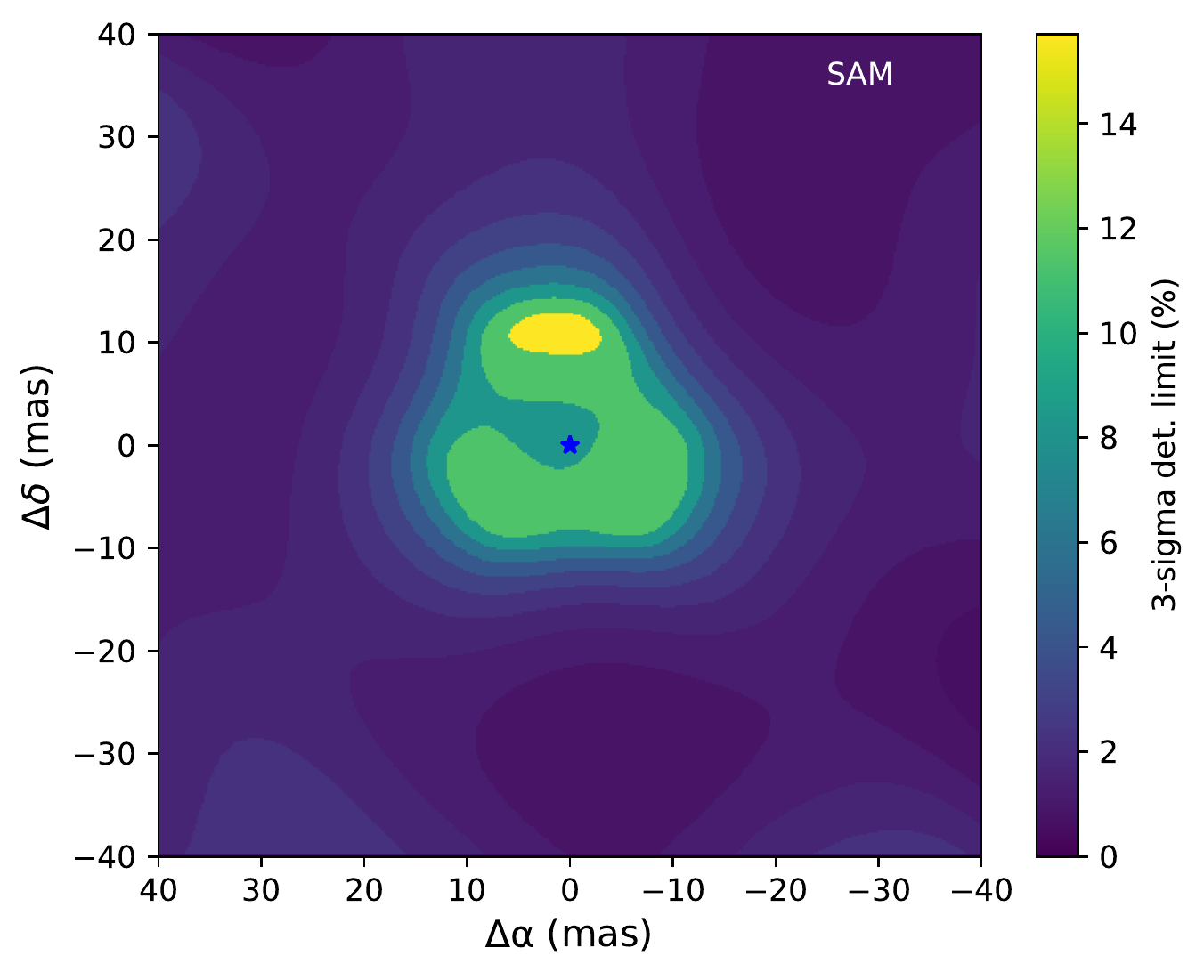}
    \includegraphics[width=4.2cm]{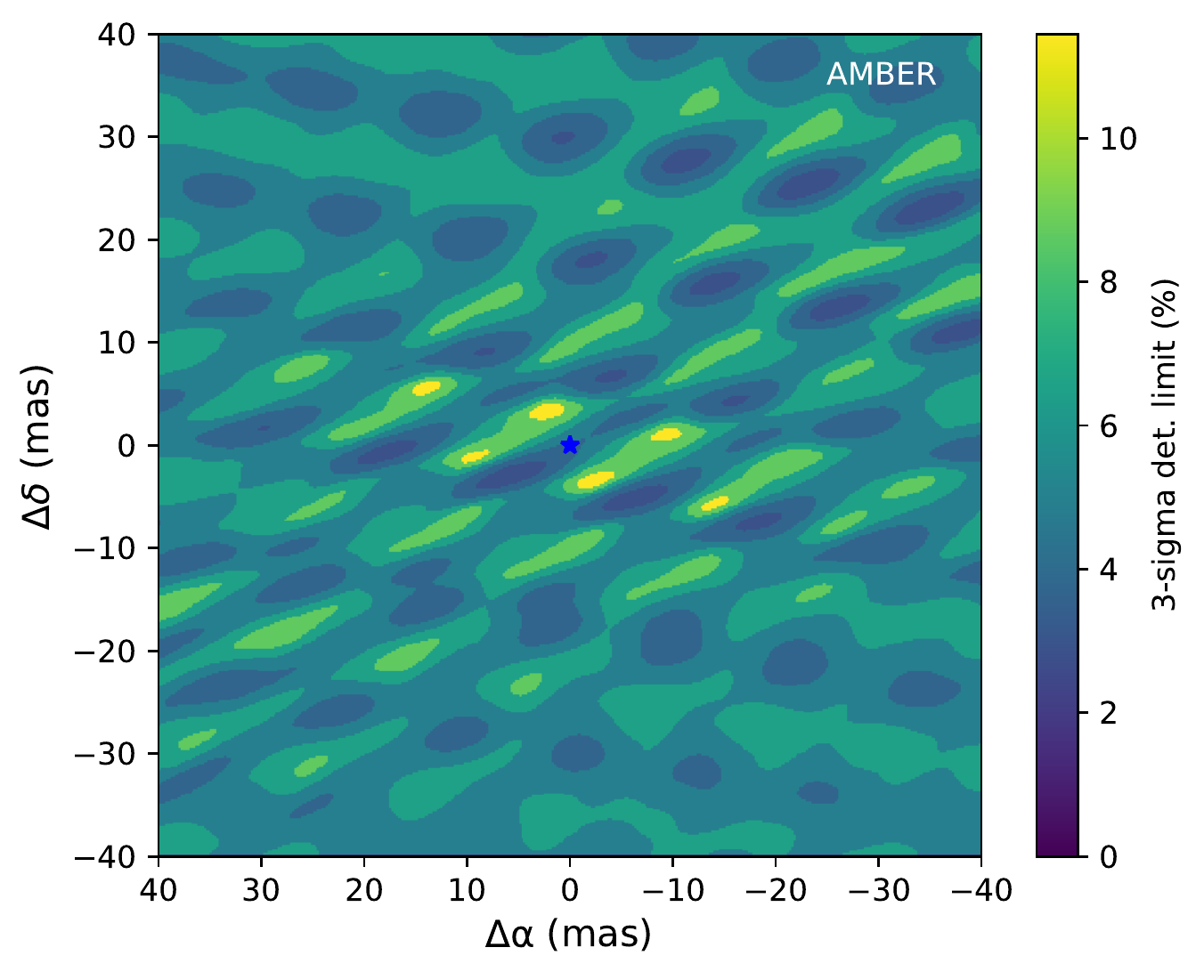}
    \includegraphics[width=4.2cm]{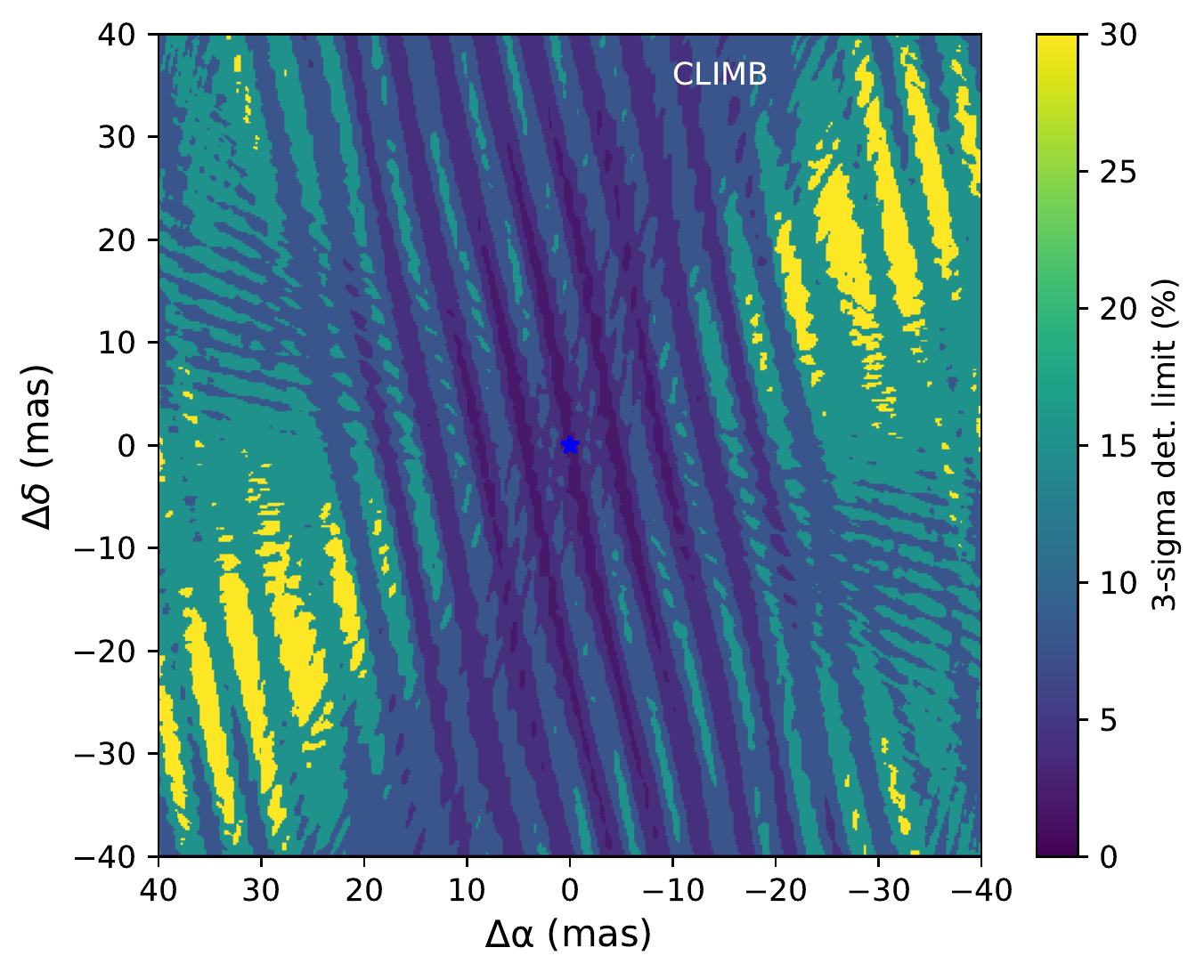}
    \caption{\modif{2D maps of 3-$\sigma$ detection limits from interferometric data in percentage of flux contribution in the respective bands. Top-left: SAM, top-right: PIONIER, bottom-left: AMBER, bottom-right: CLIMB}}
    \label{fig:detlim}
\end{figure}

\begin{figure}
    \centering
    \includegraphics[width=4.2cm]{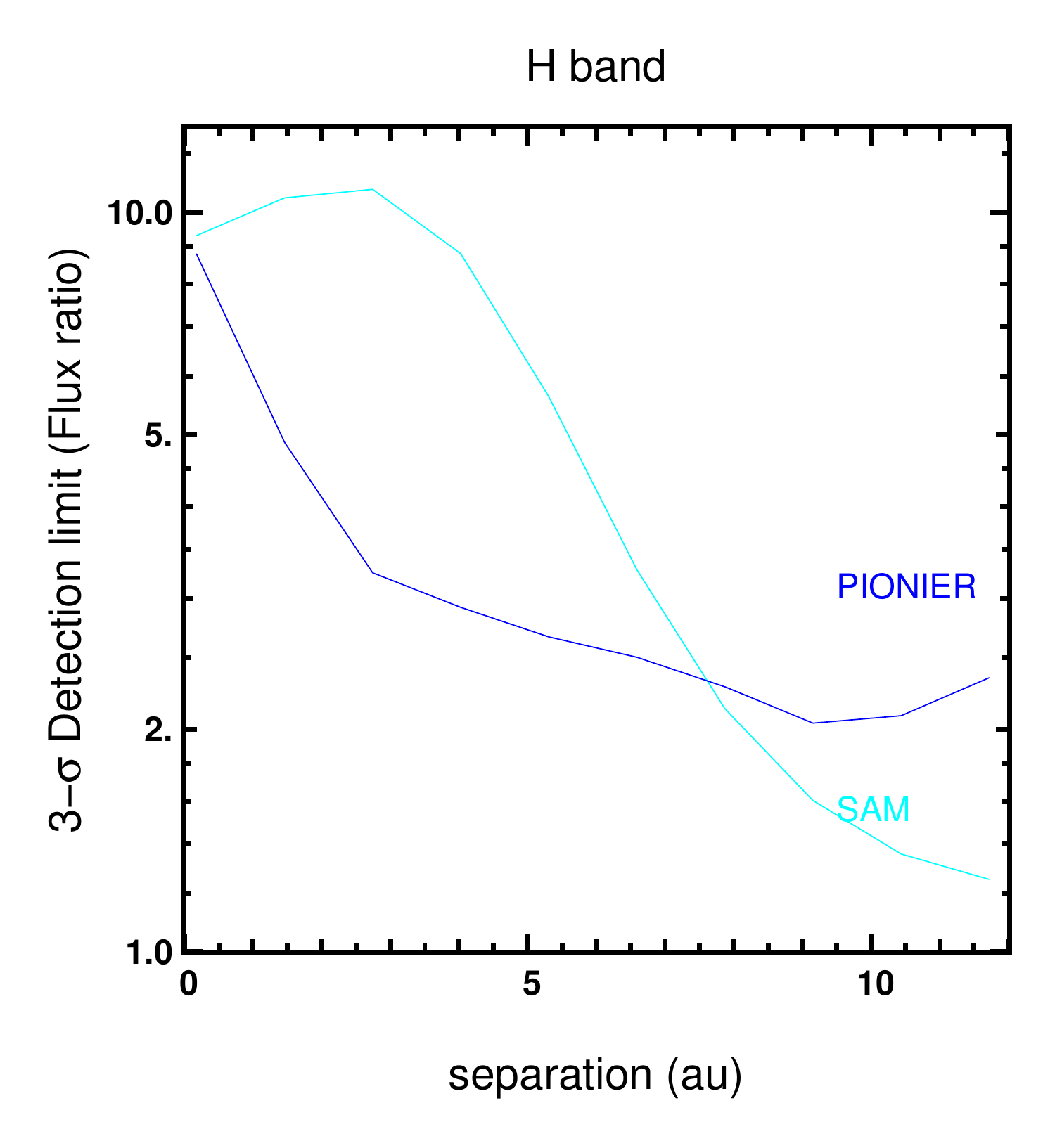}
    \includegraphics[width=4.2cm]{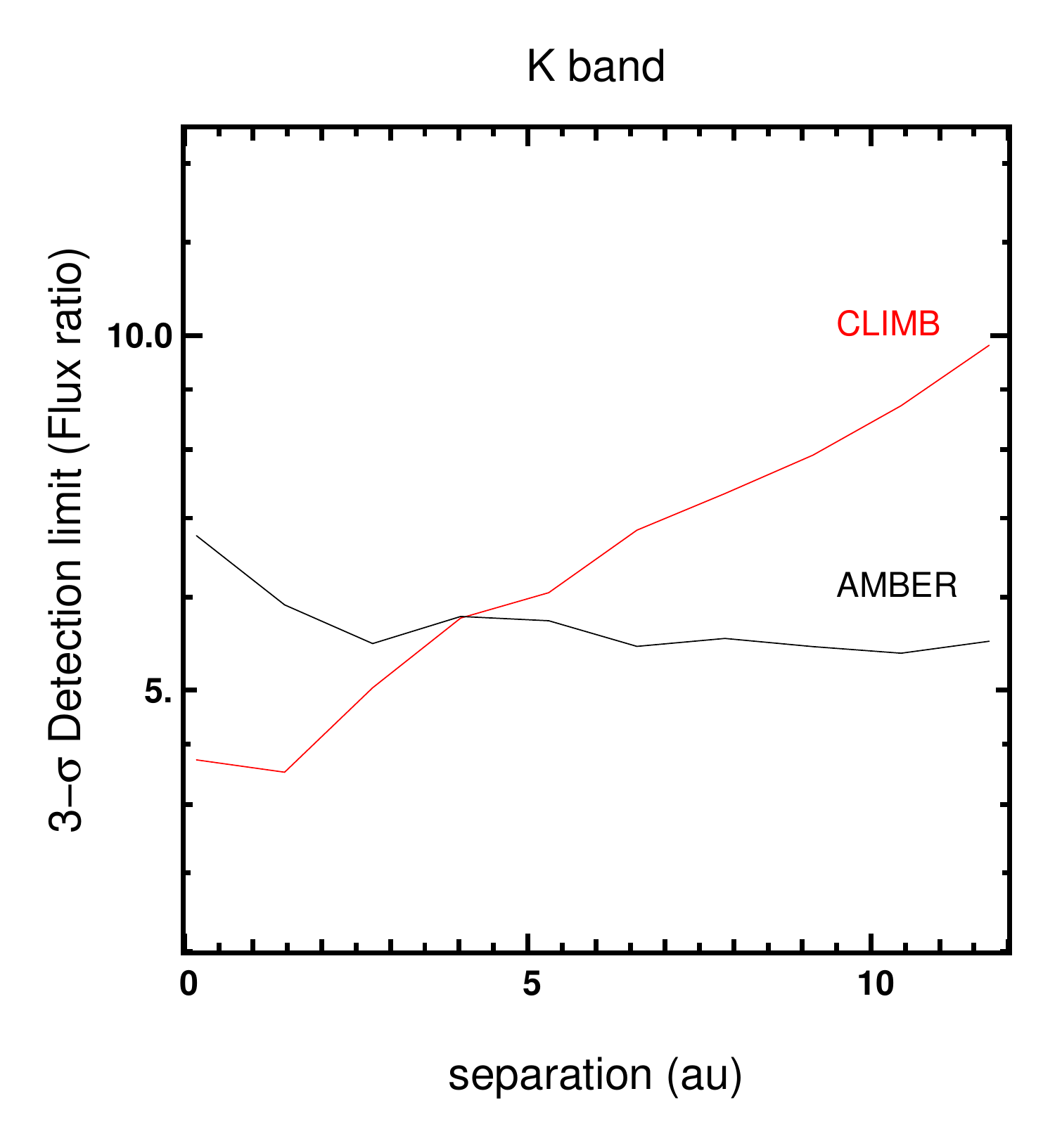}
    \caption{\modif{3-$\sigma$ detection limits averaged on rings oriented in the same way as the disk. Left: $H$-band. Right: $K$-band.}}
    \label{fig:detlim2}
\end{figure}

\modif{In order to understand the disk structure we used the near-infrared interferometric data to search for a companion inside the disk cavity.
We applied the statistical method described in} \citet{Absil2011} \modif{to derive detection limits on each of our interferometric data separately.
In our computation we took into account the bandwidth smearing effect as this effect might affect our longest-baseline data significantly.}
As we are probing the inner regions (\textless 10\,au) around the star, a companion would have moved notably between two observations that were taken several months apart. 
Therefore, we selected the CHARA dataset taken between 2011-06-12 and 2011-08-04, as these observations were taken in a sufficiently short period, but still offer a good {\uv}-plane coverage.

\modif{First, we need to compute the reference model without a companion that will serve as the null hypothesis. 
For SAM data the null scenario is the best fit model from Sect.\,\ref{sec:modelfitting}.
For the PIONIER and AMBER data we fitted a chromatic model including a star and a background with a given temperature. For the CLIMB data we fitted a monochromatic model since there is only one spectral channel.
The best fit parameters are shown in Table\,\ref{tab:nullfit}.}

\begin{table*} 
\begin{center}
\caption{\modif{Best parameters for the null hypothesis} \label{tab:nullfit}}
\begin{tabular}{l|rcl|l|rcl|rcl}
\tableline
\tableline
& \multicolumn{3}{c|}{PIONIER} & & \multicolumn{3}{c|}{AMBER}& \multicolumn{3}{c}{CLIMB}\\
\tableline
$\chi^2$ & & 7.4 & & & & 1.0 & & & 2.7 & \\
\hline
Param. & Value& $\pm$ &Err &  Param. & Value& $\pm$ &Err  & Value& $\pm$ &Err \\
\hline
$F_*^{1.65\mu\mathrm{m}}$ $[$\%$]$& 60.3 & $\pm$ &0.3 & $F_*^{2.13\mu\mathrm{m}}$ $[$\%$]$& 40.9 &$\pm$ & 0.8 & 32.7&$\pm$ &0.1   \\
$T_\mathrm{env} $[$K$]& 1680 & $\pm$ & 79 & $T_\mathrm{env} $[$K$]& 1580&$\pm$ & 144 & & - \\
\tableline
\tableline
\end{tabular}
\end{center}
\end{table*}

\modif{In the second step, we compute a grid on positions and contrasts of a potential companion and determine the significance at the global minimum of the fit to each dataset.}
The $\chi^2$ \modif{improvements compared to the null models} have low significance (\modif{1.8, 1.3, 1.6 and 2.3-$\sigma$ for PIONIER, SAM, AMBER and CLIMB respectively). 
None of the dataset contains a significant detection.}
\modif{Finally,} we derived \modif{the} detection limits at 3-$\sigma$ level for \modif{each} dataset (see Fig.\,\ref{fig:detlim}).
\modif{We also used the disk orientation derived in earlier sections} (using $i$=52.5$^\circ$ and $\theta$=26.4$^\circ$) 
\modif{to represent the detection limit in $H$ and $K$-bands as a function of the physical distance to the central star. (see Fig.\,\ref{fig:detlim2})}.
That results in upper limits of \modif{$\sim$3\% and $\sim$5\% in $H$-band and $K$-band respectively}.





\section{Discussion}
\label{sec:discussion}

Here we discuss the global structure of the disk around MWC~614 as deduced from our multi-wavelength interferometric observations.

\subsection{MIR thermal emission from the inner wall of the outer disk}

The fit of the MIDI visibilities shows that the data is compatible with a Gaussian ring with a radius of 41.8$\pm$0.9\,mas corresponding to 10.2$\pm$0.2\,au. 
This can correspond to the thermal emission from the disk wall.
The Gaussian width of the ring corresponds to 4.5$\pm$2.1\,au.
It can be interpreted as the rim scale-height.
At this distance from the star it translates into a scale-height of $z/r$ of 0.27.
This is rather high compared to other objects morphology where typical values of 0.1$\sim$0.15 are attributed \citep[e.g.][]{Benisty2010,Mulders2013,Matter2016}.
It is therefore more likely that the extension of the emission is due to a radially extended emission, possibly due to a rounded rim \citep{Mulders2013}.

\subsection{Physical origin of the extended NIR emission}
\label{sec:NIRorigin}

\begin{figure}
    \centering
    \includegraphics[width=6.9cm]{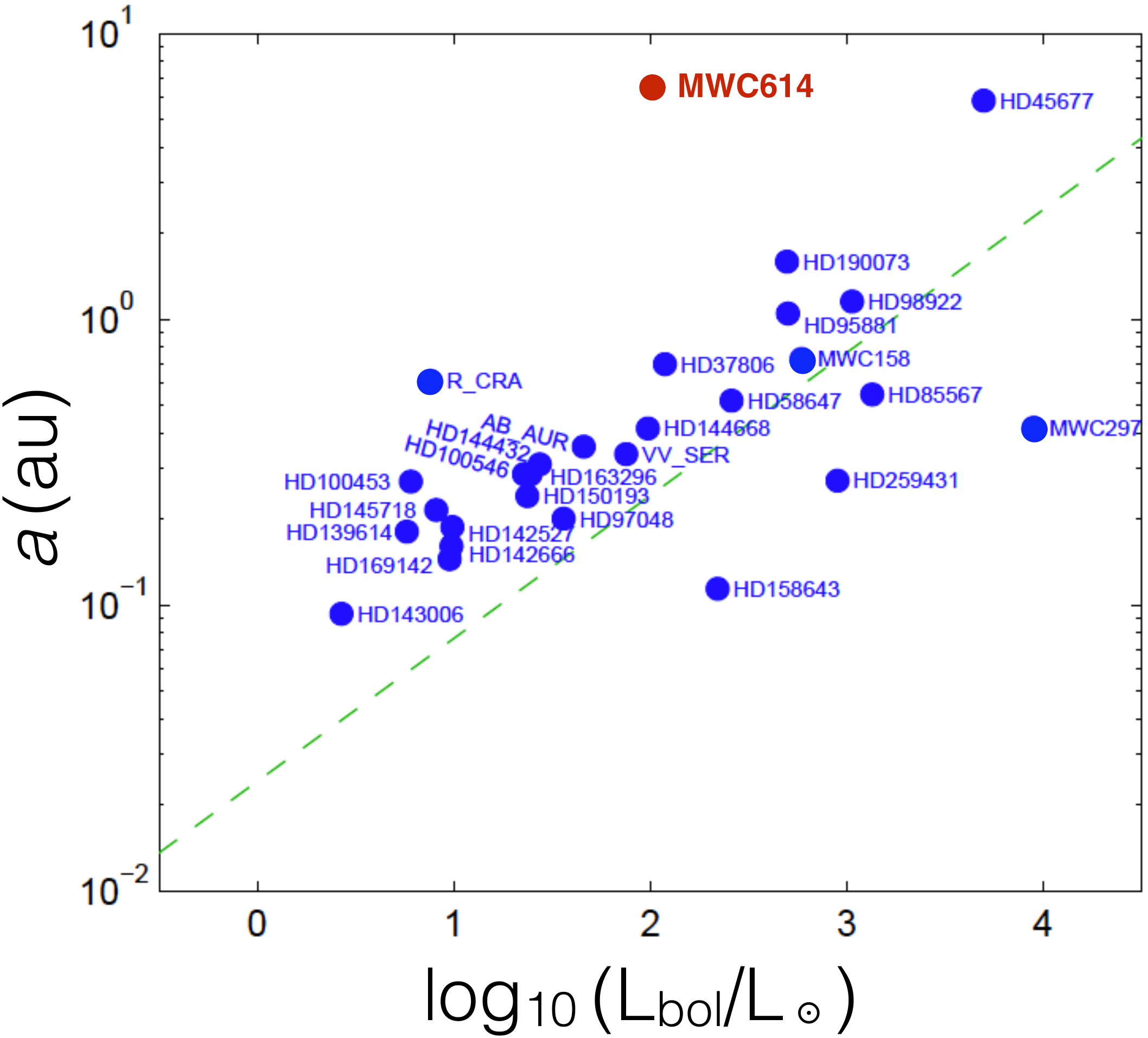}
    \includegraphics[width=6.4cm]{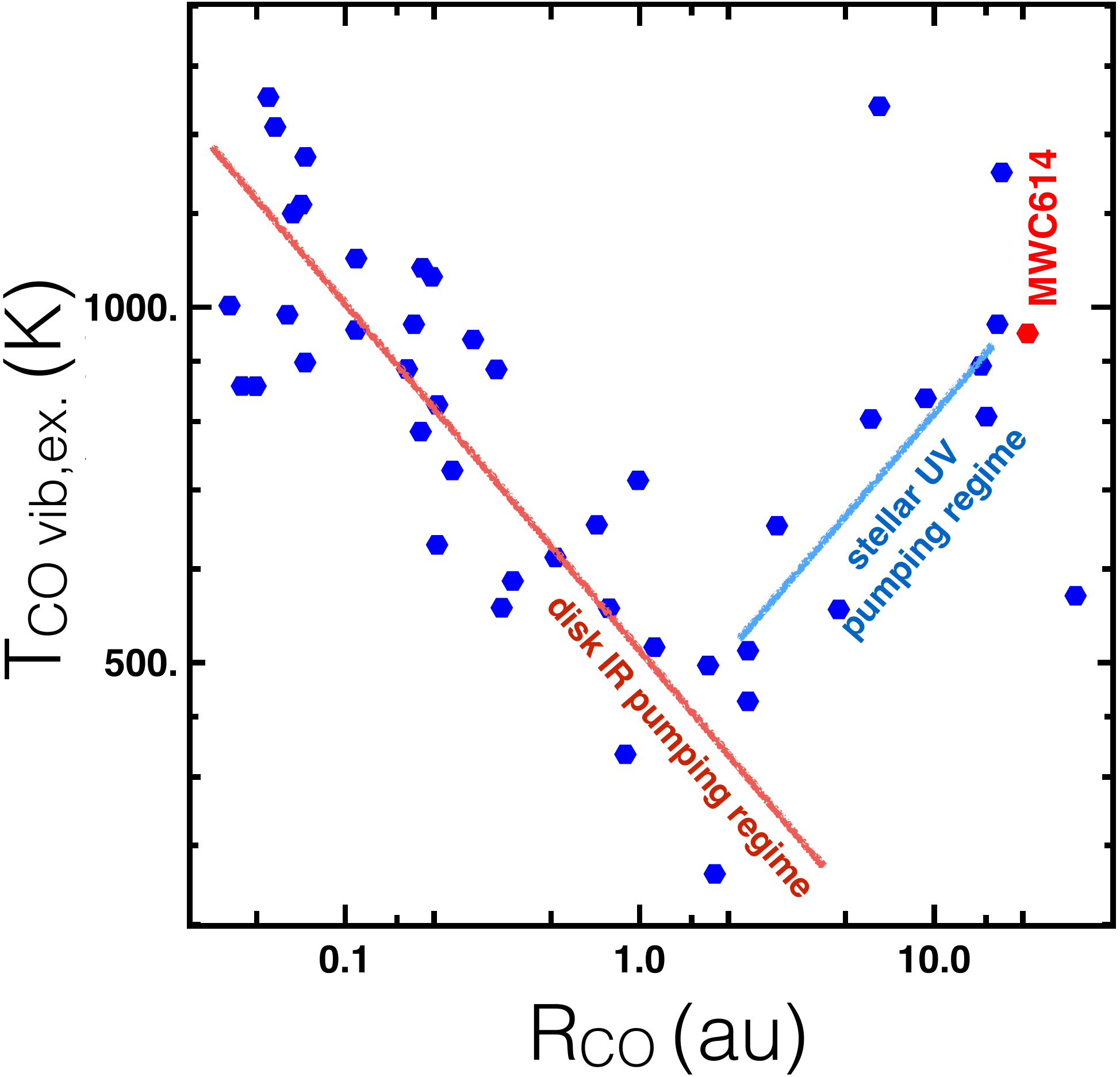}
    \caption{Left: location of MWC~614 in the near infrared size luminosity diagram. The diagram is taken from \citet{Lazareff2016}. The red filled circle is the radial extension of the near infrared flux for MWC~614. The green dashed line corresponds to the theoretical sublimation radius with a sublimation temperature of 1800\,K. Right: Figure adapted from \citet{Banzatti2015} where MWC~614 appears to have a large CO gap and a large CO excitation temperature locating it in the UV-pumping regime (see Section\,\ref{sec:NIRorigin}). }
    \label{fig:SizeLum}
\end{figure}

The FWHM of the $H$-band NIR emission is around 55\,mas, corresponding to about 16\,au. 
Using this size measurement, we plot MWC~614 in the size-luminosity diagram for young stellar objects (Fig.\,\ref{fig:SizeLum}, left).
Earlier surveys revealed that for Herbig Ae/Be stars, the NIR size of most protoplanetary disks scales roughly with the square-root of the stellar luminosity (L$_\odot$), which was interpreted as evidence that this emission traces primarily thermal emission from near the dust sublimation rim \citep[e.g.][]{Monnier2002,Monnier2005,Lazareff2016}.

On the size-luminosity diagram (Fig.\,\ref{fig:SizeLum}) MWC~614 appears as an extreme outlier.
Its radial extension ($w$/2 $\sim$ 8\,au) is 40 times larger \modif{than} the expected size of the dust sublimation radius \citep[0.2\,au, using Eq.\,14 in][assuming a sublimation temperature of $T_\mathrm{sub}$ = 1800\,K, cooling efficiency $\epsilon$ = 1 and a backwarming factor $Q_\mathrm{bw}$ = 1]{Lazareff2016}. 
Using the same equation, at 13\,au the dust should have a temperature of $\sim$350\,K which is not compatible with the temperature we derived from the SAM+PIONIER+AMBER datasets ($T = 1812\pm71$\,K, Sect.\,\ref{sec:modelfitting}).
Therefore, we argue that the NIR emission of MWC~614 is not dominated by thermal emission from the dust sublimation region. 
However, the temperature we derived is similar to a dust sublimation temperature.
As we will show later, a compelling scenario to explain this \modif{particular} feature could be the presence of small particles quantum heated by stellar UV photons up to dust sublimation \modif{temperature}.

MWC~614 belongs to a peculiar group of objects with regard to its CO ro-vibrational emission from fundamental lines near 4.7$\mu$m (mainly for the $^{12}$CO and $^{13}$CO isotopologues):
\citet{vanderPlas2015} and \citet{Banzatti2015} found a single CO component emission with high excitation temperature and estimated that it originates from stellocentric radius of 9.2$\pm$1.5\,au, far from the dust sublimation radius.
In the group of objects presenting similar CO emission line characteristics we can find objects like \object{HD100546} or \object{HD97048} where a gap was detected \citep{Banzatti2015}.
The location of the CO emission lies just inside the inner rim we estimate from the MIDI dataset (12.3$\pm$0.4\,au, Sect.\,\ref{sec:MIDIfit}).
Moreover, \citet{Banzatti2015} argue that an efficient UV pumping mechanism is needed to explain the high excitation temperatures associated with these lines and the measured line ratio between the vibrational $\nu$=1-0 and $\nu$=2-1 $^{12}$CO transitions.
\citet{Thi2013} argued that these conditions can be met in disks with an inner hole, where gas is directly exposed to the stellar radiation (see Fig.\,\ref{fig:SizeLum}, right).

Another \modif{potential} tracer of gaps is the PAH emission.
\citet{Maaskant2014} argued that the level of PAH ionization, measured by the ratio between the PAH line at 6.2$\mu$m and the one at 11.3$\mu$m, can indicate the presence of a gap in the disk.
Using the spectra by \citet{Seok2017}, we computed the PAH ionization ratio I$_{6.2}$/I$_{11.3}$ (which gives the ratio of the equivalent width of the 6.2\,$\mu$m and 11.3\,$\mu$m line) and find a value around 3.
This value is very similar to the one of \object{Oph IRS 48}, whose disk was recently found to feature an inner hole of 55\,au and PAHs emission originating between 11 and 26\,au \citep{Schworer2017}.
This object does not have a significant NIR excess and it seems to be in a more evolved state than MWC~614, being on the \modif{verge} of becoming a transition disk because it has a depleted inner region with a large disk cavity (up to 55\,au) and an emission of very small grains from 11\,au outwards \citep{Schworer2017}.

We propose that the unusually extended NIR emission associated with MWC~614 might be associated with the emission from UV-heated QHPs located in the dust cavity that was resolved with MIDI (Sect.~\ref{sec:MIDIfit}).
\citet{Klarmann2016} simulated the observational characteristics of quantum heated particles in protoplanetary disks. 
They were able to model the over-resolved disk emission associated with the transitional disk of \object{HD100453} by introducing QHP inside the disk gap between 1 and 17\,au.
These authors also found a correlation between the amount of overresolved flux in NIR interferometry data and the luminosity ratio between the PAHs and the UV ($L_\mathrm{PAH}/L_\mathrm{UV}$).
MWC~614 has a comparable luminosity ratio to HD100453 \citep[$L_\mathrm{PAH}/L_\mathrm{UV} = 5.5 \times 10^{-3}$;][]{AckevdA2004} but we show that its extended NIR emission ratio is larger (f$_\mathrm{ext}$/(1-f$_*$) = 100\% compared ot 25\% for HD100453, where f$_\mathrm{ext}$ is the extended-to-total flux fraction and f$_*$ is the stellar-to-total flux fraction).
This flux fraction is inferior to 30\% for all the other objects from the PIONIER interferometric large program \citep{Lazareff2016}.
Therefore, MWC~614 seems to be an extreme case, where the QHP emission dominates the NIR circumstellar flux.
The only object showing an extended structure in the \modif{near-}infrared is \object{Oph IRS 48} \modif{where the NIR flux between 11 and 26\,au} was also interpreted as emission from very small grains (VSGs) or PAHs.

However, for IRS48, the temperatures reached by the VSGs are of the order of few hundreds of Kelvin \citep[between 250 and 550\,K;][]{Schworer2017} far from the temperatures we derived for MWC614.
But the VSGs are located outside the first 10\,au from the central star and IRS48 has a luminosity of 48\,L$_\odot$.
In the case of HD100453, the model assumed that QHPs are distributed above the disk surface (the disk has an inner component) between 1 and 10\,au with a star luminosity of 8.04\,L$_\odot$ \citep[][]{Klarmann2016}.
The QHPs reach temperatures between few hundreds to 2400\,K.
Comparing with MWC614, where the emission is located from the star up to 8\,au and with a stellar luminosity of 100\,L$_\odot$, we can expect even higher temperatures. 
The effective temperatures are therefore likely to reach the 1800\,K we derived from our interferometric measurements.
A full modelling of this emission is needed to confirm our scenario.

\subsection{Ruling out any dust material at the dust sublimation radius}

    \begin{figure}
    \centering
    \includegraphics[width=6.6cm]{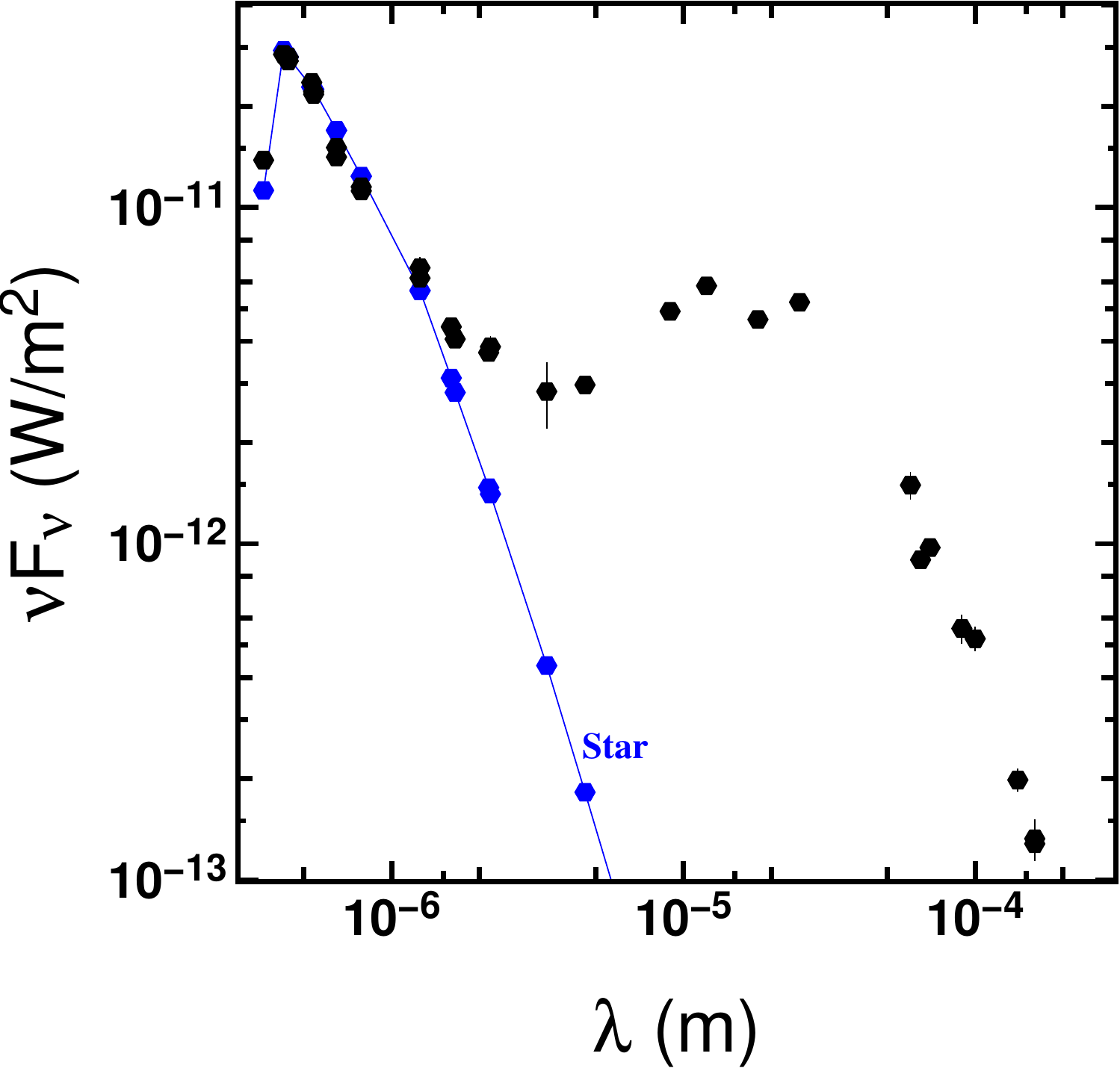}
    \includegraphics[width=6.25cm]{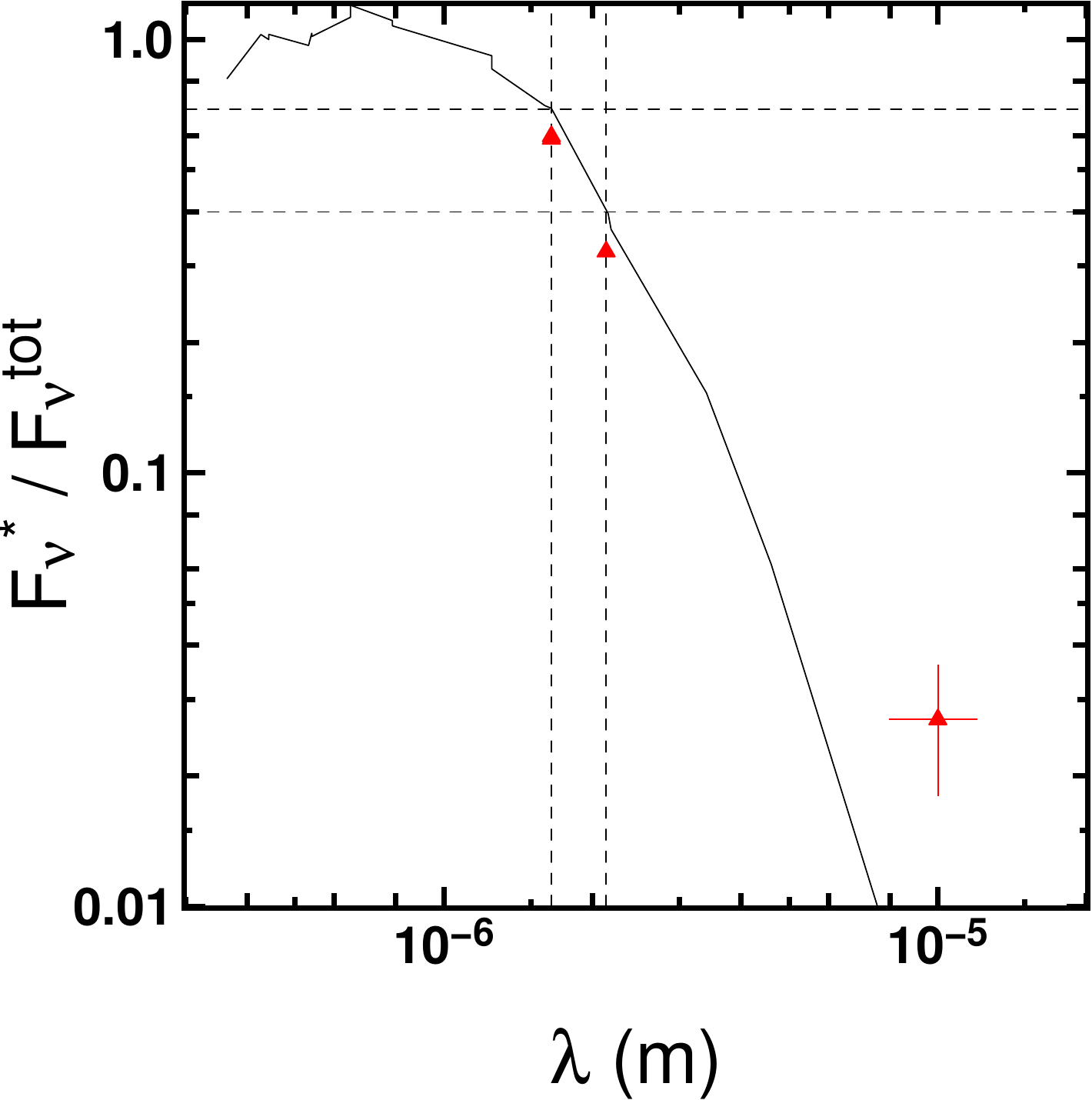}
      \caption{Left: Spectral Energy Distribution (SED) of MWC~614 with the photometric dataset (black) and the stellar photosphere (blue). Right: stellar-to-total flux ratio from the SED (black) and from the fit to the interferometric datasets (red triangles) at 1.65$\mu$m, 2.13$\mu$m and 10$\mu$m. }
         \label{fig:SED}
   \end{figure}

In Sect.\,\ref{sec:NIRorigin}, we estimated the theoretical location of the dust sublimation radius for MWC~614 to be 0.2\,au, which corresponds to an angular scale of around 1\,mas.
This should be barely resolved with the highest spatial frequencies with the VLTI baselines ($\lambda/2B$ = 1.1\,mas) and \modif{well} resolved with the CHARA \modif{baselines} ($\lambda/2B$ = 0.7\,mas).
However, no clear evidence of decreasing visibility with baseline length is found.

An independent way to determine if there is still unresolved circumstellar emission in the interferometric data is to derive the stellar-to-total flux from photometric measurements and compare it with the stellar-to-total flux ratio derived from our interferometric measurements. 

We used photometric measurements from the literature (see Table\,\ref{tab:SED}).
For the photosphere effective temperature we adopted 9500\,K \citep{Montesinos2009}.
We then fitted the photometry in the visible with Kurucz models \citep[][]{Kurucz}.
We found an A$_V$ of 0.45 \citep[using][with R$_V$=3.1]{Cardelli89} that is comparable to previous determinations \citep[A$_V$=0.54;][]{vandenAncker1998}.
The SED and the contribution of the stellar-to-total flux ratio at each wavelength are reported in Fig.\,\ref{fig:SED}.
We found the stellar-to-total flux ratio to be $\sim$69\% at 1.65$\mu$m, and $\sim$40\% at 2.13$\mu$m.

The stellar-to-total flux ratios derived from NIR interferometry are lower than the ones from the SED.
This is likely due to the fact that the fit to the SED includes stellar \modif{light} that is scattered from the disk.
Given that the stellar-to-total flux ratios from the SED are higher that the ones determined from interferometric measurements there is no evidence for the presence of an additional circumstellar component at the dust sublimation radius. 

For the MIR flux, the stellar-to-total flux ratio from the SED is negligible whereas the unresolved flux from the fit to the MIDI data indicates 2.7$\pm$0.9\%. 
However, the fit is not entirely satisfactory for the longest baselines (most sensitive to the contribution of the unresolved flux) and differ\modif{s} from the 0$\%$ flux ratio at a significance of 3-$\sigma$.

Considering all these arguments, \modif{we conclude} that we do not detect a significant contribution to the NIR flux from the dust sublimation region.

\subsection{\modif{Upper mass limits on a companion inside the disk cavity}}
\label{sec:compdiscussion}

\modif{The detection limits we found in Sect.\,\ref{sec:detlim} are defined in proportion of the flux in a given observational band.
Let us translate these values into absolute magnitudes and onto masses using evolutionary tracks.
In $H$-band the detection limit varies between 1 and 3\% between 2.5 and 12\,au from the star} \modif{which translate into absolute magnitudes between 4.3 and 3.2.
In $K$-band the detection limits are flatter and are around 5\% of the $K$-band flux which means an absolute magnitude of 2.0 between 0 and 12\,au from the central star}.

\modif{Using the evolutionary tracks of \citet{Bressan2012} \citep[assuming a solar metallicity and an age of $10^6$\,years;][]{Seok2017} the mass limits we derive are between 0.14 and 0.34\,M$_\odot$ and 0.8\,M$_\odot$ for $H$ and $K$ bands respectively (assuming an A$_V$ of 0.45).
Based on our dataset, we can therefore rule out companions with a mass larger than 0.34\,M$_\odot$ inside the cavity.}

\subsection{Disk clearing mechanism}

\begin{figure}
    \centering
    \includegraphics[width=6cm]{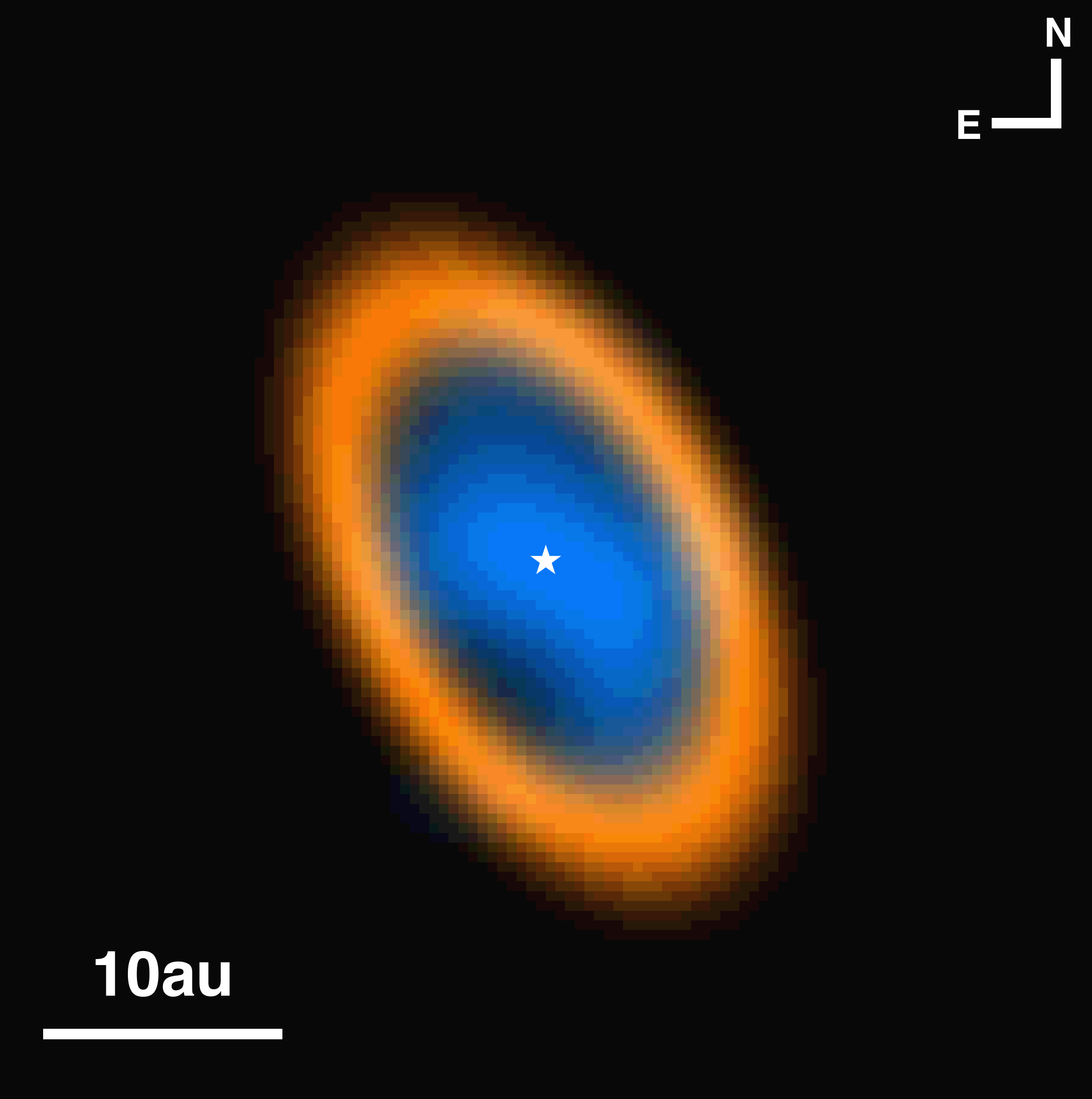}
    \caption{Composite image, including the best-fit MIR $N$-band model image (orange) and the image reconstructed from the NIR $H$-band data (blue). The star is indicated by the white star.  }
    \label{fig:RGB}
\end{figure}

In Fig.\,\ref{fig:RGB} we combine the best-fit model derived from our MIR interferometric data (orange) and with our NIR aperture synthesis image (blue).
It is clear that the NIR flux originates from inside the MIR-emitting ring-like disk structure at $\sim$12.3\,au.

One mechanism that is able to open a disk gap is photo-evaporation.
Photo-evaporation happens when the disk is directly illuminated by UV or X-rays photons from the star.
These disks have low accretion rates as the inner disk is not replenished by the dust from the outer parts \citep{OwenClarke2012}.
The accretion rate for MWC~614 has been estimated to be around 10$^{-7}M{_\odot}$/yr, based on the Balmer break and the Br$_\gamma$ line luminosity \citep{GarciaLopez2006,Donehew2011,Mendigutia2011}.
This high accretion rate is hardly explained by either UV or X-rays photo-evaporation theories alone.

The cavity in the disk around MWC~614 may \modif{also} have been opened by a low-mass companion.
\modif{The presence of a reasonably sized planet \citep[$\sim$5\,M$_J$;][]{Pinilla2012,Owen2014,Zhu2014,Pinilla2016} in a disk gap can lead to dust filtering through the disk.}
The strong pressure gradient at the edge of a planet-opened gap can stop the inward-migration of large dust grains, while allowing smaller particles (\textless1$\mu$m, such as PAH) to pass through \citep{Owen2014}.
As the NIR emission is likely coming from QHPs, this scenario might happen around MWC~614.
The relatively high accretion rate (10$^{-7}\dot{M}$/yr) can replenish the \modif{cavity} and \modif{a 5\,M$_J$} companion could filter the dust sizes letting small particles in the \modif{cavity}.
These particles are directly exposed to stellar UV light or X-rays and are quantum heated resulting in the strong NIR emission that we detect.
A quantitative validation of this scenario requires full radiative-transfer modeling, which is outside the scope of this paper.





%

\section{Conclusion}
\label{sec:conclusion}

\begin{figure}
    \centering
    \includegraphics[width=8.4cm]{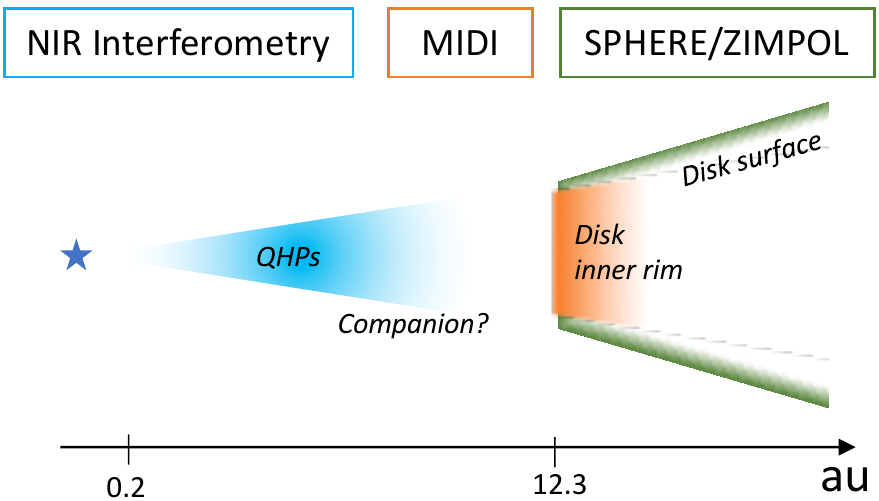}
    \caption{Sketch of MWC~614.
      }
    \label{fig:sketch}
\end{figure}

Our extensive set of high-angular resolution observations revealed the following properties of the surroundings of MWC~614 (see Fig.\,\ref{fig:sketch}):

\begin{itemize}

\item At visual wavelengths, our SPHERE/ZIMPOL polarimetry reveals scattered light from the disk, but does not resolve the inner disk cavity. 
The scattered light geometry is asymmetric, likely tracing the disk inclination.

\item The MIR emission (8-13\,$\mu$m) is confined in a ring with a radius of $12.3 \pm 0.4$\,au from the star, tracing the thermal emission of large dust grains. 
The ring features a relatively sharp inner edge, as indicated by the \modif{pronounced} lobes that we see in the MIDI visibilities.

\item The NIR emission (1.2-2.5\,$\mu$m) is unusually extended (out to $\lesssim 10$\,au) and fills the region inside of the inner disk wall rather homogeneously.
The emission does not trace thermal emission from material at the dust sublimation radius, as found in most other T\,Tauri and Herbig~Ae/Be stars.
Instead, the emission extends over a forty times larger area, indicating that the emission traces a fundamentally different mechanism.
This conclusion is also supported by the high temperature of ($1812\pm 71$\,K) that we deduce for this extended NIR component.
We propose that this emission could trace a population of small quantum-heated particles that might be able to filter through the pressure bump at the inner disk wall revealed by our MIR observations.
A detailed radiative transfer study will be needed to confirm this hypothesis and to develop it further.

\item Our interferometric image reveals an S-shaped asymmetry in the NIR $H$-band emission, indicating that the quantum-heated particles might be dynamically disturbed by a disk-clearing companion.

\item We \modif{determined an upper limit on the mass of a potential companion inside the cavity to 0.34\,M$_\odot$ (between 2 and 12\,au).}

\end{itemize}

\modif{Our study indicates that} the transitional disk around MWC~614 \modif{is} in a very special evolutionary state, where \modif{a} low-mass companion opened a gap in the disk.
The inner disk could have already been accreted onto the star exposing the PAHs of the \modif{cavity} to direct stellar light.
Further observations and characterization of the \modif{disk cavity (precise structure of the near-infrared emission, presence of a companion?)} would confirm such a scenario.

\acknowledgments

The authors acknowledge support from an Marie Sklodowska-Curie CIG grant (Grant No.\ 618910), Philip Leverhulme Prize (PLP-2013-110), STFC Rutherford Fellowship (ST/J004030/1), and ERC Starting Grant (Grant Agreement No.\ 639889). A.A. and \modif{J.D.M.} acknowledge support from NSF AAG 1311698.

The authors wish to recognize and acknowledge the very significant cultural role and reverence that the summit of Mauna Kea has always had within the indigenous Hawaiian community. We are most fortunate to have the opportunity to conduct observations from this mountain.

This work was supported by a NASA Keck PI Data Award, administered by the NASA Exoplanet Science Institute (PID 69/2013B\_N104N2). Data presented herein were obtained at the W.\ M.\ Keck Observatory from telescope time allocated to the National Aeronautics and Space Administration through the agency's scientific partnership with the California Institute of Technology and the University of California. The Observatory was made possible by the generous financial support of the W.\ M.\ Keck Foundation.

This work is based in parts upon observations obtained with the Georgia State University Center for High Angular Resolution Astronomy Array at Mount Wilson Observatory.  The CHARA Array is supported by the National Science Foundation under Grant No. AST-1211929.  Institutional support has been provided from the GSU College of Arts and Sciences and the GSU Office of the Vice President for Research and Economic Development.

This research has made use of the SIMBAD database and the VizieR catalogue access tool, operated at CDS, Strasbourg, France.



Facilities: \facility{VLT/SPHERE}, \facility{VLTI/PIONIER}, \facility{VLTI/AMBER}, \facility{VLTI/MIDI}, \facility{Keck-II/NIRC2}, \facility{CHARA/CLASSIC}, \facility{CHARA/CLIMB}.



\bibliography{biblio}

\appendix

\section{Appendix material}

\subsection{Geometric model to fit the SPHERE image}
\label{app:modelSPHERE}

\modif{Here} we describe the models used in Sec.\,\ref{sec:SPHEREfit}.
To build them we used one general equation and then set some parameters to 0 to obtain the \modif{desired} model:
\begin{eqnarray}
I_\mathrm{G}(x,y) = \frac{f}{\sigma \sqrt{2\pi}}\exp\Big(-\frac{(x'^2+y'^2) - R}{2\sigma^2}\Big) \big[1 + s \sin (\arctan \frac{y'}{x'}) \big],
\end{eqnarray}
where $f$ is the flux in arbitrary units, $R$ is the radius of the ring, $s$ the ring modulation amplitude (between -1 and 1), $\sigma = \frac{w}{2\sqrt{2\log2}}$ with $w$ being the full width half maximum and $x'$ and $y'$ defined as:
\begin{eqnarray}
x" &=& x - x_*\\
y" &=& y - y_*\\
x' &=& x" \cos \theta + y" \sin \theta \\
y' &=& (y" \cos \theta - x" \sin \theta) / \cos i
\end{eqnarray}
with $x_*$ and $y_*$ being the coordinates of the \modif{center} of the Gaussian, $\theta$ being the position angle of the Gaussian and $i$ the inclination with respect to the line of sight.

The Gaussian \modif{is centered} when $R$, $x_*$, $y_*$ and $s$ are set to 0, the Gaussian is \modif{off-center} when $R$ and $s$ are set to 0. \modif{We obtain a} Gaussian ring when $x_*$ and $y_*$ are set to 0.

\subsection{Visibility normalization}
\label{sec:V2cal}

The SAM data shows a relative drop in the visibilities but their absolute value is difficult to calibrate.
Having the VLTI/PIONIER dataset in-hand we calibrated the SAM data to the long-baseline interferometric data by minimizing the $\chi^2$ with model fitting in order to account for the differences in spectral channels and the orientation of the observed object.
This method assumes that there is continuity between the V$^2$ from aperture masking measurements and long baselines interferometry.

Because the visibility curve does not show any Bessel lobes with spatial frequency, we used a simple model of a star and an oriented Gaussian with 5 free parameters, namely the stellar-to-total flux ratio ($f$), the FWHM size of the Gaussian ($w$), its inclination ($i$) and position angle ($\theta$).
The star can be shifted with respect to the Gaussian using two parameters (x$_s$ and y$_s$).
The NIRC2 data was recorded with a $H$ broadband filter, while the PIONIER data is spectrally dispersed, covering the $H$-band with 3 channels.
Therefore, we need to parametrize the spectral dependence between the PIONIER channels in our model, which we implement by associating the circumstellar Gaussian component with a temperature $T$.
A complete description of the model can be found in the Appendix\,\ref{app:modelVis} (here $x_s$ and $y_s$ equal 0).
We performed the model fitting on the squared visibilities only.

In the rest of the paper we will use the renormalised aperture masking visibilities. 

\subsection{Geometric model used in interferometric data modeling}
\label{app:modelVis}

The model is described with two components: the disk which is an inclined Gaussian and a star which is a point source.
The inclined Gaussian is geometrically described by its FWHM ($w$), its inclination ($i$), Position Angle ($\theta$) as follows:
\begin{eqnarray}
V^\mathrm{g} (\nu') &=& \exp\Big(-\frac{(\pi w \nu')^2}{4 \log2}\Big),
\end{eqnarray}
where $V^\mathrm{g} (\nu')$ is the visibility of the Gaussian and $\nu' = \sqrt{u'^2+\mathrm{v}'^2}$ are the spatial frequencies from the \uv points oriented with respect to the object such as :
\begin{eqnarray}
u' &=& u \cos \theta + \mathrm{v} \sin \theta\\
\mathrm{v}' &=& (-u \sin \theta + \mathrm{v} \cos \theta) \cos i,
\end{eqnarray}
with $\nu=\sqrt{u^2 + \mathrm{v}^2}$ the spatial frequencies from the \uv.

The star is unresolved and can be shifted w.r.t. the center of the Gaussian by $x_*$ and $y_*$: 
\begin{eqnarray}
V^\mathrm{s} (\nu'') & = & \exp(-2 i \pi \nu''),
\end{eqnarray}
where $V^\mathrm{s} (\nu'')$ is the visibility of the star and $\nu'' = u''+\mathrm{v}''$ are the spatial frequencies from the \uv points shifted with respect to the Gaussian such as :
\begin{eqnarray}
\mathrm{u}'' &=& \mathrm{u} x_*  \\
\mathrm{v}'' &=& \mathrm{v} y_*
\end{eqnarray}

\modif{Therefore,} the total visibilities are:
\begin{eqnarray}
V^\mathrm{tot}(\nu, \lambda) = \frac{f^\mathrm{g}_{\lambda} V^\mathrm{g} (\nu) + f^\mathrm{s}_{\lambda} V^\mathrm{s} (\nu)}{f^\mathrm{g}_{\lambda} +f^\mathrm{s}_{\lambda}},
\end{eqnarray}
with $f^\mathrm{s}_{\lambda}$ the stellar to total flux ratio and $f^{g}_{\lambda}$ the Gaussian-to-total flux ratio.
To define them, we have taken a reference wavelength $\lambda_0 = 1.65\mu$m (center of the $H$-band). 
We assumed the star ($f^\mathrm{s}_{\lambda}$) is in the Rayleigh-Jeans regime ($f^\mathrm{s}_{\lambda} = sfr0 (\frac{\lambda}{\lambda_0})^{-4}$) and that the Gaussian ($f^\mathrm{g}_{\lambda}$) has its own temperature so that its flux is scaled accordingly : $f_{\lambda} = f_0 \frac{\mathrm{B}(\lambda,T)}{\mathrm{B}(\lambda_0,T)}$. 
At $\lambda_0$, the sum of the stellar flux ratio (sfr0), the extended flux ratio and the inclined Gaussian flux ratio (dfr0) equals unity.

\subsection{Parametric model results}
\label{app:paramModel}

\begin{figure*}
   \centering
   \includegraphics[width=6cm]{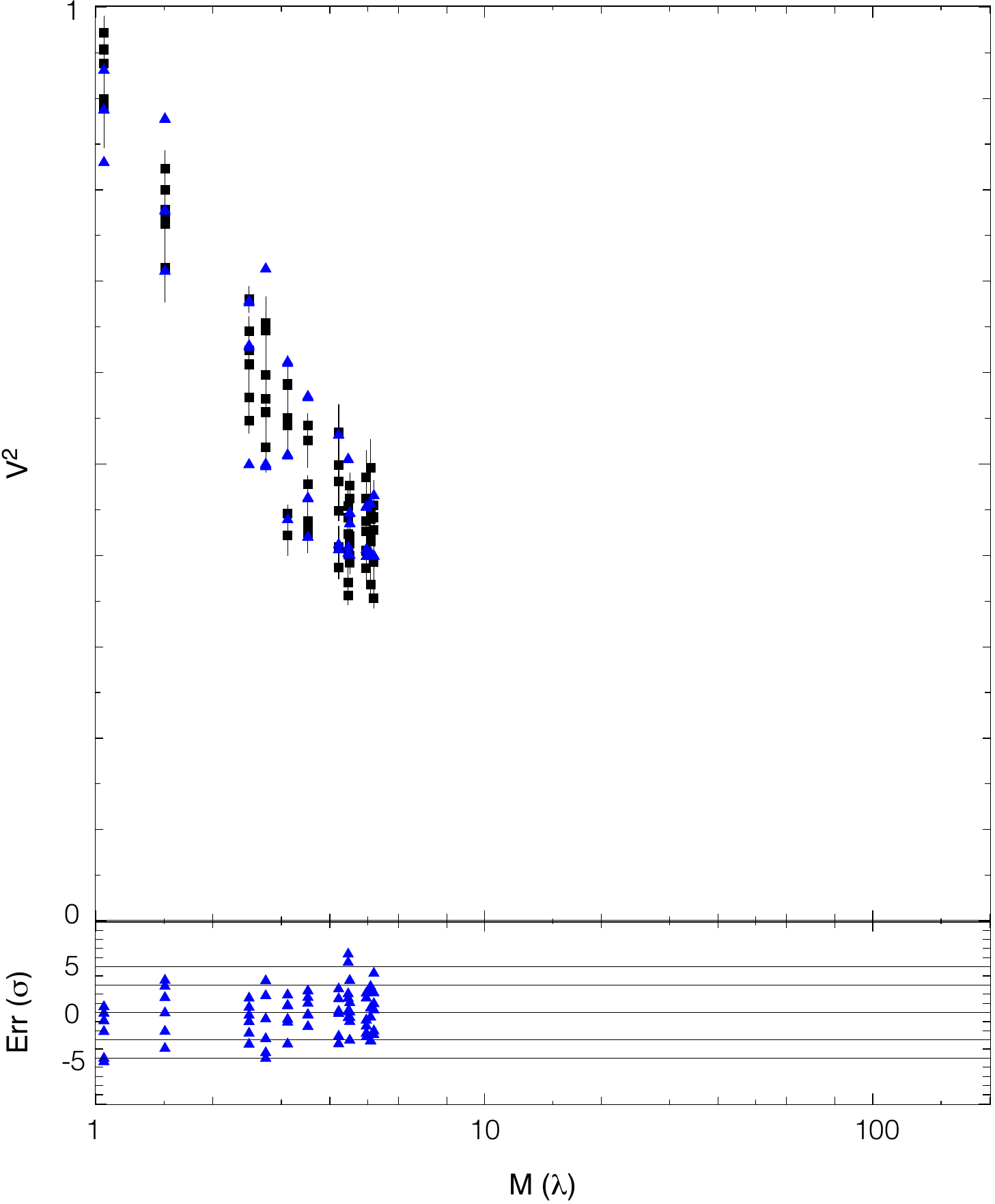}
    \includegraphics[width=6cm]{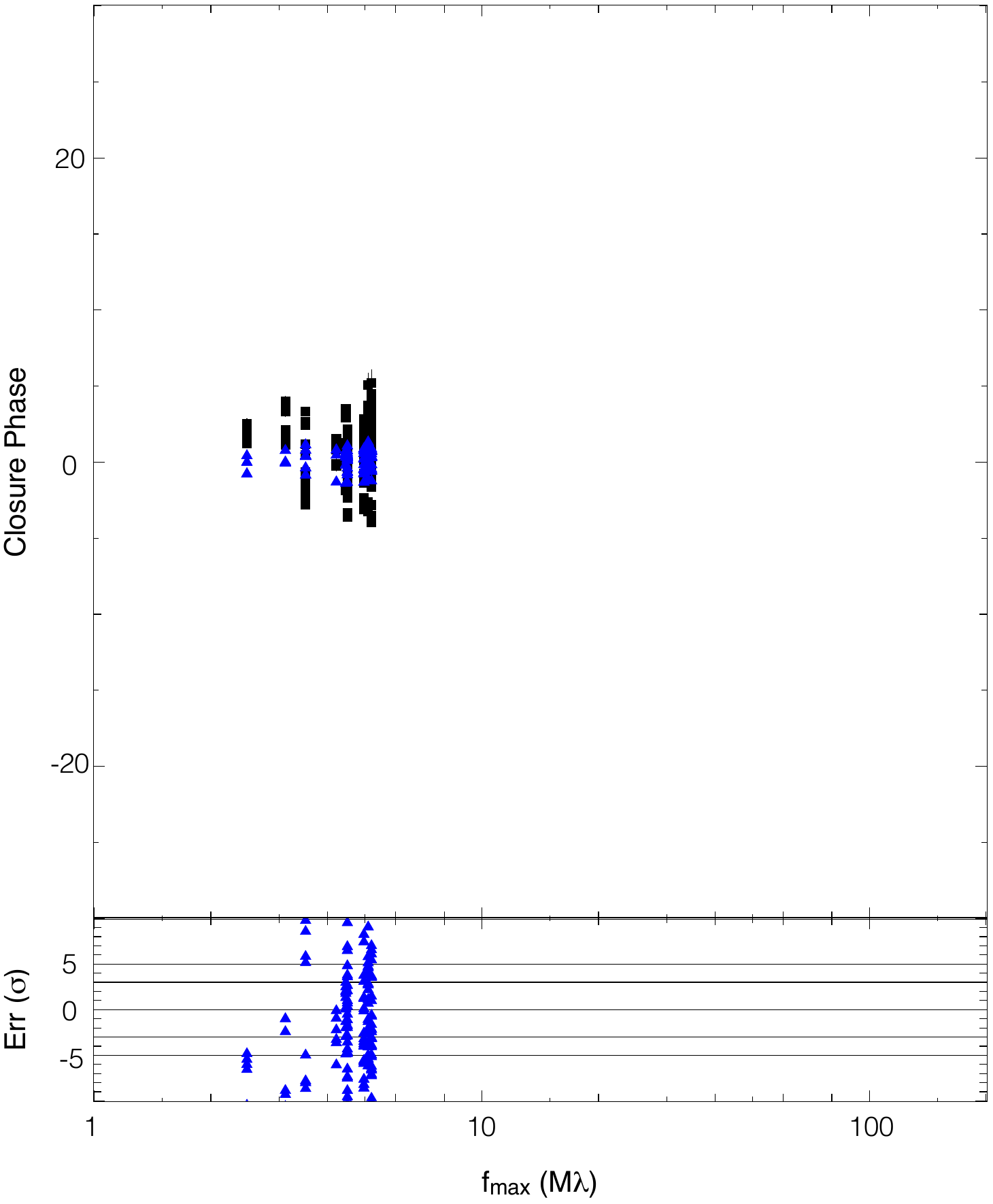}
    \includegraphics[width=6cm]{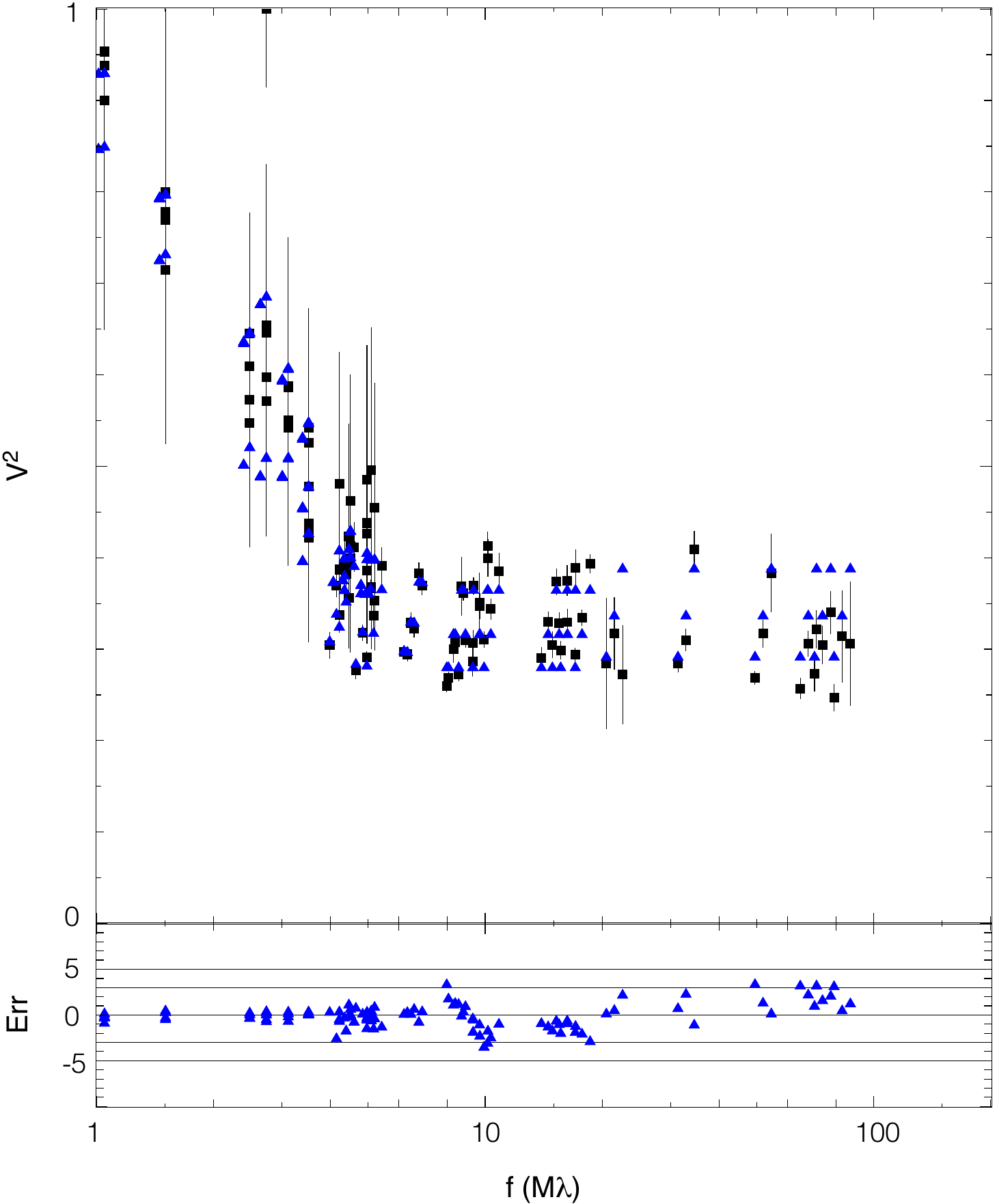}
    \includegraphics[width=6cm]{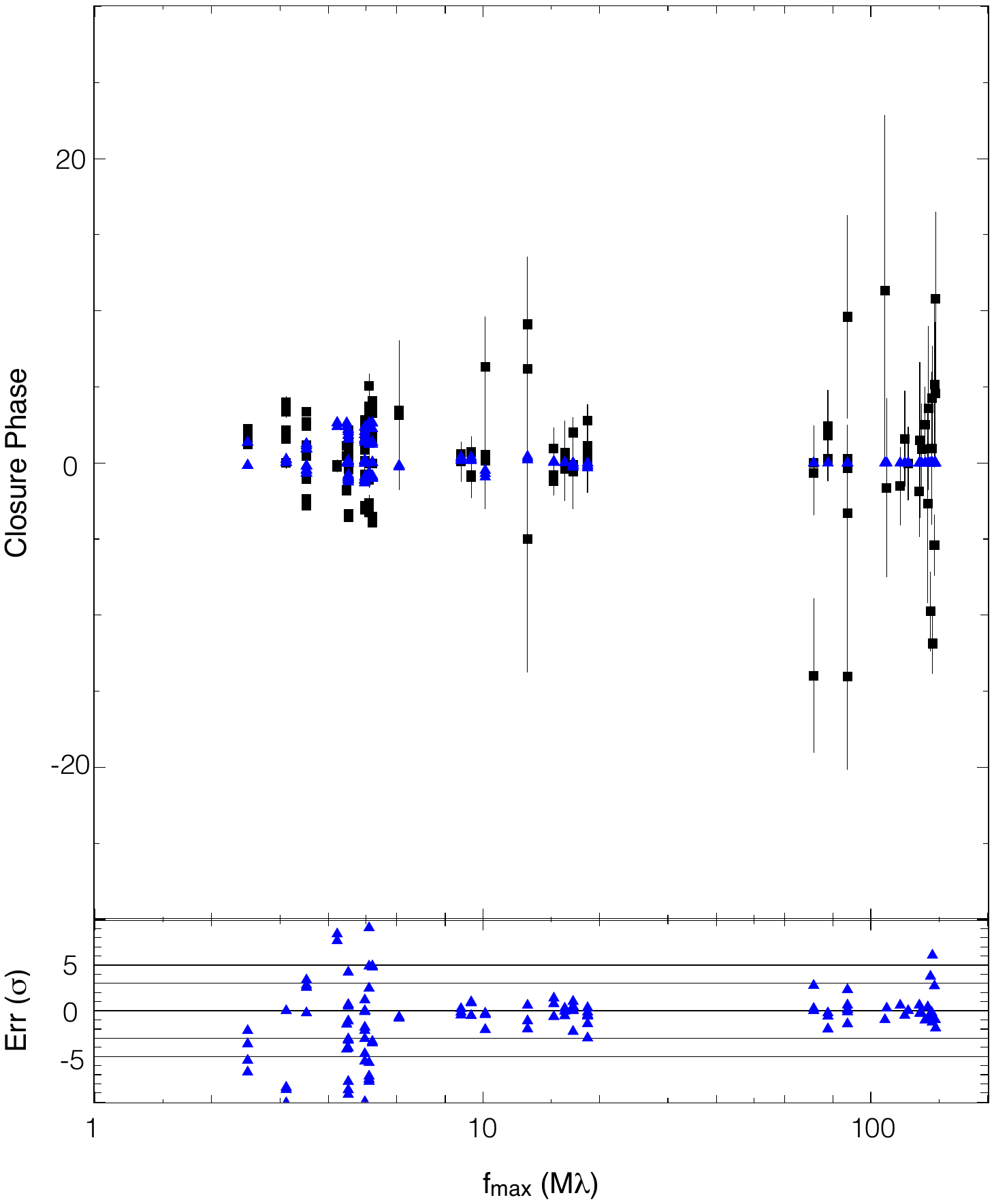}
    \includegraphics[width=6cm]{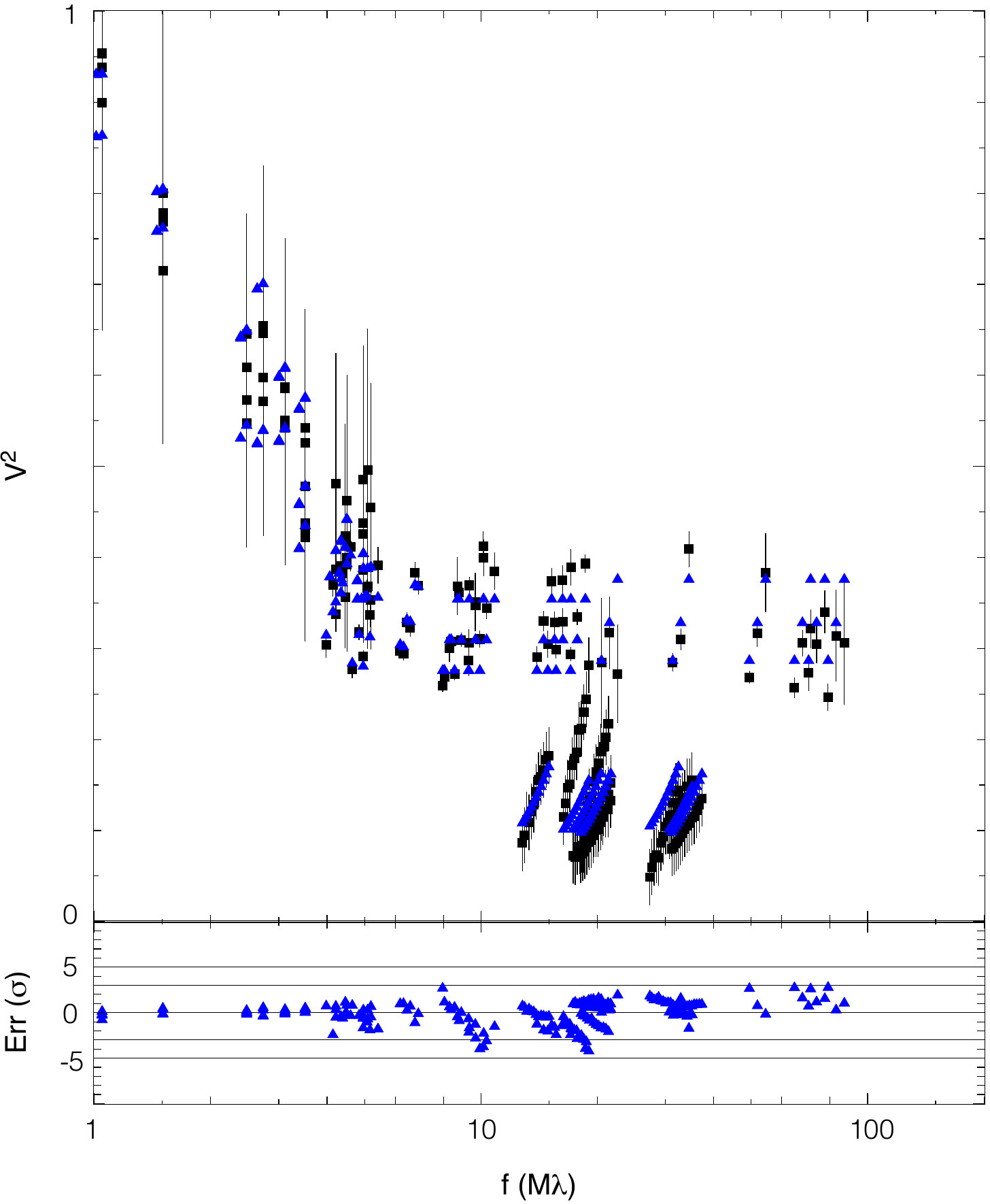}
    \includegraphics[width=6cm]{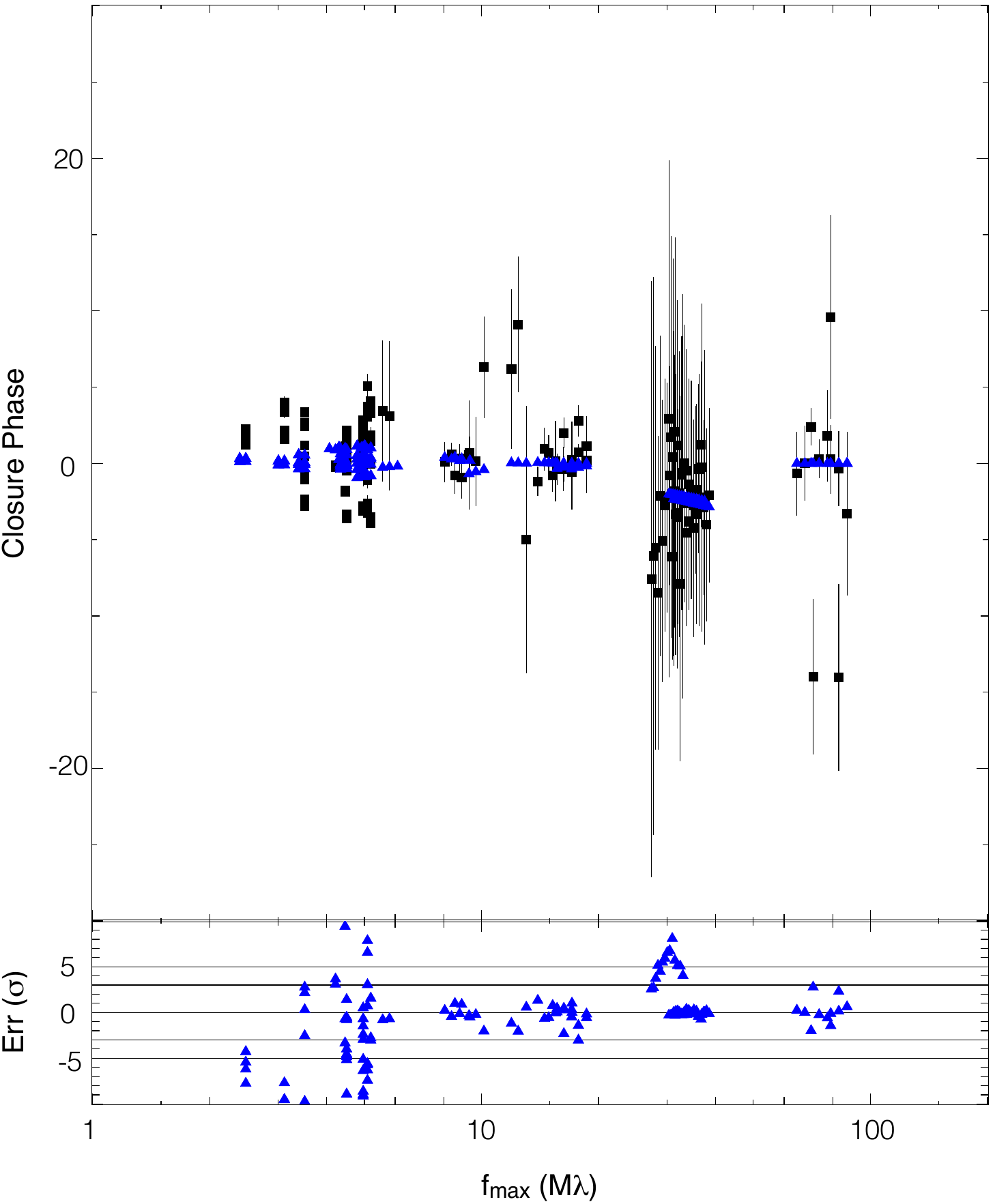}
      \caption{Best fit models to SAM (top), SAM+PIONIER (middle), SAM+PIONIER+AMBER (bottom) datasets. Left: the fit to the V$^2$. Right: Fit to the CPs. The dataset is in black squares, the model is the blue triangles. Below each plot there is the plot of the residuals.
         \label{fig:fitresV2CP}}
   \end{figure*}
   
   \begin{figure*}
   \centering
    \includegraphics[width=6cm]{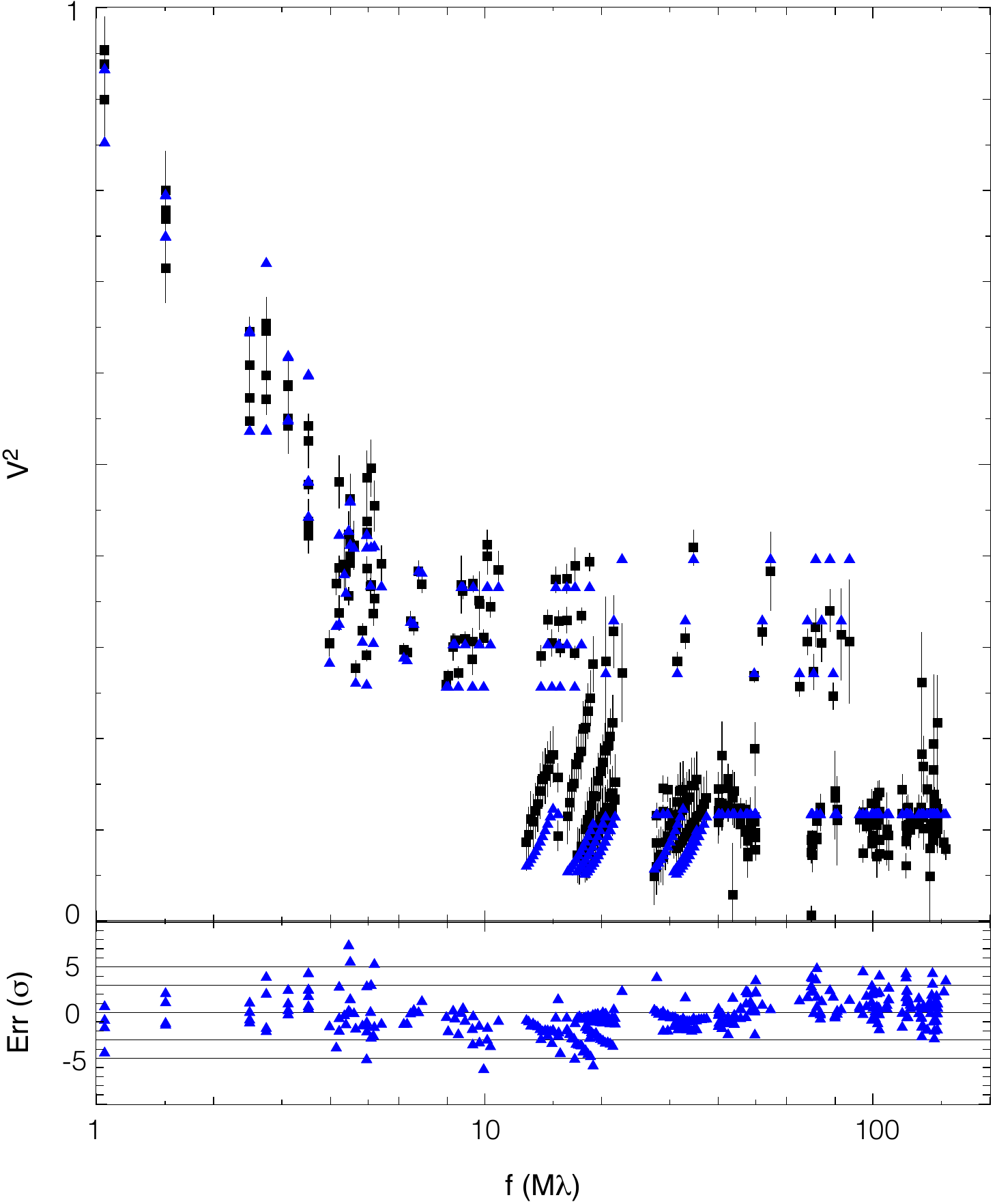}
    \includegraphics[width=6cm]{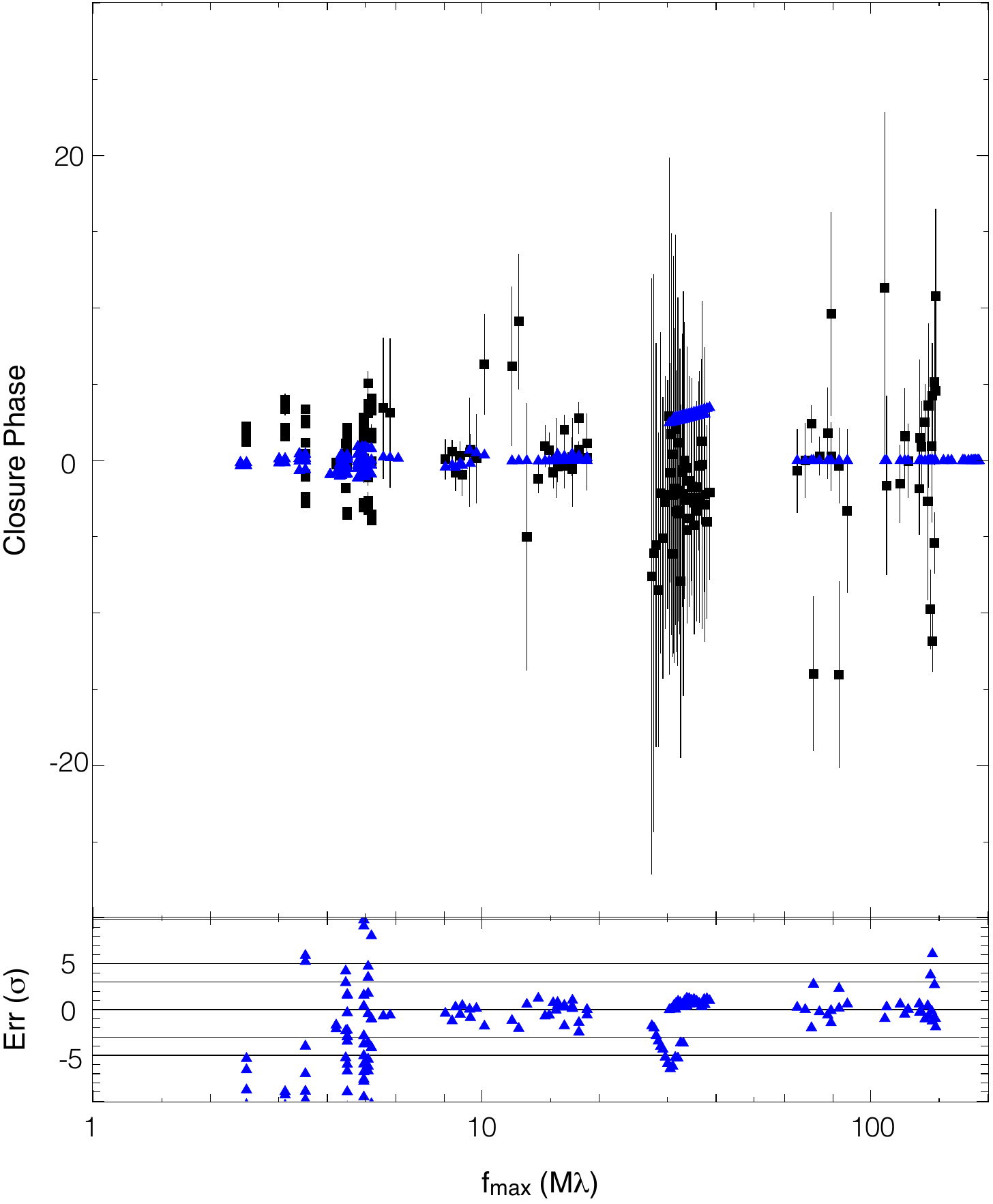}
      \caption{Best fit models to the SAM+PIONIER+AMBER+CHARA dataset. Left: the fit to the V$^2$. Right: Fit to the CPs. The dataset is in black squares, the model is the blue triangles. Below each plot there is the plot of the residuals.
         \label{fig:fitresV2CPCHARA}}
   \end{figure*}

\begin{table}[ht]
\caption{Photometry of MWC~614 \label{tab:SED}}
\begin{center}
\begin{tabular}{rrrrrr}
$\lambda_\mathrm{eff}$ & Photometric band & $\lambda$ F$_\mathrm{lambda}$ & error & $F_*^{\lambda_\mathrm{eff}}$ &Ref. \\
m & - & W.m$^{-2}$ & W.m$^{-2}$ & [\%] &- \\
\hline
3.64e-07 & Johnson:U & 1.3829e-11 & & 168 & \citet{2003AJ....126.2971V}\\
4.26e-07 & HIP:BT & 2.8533e-11 & 4.21e-13  & 106 & \citet{2000AA...355L..27H}\\
4.42e-07 & Johnson:B & 2.8024e-11 & & 107 & \citet{2003AJ....126.2971V}\\
4.42e-07 & Johnson:B & 2.7260e-11 & 1.26e-12 & 104 & \citet{2008ApJ...689..513T}\\
5.32e-07 & HIP:VT & 2.3532e-11 & 2.38e-13  & 93 & \citet{2000AA...355L..27H}\\
5.4e-07 & Johnson:V & 2.16986e-11 & & 96 & \citet{2003AJ....126.2971V}\\
5.4e-07 & Johnson:V & 2.2102e-11 & 8.14e-13  & 99 & \citet{2008ApJ...689..513T}\\
6.47e-07 & Cousins:Rc & 1.5070e-11 & & 108 & \citet{2003AJ....126.2971V}\\
6.47e-07 & Cousins:Rc & 1.4129e-11 & 7.81e-13 & 115 & \citet{2008ApJ...689..513T}\\
7.865e-07 & Cousins:Ic & 1.1202e-11 & & 98 & \citet{2003AJ....126.2971V}\\
7.865e-07 & Cousins:Ic & 1.1516e-11 & 4.24e-13  & 95 & \citet{2008ApJ...689..513T}\\
1.25e-06 & 2MASS:J & 6.1665e-12 & 1.14e-13  & 68 & \citet{2003yCat.2246....0C}\\
1.25e-06 & Johnson:J & 6.616e-12 & 4.88e-13  & 72 & \citet{2008ApJ...689..513T}\\
1.6e-06 & Johnson:H & 4.416e-12 & 2.85e-13  & 54 & \citet{2008ApJ...689..513T}\\
1.65e-06 & 2MASS:H & 4.0597e-12 & 9.72e-14  & 55 & \citet{2003yCat.2246....0C}\\
2.15e-06 & 2MASS:K & 3.7031e-12 & 6.14e-14 & 29 & \citet{2003yCat.2246....0C}\\
2.18e-06 & Johnson:K & 3.8522e-12 & 2.84e-13  & 27 & \citet{2008ApJ...689..513T}\\
3.4e-06 & WISE:W1 & 2.8356e-12 & 6.231e-13 & 11 & \citet{2012yCat.2311....0C} \\
4.6e-06 & WISE:W2 & 2.967e-12 & 1.4542e-14 & 5 & \citet{2012yCat.2311....0C} \\
9e-06 & AKARI:S9W & 4.910e-12 & 4.6634e-14  & 0 & \citet{2010AA...514A...1I}\\
1.2e-05 & IRAS:12 & 5.846e-12 & 2.9229e-13 & 0 & \citet{1994yCat.2125....0J}\\
1.8e-05 & AKARI:L18W & 4.6434e-12 & 4.913e-14  & 0  & \citet{2010AA...514A...1I}\\
2.5e-05 & IRAS:25 & 5.2284e-12 & 2.091e-13 & 0 & \citet{1994yCat.2125....0J}\\
6e-05 & IRAS:60 & 1.4940e-12 & 1.345e-13 & 0 & \citet{1994yCat.2125....0J}\\
6.5e-05 & AKARI:N60 & 8.9615e-13 & 4.109e-14  & 0 & \citet{2010yCat.2298....0Y}\\
7e-05 & Herschel:PACS:F70 & 9.7389e-13 & 4.882e-14  & 0 & \citet{2016AA...586A...6P}\\
9e-05 & AKARI:WIDE-S & 5.5961e-13 & 5.1964e-14  & 0 & \citet{2010yCat.2298....0Y}\\
0.0001 & IRAS:100 & 5.2164e-13 & 4.173e-14 & 0 & \citet{1994yCat.2125....0J}\\
0.00014 & AKARI:WIDE-L & 1.9836e-13 & 1.477e-14  & 0 & \citet{2010yCat.2298....0Y}\\
0.00016 & AKARI:N160 & 1.3262e-13 & 1.810e-14  & 0 & \citet{2010yCat.2298....0Y}\\
0.00016 & Herschel:PACS:F160 & 1.2816e-13 & 6.371e-15  &0 & \citet{2016AA...586A...6P}
\end{tabular}
\end{center}
\end{table}

\clearpage

\end{document}